\documentclass[12pt]{article}
\usepackage{eqsection,subeqnarray,indent,amsfonts,amssymb}
\usepackage{pstricks,pst-node,pst-text,pst-3d,pst-plot}
\usepackage{amsmath}     
\usepackage{bm}          
\usepackage{graphicx}
\usepackage{cite}
\usepackage{url}

\def\today{March 21, 2013 \\
           revised July 11, 2013} 
\begin{document}

\bibliographystyle{plain}

\date{\today}

\title{A generalized Beraha conjecture for non-planar graphs}

\author{
  {\small Jesper Lykke Jacobsen${}^{1,2}$
          and Jes\'us Salas${}^{3,4}$}                     \\[1mm]
  {\small\it ${}^1$Laboratoire de Physique Th\'eorique,
                          \'Ecole Normale Sup\'erieure}  \\[-0.2cm]
  {\small\it 24 rue Lhomond, 75231 Paris, FRANCE}        \\[-0.2cm]
  {\small\tt JESPER.JACOBSEN@ENS.FR}                     \\[1mm]
  {\small\it ${}^2$Universit\'e Pierre et Marie Curie,
             4 place Jussieu, 75252 Paris, FRANCE}       \\[1mm]
  {\small\it ${}^3$Grupo de Modelizaci\'on, Simulaci\'on Num\'erica 
                   y Matem\'atica Industrial}  \\[-0.2cm]
  {\small\it Universidad Carlos III de Madrid} \\[-0.2cm]
  {\small\it Avda.\  de la Universidad, 30}    \\[-0.2cm]
  {\small\it 28911 Legan\'es, SPAIN}           \\[-0.2cm]
  {\small\tt JSALAS@MATH.UC3M.ES}              \\[1mm]
  {\small\it ${}^4$Grupo de Teor\'{\i}as de Campos y F\'{\i}sica
             Estad\'{\i}stica}\\[-2mm]
  {\small\it Instituto Gregorio Mill\'an, Universidad Carlos III de
             Madrid}\\[-2mm]
  {\small\it Unidad Asociada al IEM-CSIC}\\[-2mm]
  {\small\it Madrid, SPAIN}           \\[-2mm]
  {\protect\makebox[5in]{\quad}}  
  \\
}

\maketitle
\thispagestyle{empty}   

\begin{abstract}
We study the partition function $Z_{G(nk,k)}(Q,v)$ of the $Q$--state 
Potts model on the family of (non-planar) generalized Petersen graphs 
$G(nk,k)$. We study its zeros in the plane $(Q,v)$ for $1\le k \le 7$.  
We also consider two specializations of $Z_{G(nk,k)}$, namely
the chromatic polynomial $P_{G(nk,k)}(Q)$ (corresponding to $v=-1$), 
and the flow polynomial $\Phi_{G(nk,k)}(Q)$ (corresponding to $v=-Q$).  
In these two cases, we study their zeros in the complex $Q$--plane for  
$1 \le k\le 7$.  
We pay special attention to the accumulation loci of the corresponding 
zeros when $n\to\infty$. We observe that the Berker--Kadanoff phase that is
present in two--dimensional Potts models, also exists for non-planar recursive
graphs. Their qualitative features are the same; but the main difference 
is that the role played by the Beraha numbers for
planar graphs is now played by the non-negative integers for non-planar graphs.
At these integer values of $Q$, there are massive eigenvalue cancellations,
in the same way as the eigenvalue cancellations that happen at the Beraha 
numbers for planar graphs. 
\end{abstract} 

\medskip
\noindent
{\bf Key Words:}
Potts model; 
non-planar graphs;
Beraha conjecture; 
generalized Petersen graphs; 
transfer matrix; 
Berker-Kadanoff phase.  

\clearpage

\newcommand{\be}{\begin{equation}}
\newcommand{\ee}{\end{equation}}
\newcommand{\ba}{\begin{subeqnarray}}
\newcommand{\ea}{\end{subeqnarray}}
\newcommand{\<}{\langle}
\renewcommand{\>}{\rangle}
\newcommand{\widebar}{\overline}
\def\reff#1{(\protect\ref{#1})}
\def\spose#1{\hbox to 0pt{#1\hss}}
\def\ltapprox{\mathrel{\spose{\lower 3pt\hbox{$\mathchar"218$}}
 \raise 2.0pt\hbox{$\mathchar"13C$}}}
\def\gtapprox{\mathrel{\spose{\lower 3pt\hbox{$\mathchar"218$}}
 \raise 2.0pt\hbox{$\mathchar"13E$}}}
\def\textprime{${}^\prime$}
\def\proof{\par\medskip\noindent{\sc Proof.\ }}
\def\qed{\hbox{\hskip 6pt\vrule width6pt height7pt depth1pt \hskip1pt}\bigskip}
\def\proofof#1{\bigskip\noindent{\sc Proof of #1.\ }}
\def\half{ {1 \over 2} }
\def\third{ {1 \over 3} }
\def\twothird{ {2 \over 3} }
\def\smfrac#1#2{\textstyle\frac{#1}{#2}}
\def\smhalf{ \smfrac{1}{2} }
%
%
\newcommand{\card}{\mathop{\rm card}\nolimits}
\renewcommand{\dim}{\mathop{\rm dim}\nolimits}
\newcommand{\Tr}{\mathop{\rm Tr}\nolimits}
\newcommand{\real}{\mathop{\rm Re}\nolimits}
\renewcommand{\Re}{\mathop{\rm Re}\nolimits}
\newcommand{\imag}{\mathop{\rm Im}\nolimits}
\renewcommand{\Im}{\mathop{\rm Im}\nolimits}
\newcommand{\sgn}{\mathop{\rm sgn}\nolimits}
\newcommand{\tr}{\mathop{\rm tr}\nolimits}
\newcommand{\diag}{\mathop{\rm diag}\nolimits}
\newcommand{\Gal}{\mathop{\rm Gal}\nolimits}
\newcommand{\mycup}{\mathop{\cup}}
\newcommand{\Arg}{\mathop{\rm Arg}\nolimits}
\def\hboxscript#1{ {\hbox{\scriptsize\em #1}} }
\def\zhat{ {\widehat{Z}} }
\def\phat{ {\widehat{P}} }
\def\qtilde{ {\widetilde{q}} }
\renewcommand{\mod}{\mathop{\rm mod}\nolimits}
\renewcommand{\emptyset}{\varnothing}

\def\scra{\mathcal{A}}
\def\scrb{\mathcal{B}}
\def\scrc{\mathcal{C}}
\def\scrd{\mathcal{D}}
\def\scrf{\mathcal{F}}
\def\scrg{\mathcal{G}}
\def\scrl{\mathcal{L}}
\def\scro{\mathcal{O}}
\def\scrp{\mathcal{P}}
\def\scrq{\mathcal{Q}}
\def\scrr{\mathcal{R}}
\def\scrs{\mathcal{S}}
\def\scrt{\mathcal{T}}
\def\scrv{\mathcal{V}}
\def\scrz{\mathcal{Z}}

\def\q{{\sf q}}

\def\Z{{\mathbb Z}}
\def\R{{\mathbb R}}
\def\C{{\mathbb C}}
\def\Q{{\mathbb Q}}
\def\N{{\mathbb N}}

\def\T{{\mathsf T}}
\def\H{{\mathsf H}}
\def\V{{\mathsf V}}
\def\D{{\mathsf D}}
\def\J{{\mathsf J}}
\def\P{{\mathsf P}}
\def\QQ{{\mathsf Q}}
\def\RR{{\mathsf R}}

\def\bsigma{{\boldsymbol{\sigma}}}
\def\bone{\bm{1}}
\def\vv{\bm{v}}
\def\uu{\bm{u}}
\def\ww{\bm{w}}

\newtheorem{theorem}{Theorem}[section]
\newtheorem{definition}[theorem]{Definition}
\newtheorem{proposition}[theorem]{Proposition}
\newtheorem{lemma}[theorem]{Lemma}
\newtheorem{corollary}[theorem]{Corollary}
\newtheorem{conjecture}[theorem]{Conjecture}
\newtheorem{result}[theorem]{Result}


\newenvironment{sarray}{
          \textfont0=\scriptfont0
          \scriptfont0=\scriptscriptfont0
          \textfont1=\scriptfont1
          \scriptfont1=\scriptscriptfont1
          \textfont2=\scriptfont2
          \scriptfont2=\scriptscriptfont2
          \textfont3=\scriptfont3
          \scriptfont3=\scriptscriptfont3
        \renewcommand{\arraystretch}{0.7}
        \begin{array}{l}}{\end{array}}

\newenvironment{scarray}{
          \textfont0=\scriptfont0
          \scriptfont0=\scriptscriptfont0
          \textfont1=\scriptfont1
          \scriptfont1=\scriptscriptfont1
          \textfont2=\scriptfont2
          \scriptfont2=\scriptscriptfont2
          \textfont3=\scriptfont3
          \scriptfont3=\scriptscriptfont3
        \renewcommand{\arraystretch}{0.7}
        \begin{array}{c}}{\end{array}}

%
%
\newcommand{\stirlingsubset}[2]{\genfrac{\{}{\}}{0pt}{}{#1}{#2}}
\newcommand{\stirlingcycle}[2]{\genfrac{[}{]}{0pt}{}{#1}{#2}}
\newcommand{\assocstirlingsubset}[3]{%
{\genfrac{\{}{\}}{0pt}{}{#1}{#2}}_{\! \ge #3}}
\newcommand{\assocstirlingcycle}[3]{{\genfrac{[}{]}{0pt}{}{#1}{#2}}_{\ge #3}}
\newcommand{\euler}[2]{\genfrac{\langle}{\rangle}{0pt}{}{#1}{#2}}
\newcommand{\eulergen}[3]{{\genfrac{\langle}{\rangle}{0pt}{}{#1}{#2}}_{\! #3}}
\newcommand{\eulersecond}[2]{\left\langle\!\! \euler{#1}{#2} \!\!\right\rangle}
\newcommand{\eulersecondgen}[3]{%
{\left\langle\!\! \euler{#1}{#2} \!\!\right\rangle}_{\! #3}}
\newcommand{\binomvert}[2]{\genfrac{\vert}{\vert}{0pt}{}{#1}{#2}}

\clearpage

%
%
\section{Introduction} \label{sec.intro}

The two--dimensional (2D) $Q$--state Potts model \cite{Potts52,Wu_82} is one
of the most studied models in Statistical Mechanics. Despite many
efforts over more than 60 years, its {\em exact} free energy and phase
diagram are still unknown.  The ferromagnetic regime of the Potts
model is the best understood case: exact (albeit not always rigorous)
results have been obtained for the ferromagnetic-paramagnetic phase
transition temperature $T_{\rm c}(Q)$ for several regular
lattices, the order of the transition (continuous for $0 \le Q \le
4$, and first order for $Q>4$), the phase diagram, and the 
characterization in terms of conformal field theory (CFT) of the 
corresponding universality classes. (See e.g., Ref.~\cite{Baxter_book}.)

The antiferromagnetic (AF) regime is less understood. This is partly because,
in contrast with the ferromagnetic regime, universality cannot be expected to
hold in general. Investigations must therefore proceed on a case-by-case basis.
For instance, the free energy is known exactly along some curves of the phase 
diagram $(Q,T)$ (where $T$ is the temperature), for certain regular 2D lattices 
\cite{Baxter82,Baxter_book}.
One of these curves belongs to the ferromagnetic regime, and it can be 
identified with the ferromagnetic-paramagnetic phase-transition curve.
It might be tempting to infer the very existence of a phase transition
from this (partial) solubility of the model along that curve. One well-known 
example of the invalidity of such an inference is provided by the 
zero-temperature limit of the triangular-lattice $Q$--state Potts 
antiferromagnet \cite{Baxter86,Baxter87}. Although its free energy is exactly 
known for all values of  $Q\in\R$,\footnote{
  The $Q$--state Potts model can be defined for non-integer values of $Q$ 
  using the Fortuin--Kasteleyn representation explained below 
  [cf.,~\reff{def_Z_Potts_FK}].   
} 
the system is known to be critical only in the interval $Q\in [0,4]$, and 
disordered for $Q\in (-\infty,0)\cup(4,\infty)$. 

In three dimensions (3D) there are no known exact results for the 
$Q$--state Potts model. Most numerical results come from series expansions 
and Monte Carlo simulations: see e.g., 
Refs.~\cite[and references therein]{Guttmann_94,Janke_97}. 
In the ferromagnetic regime, we expect a critical curve, which is second 
order for $Q=2$, and first-order for $Q=3$. There was an important 
controversy in the late 80s and early 90s about the precise nature of 
the $Q=3$ transition because of its relation with QCD: the four-dimensional
SU(3) lattice gauge theory should be in the same universality class as the
3D ferromagnetic 3--state Potts model \cite{Janke_97}.

The $Q$--state Potts model at temperature $T$ can be defined on any
(undirected) finite graph $G=(V,E)$ with vertex set $V$ and edge set $E$. 
On each vertex $i \in V$, we place a spin that can take $Q$ distinct values:
$\sigma_i \in \{1,2,\ldots,Q\}$. These spins interact through the Hamiltonian 
\cite{Potts52}
\be
   \mathcal{H}(\{\sigma\})  \;=\;
   - J \sum_{\< ij \> \in E} \delta_{\sigma_i \sigma_j} \; ,
\label{def_H_Potts}
\ee
where $\delta_{ij}$ is the usual Kronecker delta, and $J$ is a {\em real}
coupling constant that is proportional to $1/T$. The partition function 
is defined as usual as:
\be
 Z_G(Q,v) \;=\; \sum\limits_{\{\sigma\}} e^{-{\mathcal H}} \,. 
\label{def_Z_Potts}
\ee
Notice that initially, $Q$ is a positive integer $Q\ge 2$, and $J$ is a real
number. The ferromagnetic (resp.\/ antiferromagnetic) regime corresponds 
to $J\ge 0$ (resp.\/ $J \le 0$). 

Fortuin and Kasteleyn \cite{FK} have shown that
the partition function \reff{def_Z_Potts} can be rewritten as
\be
Z_G(Q,v) \;=\; \sum\limits_{E' \subseteq E} v^{|E'|} Q^{k(E')} \,,
\label{def_Z_Potts_FK}
\ee
where the sum runs over the $2^{|E|}$ subsets $E' \subseteq E$, with $k(E')$ 
being the number of connected components (including isolated vertices) 
in the spanning subgraph $(V,E')$, and $v$ is the temperature-like parameter
\be
v \;=\; e^{J} -1 \,.
\label{def_v}
\ee
We now can promote \reff{def_Z_Potts_FK} to the definition of the model, 
which permits us to consider the parameters $Q$ and $v$ as arbitrary complex 
numbers, because \reff{def_Z_Potts_FK} is a polynomial in both.
The Fortuin--Kasteleyn (FK) representation of the $Q$--state Potts model
\reff{def_Z_Potts_FK} is equivalent to the Tutte polynomial
studied by graph theorists \cite{Tutte54} after a change of variables. 
In terms of $v$, the ferromagnetic 
(resp.\/ antiferromagnetic) regime corresponds to $v\ge 0$ 
(resp.\/ $-1\le v \le 0$). The unphysical regime corresponds to $v<-1$ (i.e.,
complex $J$), and the behavior of the Potts model in this regime is less well
understood than that of the AF regime. 

There are several interesting particular cases of $Z_G(Q,v)$. One first example 
is the chromatic polynomial 
\be
P_G(Q) \;=\; Z_G(Q,-1)\,,
\label{def_P_G}
\ee
which corresponds to the zero-temperature limit in the AF
regime $J\to-\infty$. The restriction of this polynomial to $Q$ a positive 
integer has indeed an interpretation as a coloring problem \cite{BL46}: 
$P_G(Q)$ gives the number of proper $Q$-colorings of the graph $G$. A proper 
$Q$-coloring of a graph $G=(V,E)$ is a coloring of the vertices in $V$ 
such that for each edge $e=\<i,j\> \in E$ its endpoints $i,j$ are not colored 
alike.  

The second example is the flow polynomial 
\be
\Phi_G(Q) \;=\; (-1)^{|E|} Q^{-|V|} Z_G(Q,-Q)\,.
\label{def_Phi_G}
\ee
Again, the restriction of this polynomial to integer $Q$ has a combinatorial 
interpretation: it gives the number of nowhere $\Z_Q$-flows on $G$. 
If $\Gamma$ is an additive Abelian group of order $Q$, a $\Gamma$-flow on $G$ 
is a function $\phi \colon E \mapsto \Gamma$, so that each edge $e\in E$ 
is associated to a variable $\phi(e)$, subject to the constraint that these 
variables are conserved at each vertex $i\in V$, given an arbitrary 
orientation of the edges in $E$. 
A no-where zero $\Gamma$-flow is a $\Gamma$-flow $\phi$ such that
$\phi(e)\neq 0$ for all edges $e\in E$ \cite{Tutte_book,Jaeger,Zhang}.
If $\Gamma$ is a {\em finite}\/ Abelian group of order $Q$,
then the number of nowhere zero $\Gamma$-flows depends only
on $Q$ (not on the {\em specific}\/ structure of the group $\Gamma$), and
it is in fact the restriction to $Q\in\N$ of a polynomial in $Q$ called
the flow polynomial $\Phi_G(Q)$ \cite{Tutte50}. We choose $\Gamma = \Z_Q$ as
our group of order $Q$. 

These two polynomials are related for planar graphs. This relation is based
on the duality transformation for the $Q$-state Potts model
on a planar graph $G$ \cite{Wu_Wang76}:
\be
 Z_G(Q,v) \;=\; Q^{|V|-|E|-1} \, v^{|E|} \,  Z_{G^*}(Q,v^*) \,,
 \label{duality_Z_G}
\ee
where $G^*$ is the dual graph of $G$, and $v^*$ is the dual of $v$
\be
 v \, v^* \;=\; Q \,.
 \label{def_v_star}
\ee
Then for $v=-1$, $v^* = -Q$, so that \reff{duality_Z_G} reduces to 
\be
 P_{G^*}(Q) \;=\; Q \, \Phi_{G}(Q) \,,
 \label{relation_P_Phi}
\ee
The duality relation \reff{relation_P_Phi} can also be understood
combinatorially: the flow along an edge $e \in E$ of $G$ determines
the color difference between the two faces (vertices of $G^*$) that
are adjacent to $e$.  The conservation of the $\Z_Q$-flow at each
vertex $i \in V$ ensures that the color differences thus defined add
up to zero upon encircling $i$. A definite $Q$-coloring of $G^*$
is obtained from these color differences upon fixing the color of one
reference vertex; this is responsible for the factor of $Q$ appearing
in \reff{relation_P_Phi}.

For non-planar graphs there is no known relation whatsoever between
$\Phi_{G}(Q)$ and $P_{G^*}(Q)$.

The evaluation of $Z_G(Q,v)$ for a general graph is a hard problem,
viz., at least as demanding as the determination of the chromatic or the flow
polynomials.
In particular, the determination of the coefficients of the chromatic
or flow polynomials of a general graph (including bipartite planar graphs) is 
\#P-hard \cite[Proposition~2.1]{Oxley_02}, and the same thus holds in
general for the computation of the coefficients of the full
partition function $Z_G(Q,v)$.
For recursive families of graphs, the partition function
$Z_{G_n}(Q,v)$ of any member of the family $G_n$ (composed by $n$ identical 
layers of size $m$ each) can be computed from a transfer matrix 
$\mathsf{T}$ and certain boundary condition vectors $\bm{u}$ and $\bm{v}$:
\be
  Z_{G_n}(Q,v) \;=\; \bm{u}(m)^{t} \cdot \mathsf{T}(m)^n 
                                        \cdot \bm{v}(m) \,,
 \label{def_Z_TM}
\ee
where $t$ denotes the transpose.
The computation time thus grows as a polynomial in $n$; but it does however
still grow exponentially in $m$. Indeed, there are similar formulas for 
computing $P_G$ and $\Phi_G$ directly from the corresponding transfer matrices. 
The partition function \reff{def_Z_TM} can be written as a sum over the 
eigenvalues $\lambda_j$ of the transfer matrix $\mathsf{T}$ with appropriate 
amplitudes $\alpha_j$:
\be
  Z_{G_n}(Q,v) \;=\; \sum\limits_{j=1}^{\dim \mathsf{T}}
                     \alpha_j \, \lambda_j^n \,,
 \label{def_Z_eigenvalues}
\ee
where both the amplitudes and eigenvalues are algebraic functions of both 
$Q$ and $v$.

{}From \reff{def_Z_TM} one can compute the partition function $Z_{G_n}$ of the 
$Q$-state Potts model (or any of its specializations $P_{G_n}$ or $\Phi_{G_n}$)
on any member $G_n$ of a recursive family of graphs. In practice, this 
family of graphs will be a 2D strip graph of some regular lattice of width $m$
and length $n$ with some boundary conditions.\footnote{
   We adopt Shrock's \cite{Shrock_01_review} terminology
   for boundary conditions of 2D strip graphs:
   free ($m_{\rm F} \times n_{\rm F}$),
   cylindrical ($m_{\rm P} \times n_{\rm F}$),
   cyclic ($m_{\rm F} \times n_{\rm P}$), and
   toroidal ($m_{\rm P} \times n_{\rm P}$).
   Here the first dimension ($m$) corresponds to the transverse (``short''
   or space-like) direction, while the second dimension ($n$) corresponds
   to the longitudinal (``long'' or time-like) direction.
}
In this case, we will simplify the notation by writing
$Z_{m_\alpha\times n_\beta}$, where the subscripts $\alpha,\beta=\text{F,P}$
denote free and periodic boundary conditions in the respective
lattice directions. For 3D ``slab''
graphs, we use $Z_{m_\alpha\times p_\beta\times n_\gamma}$ for the partition
function of a slab of section $m\times p$ and thickness $n$. As the partition
function is a polynomial in $Q$ and $v$, we can study its roots, e.g., by 
fixing $Q$ (resp.\/ $v$) to a physical value and then obtaining the roots 
of the resulting one-variable polynomial in the complex $v$-plane
(resp.\/ $Q$-plane). This programme has been done for 2D strip graphs of the
square and triangular lattices with free and cylindrical boundary conditions
\cite{Tutte_sq,Tutte_tri}. Indeed, there are several ways to study the zeros
of $Z_{G_n}$: for instance, considering the plane $(Q,v)$ where both variables
take real values. This approach will produce the ``phase diagram'' of the
model for finite transverse size.  

The above study is hard to carry out in practice. In fact, most studies are 
concerned with the zeros in the complex $Q$-plane of the chromatic polynomial
$P_G$ (or chromatic zeros of $G$). In a series of papers 
\cite[and references therein]%
{transfer1,transfer2,JSStri,JScyclic,JStorus,transfer5}, we have studied 
2D strip graphs of the square and triangular lattices with  
free, cylindrical, cyclic, and toroidal boundary conditions. 
For 3D lattices, the authors of Ref.~\cite{Shrock_01} computed the chromatic 
roots of ``slabs'' with some small transverse sections (with free and periodic 
boundary conditions) and several values of the longitudinal size $n$ (with
free boundary conditions).  
Finally, the roots of the flow polynomial (or flow roots) have been studied 
for some 2D strip graphs and boundary conditions in Ref.~\cite{Chang_03}.

{}From these zeros it is very hard to obtain infinite-volume
quantities ($m,n\to\infty$), as they have strong finite--size--scaling
(FSS) corrections. Moreover, one must face the well-known difficulty
that the limits $\lim_{m=n \to \infty}$ (Fisher limit) and $\lim_{m
  \to \infty} \lim_{n \to \infty}$ (van Hove limit) may give different
results in some parts of the parameter space. In this paper we
consider the latter limit, which has two major advantages: 1) the
first limit, $\lim_{n \to \infty}$, is convenient (and accessible in
polynomial time) within the transfer matrix approach, and 2) the
resulting FSS dependence in the second limit, $\lim_{m \to \infty}$, is
much weaker than when considering the $m$ and $n$ limits
simultaneously. In other words, we first compute, for fixed (and
small) values of $m$, accumulation sets of partition function zeros in
the infinite-length limit $n\to\infty$, and study the infinite-width
limit $m \to \infty$ subsequently.

According to the Beraha--Kahane--Weiss (BKW) theorem
\cite{BKW75,BKW78,BK79,BKW80,Sokal04}, these zeros accumulate along 
certain limiting curves $\mathcal{B}_m$ (when the two eigenvalues that
are largest in modulus become equimodular), and around isolated limiting points 
(when there is a unique dominant eigenvalue and its amplitude vanishes).
By computing the eigenvalues $\lambda_j$ and amplitudes $\alpha_j$ of 
the transfer matrix $\mathsf{T}$ \reff{def_Z_eigenvalues},
we are able to obtain the {\em exact} values of the relevant physical
quantities in the limit $n\to\infty$. Their FSS corrections are found to be
smaller than when both $m$ and $n$ are finite. The extrapolation to the
true infinite-volume limit $m\to\infty$, using standard FSS
techniques, therefore becomes more precise by employing this order of limits.

When looking at the {\em real} chromatic zeros for 2D strip graphs of the
square and triangular lattices with free, cylindrical and cyclic boundary
conditions 
\cite[and references therein]{transfer1,transfer2,JSStri,JScyclic,transfer5},
we observe (as many authors in the literature did in the past!) that there 
exist real accumulation points of these real chromatic roots.\footnote{
    The same is true for the full partition function zeros of strip graphs 
    of the square and triangular lattices 
    with free and cylindrical boundary conditions for fixed physical values 
    of $v\ge -1$ \cite{Tutte_sq,Tutte_tri}.
}
This observation dates back to Beraha \cite{Beraha_thesis} who made a 
conjecture about the possible values of $Q$ that can be accumulation points
for real chromatic roots.  As explained by Saleur \cite{Saleur90} there are 
two statements of the Beraha conjecture that do not seem to be fully 
equivalent: The first one is contained in Beraha's PhD Thesis 
\cite{Beraha_thesis} (the statement is taken from Ref.~\cite{BKW75}):
\begin{conjecture}[Beraha v1] \label{conj.beraha_v1}
Among the limits of all (not necessarily recursive) families of chromatic 
polynomials are found the numbers of the form:
\be
B_n \;=\; 2 + 2 \cos \frac{2\pi}{n} \;=\; 4 \cos^2 \frac{\pi}{n} \,, 
\label{def_Bn}
\ee
for any positive integer $n$. Notice that $B_1 = \lim_{n\to\infty} B_n = 4$.
\end{conjecture}
This conjecture has a slightly different form in Baxter's book 
\cite{Baxter_book} (the statement is taken from Ref.~\cite{Baxter87}):
\begin{conjecture}[Beraha v2] \label{conj.beraha_v2} 
Some of the real zeros of chromatic polynomials of planar graphs should, 
in the limit of the graph becoming large, occur at points in the sequence 
$B_n$ \reff{def_Bn} with integer $n\ge 2$. 
\end{conjecture}
Finally, Jackson \cite{Jackson_03} states the Beraha conjecture in another
slightly different way:
\begin{conjecture}[Beraha v3] \label{conj.beraha_v3}
There exists a plane triangulation with a real chromatic root in 
$(B_n - \epsilon,  B_n + \epsilon)$  for all $n \ge 2$ and all $\epsilon > 0$.
\end{conjecture}

\medskip

\noindent
{\bf Remark}. In the context outlined above, our understanding is that
the Beraha conjecture is a statement about the accumulation points of
chromatic roots in the infinite-size limit of planar regular graphs.
Conjecture~\ref{conj.beraha_v2} claims that generically 
the Beraha numbers are limiting points for the chromatic zeros of families 
of planar graphs. It is clear that integer Beraha numbers $B_2,B_3,B_4,B_6$ can
actually be chromatic zeros of planar graphs (e.g., $K_4$ has all these
numbers as chromatic roots: $P_{K_4}(Q)=Q(Q-1)(Q-2)(Q-3)$). However,
non-integer Beraha numbers (except perhaps $B_{10}$) cannot be chromatic
zeros of any planar graph \cite[Corollary~2.4]{transfer1}. Furthermore,
$B_{10}$ is not a chromatic zero of any {\em plane near-triangulation}
\cite[Proposition~2.3(c)]{transfer1}, nor of any plane triangulation.

\bigskip

Let us now consider a strip graph of a (not necessarily planar)
regular lattice with {\em periodic} boundary conditions along the
longitudinal direction. To treat this case within the transfer matrix
formalism, we have to keep track of the bottom- and top-row
connectivities. As explained in detail in
Refs.~\cite{JScyclic,JStorus,JS_flow}, the transfer matrix only acts
on the top-row connectivity, so the full transfer matrix has a
block-diagonal form, each block corresponding to a different
bottom-row connectivity. Indeed, there are blocks of the top-row state
connected to blocks of the bottom-row state; each of these structures
will be called a link (or bridge). As the transfer matrix cannot
increase the number of links of a given connectivity state, if we
order the state appropriately, the transfer matrix takes an
upper-triangular form. Therefore, the eigenvalues can be obtained from
those diagonal blocks corresponding to a given number $\ell$ of
links. The corresponding amplitudes $\alpha_\ell$ are non-trivial and
can be inferred either from a combinatorial reasoning applied to the
upper-triangular decomposition of the transfer matrix
\cite{Chang_01,Richard06,Richard07}, from quantum field theory
\cite{Saleur91,RS01}, or from representation theoretical
considerations \cite{HR05}. We now have three cases:   
\begin{enumerate}

\item[(1)]
      The strip is planar with free boundary conditions along the transverse
      direction. This is the geometry of an annulus.  
      In this case, the links cannot interchange their positions,
      so the group acting on them is the trivial group $E$ consisting only of
      the identity. The transfer matrix commutes with the generators of the
      quantum algebra $U_qsl(2)$, with $\sqrt{Q} = q+q^{-1}$. In this case
      the amplitudes $\alpha_\ell$ are given by Chebyshev polynomials of
      the second kind \cite{Pasquier90,Saleur91,Chang_01}.   
      The numerical evidence \cite{JScyclic} shows that there are 
      accumulation points at the Beraha numbers \reff{def_Bn}.\footnote{
          We find \cite{transfer1,transfer2,JSStri,transfer5} that the 
          Beraha numbers are accumulation points also for strip graphs of 
          the square and triangular lattices with free and cylindrical 
          boundary conditions. 
      } 
      This is compatible with the CFT predictions of the pattern of
      eigenvalue dominance \cite{Saleur91}.

\item[(2)]
      The strip is planar, but periodic boundary conditions are imposed
      along the transverse direction. The resulting geometry is that of
      the torus, which is obviously non-planar.
      In this case, the links can be cyclically interchanged across the strip.
      Therefore, the group acting in this situation is the cyclic group
      $C_\ell$. The representation theory of this group leads to very
      different expressions for the amplitudes 
      \cite[Equations~(1.2)/(1.3)]{Richard07} (see also \cite{RS01})
      involving number theoretical functions. The fact that the $\ell$ links
      can now wind around the periodic transverse direction with some
      momentum $p$ implies that the amplitudes $\alpha_{\ell,p}$ depend
      on both parameters. In this case the numerical
      evidence \cite{JStorus}, and CFT arguments for the pattern of
      eigenvalue dominance, shows that there are accumulation points of
      real zeros at $Q=0,1,2,3,4$. 

\item[(3)]
      The strip graph itself is {\em non-planar}. In this case the links can be 
      interchanged in any way, so the group describing this situation is 
      the full symmetric group $S_\ell$. The amplitudes $\alpha_{\ell,\lambda}$
      obtained from the 
      representation theory of this group are certain polynomials
      that depend on $\ell$ and on the irreducible representation $\lambda$
      of $S_\ell$ \cite{HR05}. The best numerical evidence comes from the 
      study of the flow polynomial of the generalized Petersen graphs 
      \cite{JS_flow}. In this case there are accumulation points of real
      roots at non-negative integer values of $Q=0,1,2,\ldots$. The present
      paper extends this evidence to higher values of $Q$, and to the
      Potts model partition function in general.
\end{enumerate}

\noindent {\bf Remark}. Obvious a planar strip graph is a special case
on a non-planar one; yet the results (3) do not apply to the cases (1)
or (2). This is because the transfer matrix enjoys more symmetries
(i.e., has a larger commutant) in the planar case, and this must be
taken into account in the corresponding representation theory. In the
same vein, it might be that some special classes of non-planar strip graphs
lead to higher symmetries than generic non-planar strips (see
e.g.~\cite{Dasmahapatra96}) and hence to modifications of the results
(3). In this paper we shall give compelling evidence that the
generalized Petersen graphs provide an example of generic non-planar
strip graphs.

\bigskip

In practice, one does not see all the real accumulation points
expected from the above scenario. In fact, only those within the
Berker--Kadanoff (BK) phase \cite{Saleur90,Saleur91} can be
observed. This is a massless phase with algebraic decay of
correlations throughout which the temperature parameter $v$ is
irrelevant in the renormalization group (RG) sense.
In the generic case, one expects the following picture
\cite{JSboundary} of the phase diagram of the 2D Potts model on a
given regular lattice in the $(Q,v)$ plane. We start with the
ferromagnetic critical curve $v_\text{FM}(Q)>0$; the transition is
second order for $0\le Q \le 4$, and first-order for $Q>4$
\cite{Baxter73}. Along this curve, the thermal operator is relevant,
and becomes marginal in the limit $Q\to 0$ of spanning trees. The
analytic continuation of $v_\text{FM}$ into the AF 
regime is a critical curve denoted $v_\text{BK}(Q)< 0$ with $0\le Q\le
4$.  Along $v_\text{BK}$ the thermal operator is irrelevant. We can
think of this curve as a renormalization-group attractor, whose basin
of attraction is the BK phase. This phase is bounded by two curves
$v_{\pm}(Q)$. The upper one $v_{+}$ is usually identified with the
AF critical curve.~\footnote{
  By definition this is the phase transition curve in the
  $v<0$ region that is closest to the infinite-temperature limit $v=0$.
}

This picture has been verified \cite{JScyclic,JStorus} for the square
and triangular lattice. For the square lattice we know the exact form
of these curves $v_\text{FM}(Q)=+\sqrt{Q}$, $v_\text{BK}(Q) =
-\sqrt{Q}$, and $v_{\pm}(Q) = -2 \pm \sqrt{4-Q}$
\cite{Baxter73,Baxter_book,Baxter82}.  Exactly at $v=v_\pm(Q)$ we find a
different type of critical behavior \cite{AFPotts}. For the triangular
lattice we know that $v_\text{FM}$, $v_\text{BK}$, and $v_{-}$ are the
tree branches of the equation \cite{Baxter78,Baxter_book}:
\be
v^3 + 3v^2 \;=\; Q \,.
\ee

If we denote as $Q_2(-1)$ the first point where $v_{+}$ crosses the
line $v=-1$, then one can only see the accumulation points of
chromatic zeros in the interval $[0,Q_2(-1)]$. For the square lattice,
we know that $Q_2(-1) = 3$, while for the triangular lattice, $Q_2(-1)
\approx 3.81967$ \cite{Baxter86,Baxter87}.%
\footnote{However, in Ref.~\cite{JSStri} a reanalysis of Baxter
  eigenvalues given in Refs.~\cite{Baxter86,Baxter87} yielded the
  value $Q_2(-1) = B_{12} \approx 3.73205$.}
It is important to stress that for 2D Potts models the BK phase does not exist
when $Q$ is one of the Beraha numbers $B_n$ \reff{def_Bn}. This is due to the
fact that at these values of $Q$, some of the amplitudes vanish, and there
are eigenvalue cancellations, which give rise to a different physics. These
phenomena were already found in the study of the chromatic polynomial of
strip graphs of the square and triangular lattices with cyclic 
\cite[Section 6.4]{JScyclic} boundary conditions.    

If we move to lower values of $v < -1$ (i.e., in the unphysical
phase), the BK phase extends to a $v$-dependent value $Q_2(v)$. For
any given lattice, we denote by $Q_\text{max}$ the maximum value of
$Q_2(v)$ that one can attain within the BK phase; obviously then
$Q_\text{max} \ge Q_2(-1)$. One has $Q_\text{max} = 4$ for both the
square and the triangular lattices
\cite{Baxter73,Baxter78,Baxter82,JScyclic,JStorus}. Moreover, the
representation theory of $U_qsl(2)$ implies \cite{Pasquier90} that one
has $Q_\text{max} \le 4$ for any 2D Potts model.

\medskip

\noindent {\bf Remark}. The main features of the BK phase in 2D,
including the special role of the Beraha numbers, have recently been
confirmed by the independent technique of graph polynomials
\cite{Scullard12,Scullard13}.

\bigskip

To attain larger values of $Q_\text{max}$, one must thus study
families of non-planar graphs [like in point (3) above].  In this
paper, and motivated by our findings in Ref.~\cite{JS_flow}, we will
consider the generalized Petersen graphs $G(m,k)$ (see
Section~\ref{sec.petersen}). These are generically non-planar graphs
that can be built in a recursive way. We expect the generalized
Petersen graphs to be representative for generic non-planar
strip graphs. 
The main idea behind this claim is the following: as we will see 
in Section~\ref{sec.tm}, the eigenvalues of the transfer matrix for the 
generalized Petersen graphs are ``almost non-degenerate''. As we shall see
in that section, the transfer matrix for $G(nk,k)$ can be decomposed into 
sectors labeled by the number of links $\ell \in \{0,1,\ldots,k,k+1\}$,
and {\em almost all} eigenvalues are distinct functions of $Q$ and $v$.  
This means that each eigenvalue appears once in one (and only one) sector,
with two exceptions: 1) the sector $\ell=k+1$ contains a single 
eigenvalue (the trivial one) that also appears in every sector $\ell\ge 1$, 
and 2) the eigenvalues coming from the sector $\ell=k$ group into only 
$2k+1$ {\em non-trivial} distinct eigenvalues, each of them with a multiplicity 
greater than one. 
Therefore, all the amplitudes associated to the transfer-matrix eigenvalues, 
except for the last two sectors, are the generic ones for a non-planar 
graph \cite{HR05}. So in this sense, the Petersen graphs can be considered as
good representatives of the class of generic non-planar strip graphs.
Moreover, they are devised to have large values of
$Q_\text{max}$, so that we gain access to a vast swath of the BK
phase.

We find for this family of graphs that:
\begin{itemize}
  \item The qualitative picture of the phase diagram in the $(Q,v)$ 
        plane for the (non-planar) generalized Petersen graphs agrees 
        well with that of 2D Potts models. In particular we find a 
        BK phase. 
  \item The value of $Q_\text{max}$ is numerically found to be large,
        $Q_\text{max} \gtapprox 12.4(1)$.
  \item The set of non-negative integers $\{0,1,2,\ldots\}$ plays the same 
        role as the Beraha numbers did for 2D Potts models. 
  \item At these integer values of $Q$, we find amplitude
        vanishing and eigenvalue cancellations. This leads to a
        scenario that resembles that of $U_qsl(2)$ symmetric 2D Potts
        models. We give details of the inclusion/exclusion
        compensation of eigenvalues for $Q=0,1,2,3,4$.
        This phenomena should be related to the representation 
        theory of the general partition algebra \cite{HR05}, which for 
        the case of planar graphs reduces to the Temperley--Lieb algebra.
        To our knowledge, there is no systematic study of this question 
        in the literature; so this work is a first step towards 
        understanding the cancellation mechanism that takes place in 
        this general partition algebra.
  \item In the phase diagram in the $(Q,v)$ plane we find {\em regions} 
        where there
        is not a single dominant eigenvalue, but a pair of complex-conjugate
        eigenvalues. According to the BKW theorem, we expect that these 
        regions will contain a dense set of partition-function zeros.
        In previous studies all zeros were found to accumulate on points
        or curves, so this is a genuinely new feature.
\end{itemize}

The plan of this paper is as follows. In Section~\ref{sec.petersen} we 
describe the most important properties of the generalized Petersen graphs
$G(m,k)$. In Section~\ref{sec.tm} we summarize the basic properties of the 
transfer-matrix formalism applied to the family of graphs $G(nk,k)$, and 
in Section~\ref{sec.num.res} we will show our results for the 
partition-function zeros and limiting curves in the $(Q,v)$ plane. 
In Section~\ref{sec.integers}, we will analyze in detail what happens to the
eigenvalues and amplitudes when $Q$ is a non-negative integer. 
In Sections~\ref{sec.flow} and~\ref{sec.chromatic} we will consider
the zeros and limiting curves in the complex $Q$-plane of two specializations
of $Z_G(Q,v)$, namely the flow and chromatic polynomials, respectively. 
Finally in Section~\ref{sec.final} we discuss our results and make some
general conjectures about the role of Beraha (resp.\/ non-negative integers)
for planar (resp.\/ non-planar) graphs. 
In Appendix~\ref{sec.sc} we will show another example of non-planar graphs 
(the 3D simple-cubic graph of section $2\times 2$) with similar properties to 
those of the graphs $G(m,k)$.  

%
%
\section{The generalized Petersen graphs}
\label{sec.petersen}

We shall consider the family of graphs $G(m,k)$ called
{\em generalized Petersen graphs} and defined as
follows: Let $m,k$ be positive integers such that $m>k$.
Then $G(m,k)$ is a cubic graph (i.e., all vertices have degree 3) 
with $2m$ vertices denoted $i_p$ and $j_p$ for $p=1,2,\ldots,m$: i.e.,
\be
V(G(m,k)) \;=\; \{ i_1,\ldots,i_m,j_1,\ldots,j_m\} \,.
\ee
The edge set consists of $3m$ edges
$(i_p j_p)$, $(i_p i_{p+1})$, $(j_p j_{p+k})$, for $p=1,2,\ldots,m$,
and with all indices considered modulo $m$: i.e.,
\be
E(G(m,k)) \;=\; \{ (i_p,j_p), (i_p i_{p+1}), (j_p j_{p+k}) \mid 1\le p\le m \}
\,.
\ee
Note that $G(m,k)$ is simple for $m\neq 2k$; but it has double edges
when $m=2k$.
These graphs were introduced by Watkins \cite{Watkins69}.
As an example, $G(12,4)$ can be drawn as follows:
$$
%
%
\begin{pspicture}(-3.2,-3.2)(3.2,3.2)
 \pscircle[linewidth=2pt](0.0,0.0){3.04}
 \psline[linewidth=2pt](3.000,0.000)(2.000,0.000)
 \psline[linewidth=2pt](2.598,1.500)(1.732,1.000)
 \psline[linewidth=2pt](1.500,2.598)(1.000,1.732)
 \psline[linewidth=2pt](0.000,3.000)(0.000,2.000)
 \psline[linewidth=2pt](-1.500,2.598)(-1.000,1.732)
 \psline[linewidth=2pt](-2.598,1.500)(-1.732,1.000)
 \psline[linewidth=2pt](-3.000,0.000)(-2.000,0.000)
 \psline[linewidth=2pt](-2.598,-1.500)(-1.732,-1.000)
 \psline[linewidth=2pt](-1.500,-2.598)(-1.000,-1.732)
 \psline[linewidth=2pt](0.000,-3.000)(0.000,-2.000)
 \psline[linewidth=2pt](1.500,-2.598)(1.000,-1.732)
 \psline[linewidth=2pt](2.598,-1.500)(1.732,-1.000)
 \psline[linewidth=2pt](2.000,0.000)(-1.000,1.732)
 \psline[linewidth=2pt](1.732,1.000)(-1.732,1.000)
 \psline[linewidth=2pt](1.000,1.732)(-2.000,0.000)
 \psline[linewidth=2pt](0.000,2.000)(-1.732,-1.000)
 \psline[linewidth=2pt](-1.000,1.732)(-1.000,-1.732)
 \psline[linewidth=2pt](-1.732,1.000)(0.000,-2.000)
 \psline[linewidth=2pt](-2.000,0.000)(1.000,-1.732)
 \psline[linewidth=2pt](-1.732,-1.000)(1.732,-1.000)
 \psline[linewidth=2pt](-1.000,-1.732)(2.000,0.000)
 \psline[linewidth=2pt](0.000,-2.000)(1.732,1.000)
 \psline[linewidth=2pt](1.000,-1.732)(1.000,1.732)
 \psline[linewidth=2pt](1.732,-1.000)(0.000,2.000)
 \pscircle*[linecolor=gray](3.000,0.000){4pt}
 \pscircle(3.000,0.000){4pt}
 \pscircle*[linecolor=gray](2.598,1.500){4pt}
 \pscircle(2.598,1.500){4pt}
 \pscircle*[linecolor=gray](1.500,2.598){4pt}
 \pscircle(1.500,2.598){4pt}
 \pscircle*[linecolor=gray](0.000,3.000){4pt}
 \pscircle(0.000,3.000){4pt}
 \pscircle*[linecolor=gray](-1.500,2.598){4pt}
 \pscircle(-1.500,2.598){4pt}
 \pscircle*[linecolor=gray](-2.598,1.500){4pt}
 \pscircle(-2.598,1.500){4pt}
 \pscircle*[linecolor=gray](-3.000,0.000){4pt}
 \pscircle(-3.000,0.000){4pt}
 \pscircle*[linecolor=gray](-2.598,-1.500){4pt}
 \pscircle(-2.598,-1.500){4pt}
 \pscircle*[linecolor=gray](-1.500,-2.598){4pt}
 \pscircle(-1.500,-2.598){4pt}
 \pscircle*[linecolor=gray](0.000,-3.000){4pt}
 \pscircle(0.000,-3.000){4pt}
 \pscircle*[linecolor=gray](1.500,-2.598){4pt}
 \pscircle(1.500,-2.598){4pt}
 \pscircle*[linecolor=gray](2.598,-1.500){4pt}
 \pscircle(2.598,-1.500){4pt}
 \pscircle*[linecolor=gray](2.000,0.000){4pt}
 \pscircle(2.000,0.000){4pt}
 \pscircle*[linecolor=gray](1.732,1.000){4pt}
 \pscircle(1.732,1.000){4pt}
 \pscircle*[linecolor=gray](1.000,1.732){4pt}
 \pscircle(1.000,1.732){4pt}
 \pscircle*[linecolor=gray](0.000,2.000){4pt}
 \pscircle(0.000,2.000){4pt}
 \pscircle*[linecolor=gray](-1.000,1.732){4pt}
 \pscircle(-1.000,1.732){4pt}
 \pscircle*[linecolor=gray](-1.732,1.000){4pt}
 \pscircle(-1.732,1.000){4pt}
 \pscircle*[linecolor=gray](-2.000,0.000){4pt}
 \pscircle(-2.000,0.000){4pt}
 \pscircle*[linecolor=gray](-1.732,-1.000){4pt}
 \pscircle(-1.732,-1.000){4pt}
 \pscircle*[linecolor=gray](-1.000,-1.732){4pt}
 \pscircle(-1.000,-1.732){4pt}
 \pscircle*[linecolor=gray](0.000,-2.000){4pt}
 \pscircle(0.000,-2.000){4pt}
 \pscircle*[linecolor=gray](1.000,-1.732){4pt}
 \pscircle(1.000,-1.732){4pt}
 \pscircle*[linecolor=gray](1.732,-1.000){4pt}
 \pscircle(1.732,-1.000){4pt}
\end{pspicture}
$$
The graphs $G(m,k)$ are non-planar for all pairs $(m,k)$, except
for the case $(3,2)$, and the two sub-families $(p,1)$ and $(2p,2)$
with $p\ge 1$. More properties of these graphs can be found in
Ref.~\cite[and references therein]{JS_flow}.
In this paper, we will focus on the sub-family of generalized Petersen
graphs $G(nk,k)$ with $n\ge 2$. Transfer-matrix methods can be efficiently
applied for this family (see next section).

%
%
\section{Potts model transfer matrix} \label{sec.tm}

We wish to evaluate $Z_{G}(Q,v)$ \reff{def_Z_Potts}/\reff{def_Z_Potts_FK}
for $G=G(nk,k)$ by a transfer matrix construction.
Contrary to an often repeated but false statement, evaluating
$Z_G(Q,v)$ by a transfer matrix construction is possible for {\em any}
graph $G$, and does not require $G$ to consist of a number of identical
layers \cite{Bedini_10}.
However, when $G$ does have a layered structure---as is the case
here---$Z_G(Q,v)$ can be computed by the repeated application of the
{\em same} transfer matrix. 

The construction of the transfer matrix for the partition function of the
generalized Petersen graphs $G(nk,k)$ was done in detail in Ref.~\cite{JS_flow}.
So here we will just summarize the main results. The first step is to redraw 
the graph $G(nk,k)$ in such a way that it is composed by $n$ identical layers,
each one containing $2k$ vertices, and with periodic boundary conditions in the
longitudinal direction. Each layer has width $L=k+1$. 
(See \cite[Figure~1]{JS_flow}.) All edges now link vertices within the same 
layer, or in two adjacent layers. 

We will work on a space of partitions of the set 
$\{0,1,2,\ldots,k,0',1',2',\ldots,k'\}$, where $\{0,1,2,\ldots,k\}$ (resp.\/ 
$\{0',1',2',\ldots,k'\}$) belong to the top (resp.\/ bottom) row of the strip.
Given a partition, a link is a block that contains vertices belonging to both 
the top and bottom rows. The number of links in a given partition will be 
denoted by $\ell$.  

The transfer matrix $\mathsf{T}_L$ adds one layer on top of the strip.
It can be written in terms of vertical 
$\mathsf{V}_i$ and horizontal $\mathsf{H}_{ij}$ operators:
\begin{subeqnarray}
\mathsf{H}_{ij} \;=\; I + v \, \mathsf{J}_{ij} \slabel{def_Hij} \\
\mathsf{V}_i    \;=\; v\, I + \mathsf{D}_i \slabel{def_V_i} 
\label{def_H_V}
\end{subeqnarray}
where $\mathsf{J}_{ij}$ is a ``join'' operator that amalgamates the blocks 
containing the points $i,j$, and $\mathsf{D}_i$ is a ``detach'' operator that
removes point $i$ from its block and turn it into a singleton (if $i$ was 
already a singleton it provides a factor $Q$ to that term). Then the 
full transfer matrix can be written as
\be
\mathsf{T}_L \;=\; \mathsf{H}_{01} \, 
                   \left( \prod\limits_{i=k}^2 \mathsf{V}_0 \mathsf{H}_{0,i}
                   \right) \, 
                   \left( \prod\limits_{i=0}^k \mathsf{V}_i \right) 
 \;=\; \mathsf{H}_{01} \mathsf{V}_0 \mathsf{H}_{02} \mathsf{V}_0
       \mathsf{H}_{03} \ldots \mathsf{V}_0 \mathsf{H}_{0k} \mathsf{V}_k
       \mathsf{V}_{k-1} \ldots \mathsf{V}_0 \,,
\label{def_T}
\ee  
where the rightmost operators act first. 

As the transfer matrix only acts on the top-row connectivity state, it is
independent of the bottom-row state. So it can decomposed into a diagonal
block form, each block corresponding to a different bottom-row state. All 
these blocks give the same eigenvalues, so we can choose a simple one. 
As the transfer matrix cannot increase the number of links $\ell$, 
then if we order the states in an appropriate way, the transfer matrix can be
written in an upper-triangular block form. Indeed, the eigenvalues can be 
obtained from its diagonal blocks $\mathsf{T}_{L,\ell}$, each of them 
corresponding to a given value of the number of links $0\le \ell \le L=k+1$.
Finally, given a block with $\ell$ links, these links can be interchanged 
in any possible way, so the relevant group is the symmetric group $S_\ell$.
Then it turns out \cite{HR05} that the relevant eigenvalues come from the
irreducible representations $\lambda$ of $S_\ell$ of dimension $\dim\lambda$. 
If we choose as our basis the appropriate linear combination of states, 
then the transfer matrix $\mathsf{T}_{L,\ell}$ takes a diagonal block form, 
where each block $\mathsf{T}_{L,\ell,\lambda}$ corresponds to a distinct 
irreducible representation $\lambda\in S_\ell$.  
Then, the partition function for $G(nk,k)$ can then be written as a sum of 
ordinary traces:
\be
Z_{G(nk,k)}(Q,v) \;=\; \sum\limits_{\ell=0}^{k+1} 
                       \sum\limits_{\lambda\in S_\ell} \alpha_{\ell,\lambda} 
                       \tr (\mathsf{T}_{L,\ell,\lambda})^n \,.
\label{def_Z_petersen}
\ee
The amplitudes $\alpha_{\ell,\lambda}$ are polynomials in $Q$ given in 
Ref~\cite[Proposition~3.24]{HR05}: 
\be
 \alpha_{\ell,\lambda} \;=\;
   \frac{\dim \lambda}{\ell !}
   \prod\limits_{i=0}^{\ell-1} (Q-i-\lambda_{\ell-i}) \,,
 \label{eigen_amp}
\ee
where the representation $\lambda \in S_\ell$ is considered through its 
corresponding Young diagram, 
$Y(\lambda) = (\lambda_1,\lambda_2,\ldots,\lambda_\ell)$, where $\lambda_i$
is the number of boxes in the $i$'th row. If there is less than $\ell$
rows in $Y(\lambda)$ the expression is of course padded with zeros.

We have symbolically computed all 
relevant blocks $\mathsf{T}_{L,\ell,\lambda}$ for $1\le k\le 6$, 
$0\le \ell \le k+1$, and $\lambda \in S_\ell$. Therefore, we could in principle
compute the partition function for all generalized Petersen graphs $G(nk,k)$
with $1\le k\le 6$ and arbitrary $n\ge 2$. 
It is worth stressing that Eq.~\reff{def_Z_petersen} holds true irrespective 
of whether some eigenvalues happen to be identical or not. However, 
it will turn out useful in practice to have a ``{\em complete}'' 
decomposition of the partition function, in the sense that there are
no equalities among eigenvalues (except for what is accounted for in the
multiplicities $\alpha_{\ell,\lambda}$). 

As for the particular case of the flow polynomial $\Phi_{G(nk,k)}$ 
\cite{JS_flow}, we have found that for $1\le k\le 6$, the blocks 
$\mathsf{T}_{L,\ell,\lambda}$ with $L=k+1$ have an upper-triangular block form:
\be
\mathsf{T}_{k+1,\ell,\lambda} \;=\; \left( 
 \begin{array}{cl}
 \mathsf{D}_{k+1,\ell,\lambda} & \mathsf{F}_{k+1,\ell,\lambda} \\[2mm]
 0                                  & 
 \displaystyle \mathsf{T}^{(nt)}_{k+1,\ell,\lambda} 
 \end{array} \right) \,,
\label{def_T_blocks}
\ee
where $\mathsf{D}_{k+1,\ell,\lambda}$ is a diagonal matrix with all 
its diagonal terms are equal to the ``trivial'' eigenvalue 
$\mu_{k,k+1}=v^{2k}$, and $\mathsf{T}_{k+1,\ell,\lambda}^{(nt)}$ is the 
diagonal block containing the non-trivial eigenvalues (hence the superscript 
``({\em nt})''). Notice that for $\ell=0$, there are no trivial 
eigenvalues: i.e., $\dim \mathsf{D}_{k+1,0}=0$; 
while for $\ell=k+1$, all eigenvalues are trivial: i.e., 
$\dim \mathsf{T}_{k+1,k+1,\lambda}^{(nt)} = 0$. The dimension of the 
diagonal matrix $\mathsf{D}_{k+1,\ell,\lambda}$ is the same as for the 
flow-polynomial case \cite{JS_flow}. 

Finally, we have also observed that for $1\le k\le 6$, 
the block $\mathsf{T}_{k+1,k,\lambda}^{(nt)}$ contains only $2k+1$ non-trivial 
eigenvalues $\mu_{k,k,s}$ of multiplicity $\dim\lambda$ for all irreducible 
representations $\lambda\in S_k$. These eigenvalues come from a reduced 
transfer matrix $\mathsf{T}_{k+1,k}^{(nt)}$. 

Therefore, for $1\le k\le 6$ the partition function for the generalized 
Petersen graphs $G(nk,k)$ \reff{def_Z_petersen} can be rewritten as 
\be
Z_{G(nk,k)}(Q,v) \;=\; \sum\limits_{\ell=0}^{k-1} 
                       \sum\limits_{\lambda\in S_\ell} \alpha_{\ell,\lambda} 
                       \tr (\mathsf{T}_{k+1,\ell,\lambda}^{(nt)})^n + 
                       \beta_k \tr (\mathsf{T}_{k+1,k}^{(nt)})^n +
                       \gamma_{k+1} v^{2nk} \,, 
\label{def_Z_petersen2}
\ee
where the coefficient $\beta_\ell$ is given by the sum
\be
\beta_\ell \;=\; \sum\limits_{\lambda\in S_\ell} \alpha_{\ell,\lambda} 
                 \dim \lambda\,,
\label{def_beta_ell}
\ee
and the coefficients $\gamma_{k+1}$ are given by
\be
\gamma_{k+1} \;=\; \beta_{k+1} + \sum\limits_{\ell=1}^k 
             \sum\limits_{\lambda \in S_\ell} \alpha_{\ell,k} 
             \dim \mathsf{D}_{k+1,\ell,\lambda} \,.
\label{def_gamma_ell+1}
\ee
The expressions for the needed $\beta_\ell$ and $\gamma_{k+1}$ can be read
from \cite[Eqs.~(3.28)/(B.11)]{JS_flow}.

We have proven \reff{def_Z_petersen2} by explicit computation for 
$1\le k\le 6$. We conjecture that \reff{def_Z_petersen2} is also true  
for each $k\ge 7$. This conjecture indeed holds true in the particular case 
of the flow polynomial \cite{JS_flow}. 

If we denote by $\mu_{k,\ell,\lambda,s}$ (resp.\/ $\mu_{k,k,s}$)
the eigenvalues of $\mathsf{T}_{k+1,\ell,\lambda}^{(nt)}$ 
(resp.\/ $\mathsf{T}_{k+1,k}^{(nt)}$), then \reff{def_Z_petersen2} can be
written as
\be
Z_{G(nk,k)}(Q,v) \;=\; \sum\limits_{\ell=0}^{k-1}
                       \sum\limits_{\lambda\in S_\ell} \alpha_{\ell,\lambda}
                       \sum\limits_{s=1}^{N_k(\ell,\lambda)}
                       {\mu_{k,\ell,\lambda,s}^n} +
                       \beta_k \sum\limits_{s=1}^{2k+1} {\mu_{k,k,s}^n} +
                       \gamma_{k+1} v^{2nk} \,,
\label{def_Z_petersen3}
\ee
where $N_k(\ell,\lambda)$ is the number of non-trivial eigenvalues 
corresponding to $\ell$ links ($0\le \ell \le k+1$) and the irreducible
representation $\lambda\in S_\ell$. 
The expression \reff{def_Z_petersen3} can be regarded as 
a ``complete'' decomposition of the partition function in the sense that for
a fixed value of $k$, all the eigenvalues appearing in \reff{def_Z_petersen3}
(namely, $\mu_{k,\ell,\lambda,s}$, $\mu_{k,k,s}$, and $v^{2k}$) are all 
distinct. This fact was proven for $1\le k\le 6$ by explicit computation of the 
eigenvalues. Furthermore, we conjecture that \reff{def_Z_petersen3} holds 
true with all eigenvalues being distinct also for $k\ge 7$.
The total number of distinct eigenvalues is $6, 20, 113, 755, 5568, 43975$ 
for $k=1,\ldots,6$.

As each eigenvalue in \reff{def_Z_petersen3} is characterized by the number 
of links $\ell$, we can use this number as a sector label. Therefore we
will use the expression ``sector $\ell$'' (or sector of $\ell$ links)
as a shorthand for the set of {\em non-trivial} eigenvalues with number 
of links equal to $\ell$ ($0\le \ell \le k$). The sector $\ell=k+1$ will 
correspond to the trivial eigenvalue $\mu_{k,k+1}=v^{2k}$.

%
%
\section{Numerical results} \label{sec.num.res}

In this section we provide information about the practical procedure we have 
followed to compute the relevant blocks of the transfer matrix. We then
show our results concerning the phase diagram for each value of $k$. Finally,
by comparing these finite-$k$ results, we attempt to gain some insight about 
the thermodynamic limit $k\to\infty$. 

%
%
\subsection{Practical procedure} \label{sec.algorithm}

We have written a {\sc Mathematica} script to compute the symbolic 
transfer matrix $\mathsf{T}_L$ (with $L=k+1$ and for $1\le k\le 6$) 
using ideas similar to those already explained in 
\cite{JScyclic,JStorus,JS_flow}.

The first goal is to obtain the relevant diagonal blocks 
$\mathsf{T}_{k+1,\ell,\lambda}$ of the transfer matrix $\mathsf{T}_L$. 
We first fix the number of links $\ell$ ($0\le \ell\le k+1$) and a bottom-row
state compatible with the chosen value of $\ell$. We then determine the basis
of connectivity states. The result indeed does not depend on the chosen 
bottom-row state. 
We now choose an irreducible representation $\lambda\in S_\ell$ 
of dimension $\dim\lambda$. The relevant diagonal block
$\mathsf{T}_{k+1,\ell,\lambda}$ is obtained by taking as our basis vectors 
those linear combinations of the ``standard basis'' with the appropriate 
symmetries under $S_\ell$. 
To decrease the CPU time needed for the computation, we extract the trivial
eigenvalues $\mu_{k,k+1}=v$ by looking for a upper-block diagonal 
form like in \reff{def_T_blocks}.  

Once the basic blocks $\mathsf{T}_{k+1,\ell,\lambda}^{(nt)}$ are obtained, we
compute the traces $\tr (\mathsf{T}_{k+1,\ell,\lambda}^{(nt)})^n$ using again
{\sc Mathematica}. We then reconstruct the partition function using 
\reff{def_Z_petersen2}--\reff{def_gamma_ell+1}. 

We have checked the partition functions for small values of $n$ by computing 
the partition function using {\sc Maple}. Indeed, from the 
full partition function $Z_{G(nk,k)}(Q,v)$ we can obtain its flow-polynomial 
specialization $\Phi_{G(nk,k)}(Q)$. These polynomials were also compared to 
those obtained by other methods in Ref.~\cite{JS_flow}. We find a perfect 
agreement in all cases.   

We have symbolically computed all relevant blocks 
$\mathsf{T}_{k+1,\ell,\lambda}$ for $1\le k\le 6$, $0\le \ell \le k+1$, 
and $\lambda \in S_\ell$. We have also found the ``complete'' decomposition of
the partition function for these values of $k$. However, the actual computation
of the traces is quite demanding: for $k\le 5$ we were able to compute the 
partition function for values $n\le 10$. For $k=6$, we could only compute 
$n=1,2$. The partition functions for $n=3,4$ were obtained using 
{\sc Maple}. For $k=7$ we
could not obtain the transfer-matrix blocks; but we managed to get
the partition functions up to $n=4$ using again {\sc Maple}.  

%
%
\subsection{Results for the generalized Petersen graphs} \label{sec.res1} 

Our goal is to study the ``phase diagram'' of the $Q$-state Potts model on the
generalized Petersen graphs $G(nk,k)$ in the $(Q,v)$ plane. From a physical 
point of view, this is the most natural approach to the problem. We will 
focus on the antiferromagnetic/unphysical region $v < 0$, as we are 
interested in studying a possible BK phase in this model. 

%
%
\begin{figure}
  \vspace*{-1cm}
  \centering
  \begin{tabular}{cc}
  \includegraphics[width=200pt]{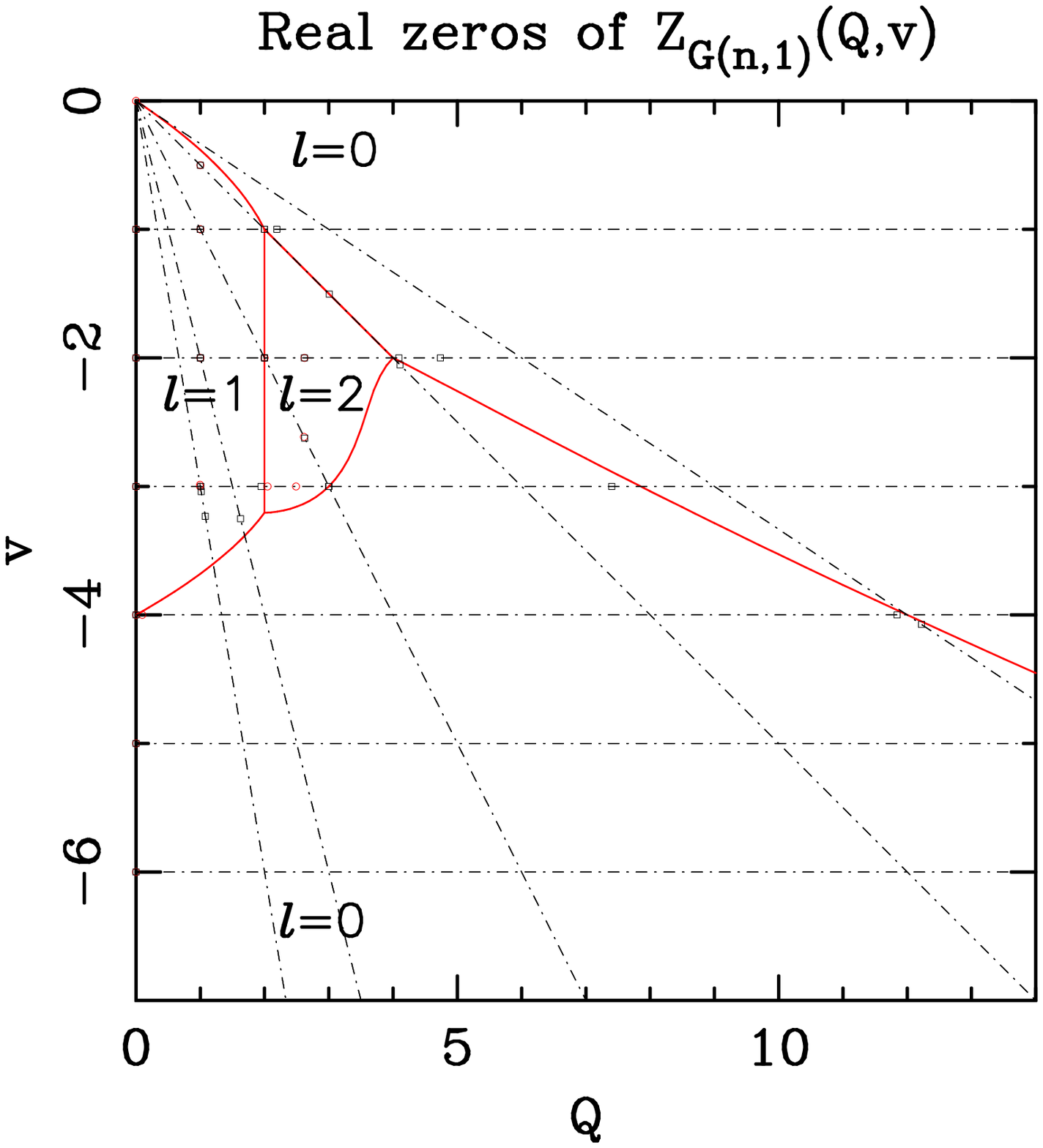} & 
  \includegraphics[width=200pt]{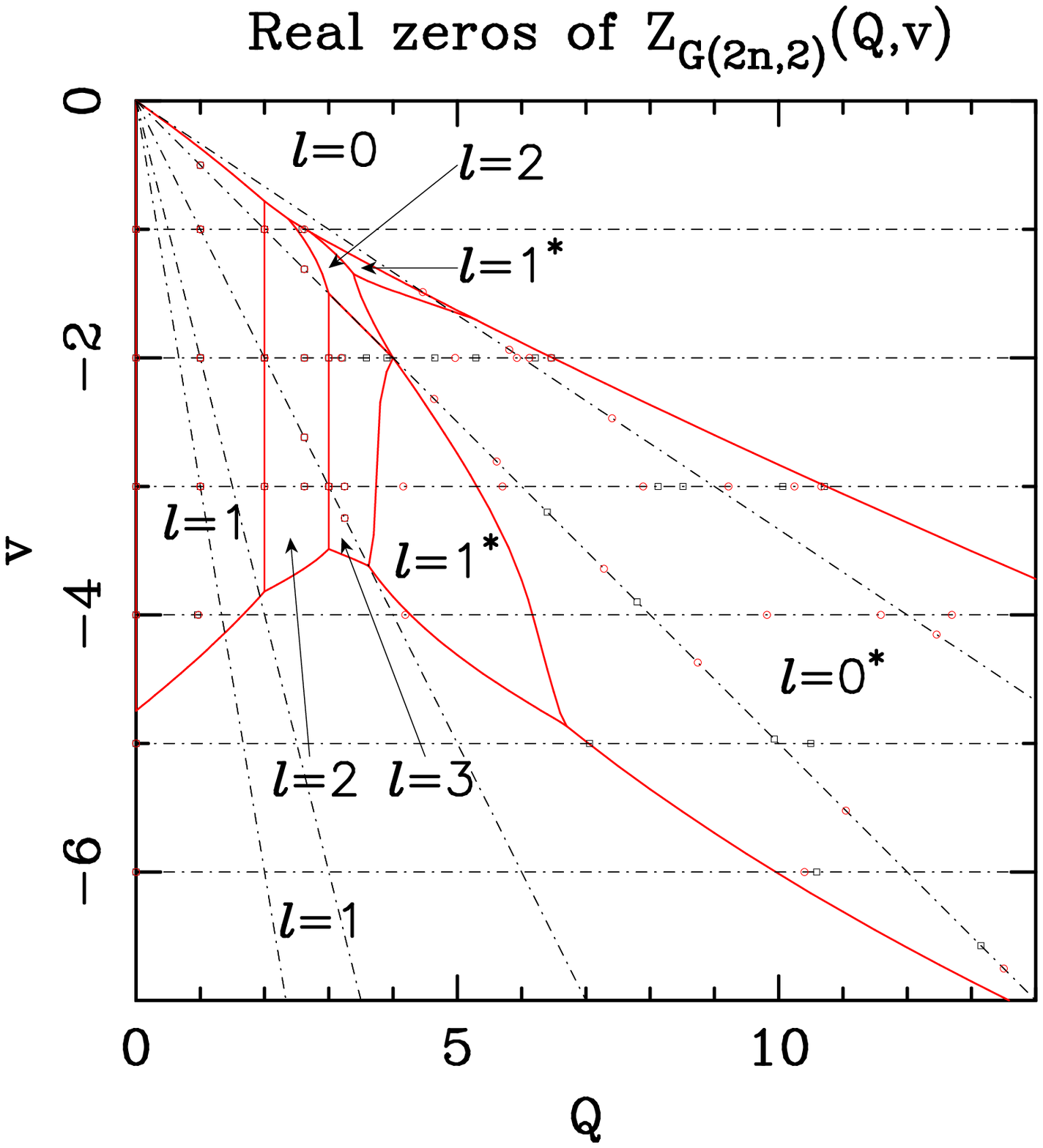} \\
  \qquad (a) & \qquad (b) \\[2mm]
  \includegraphics[width=200pt]{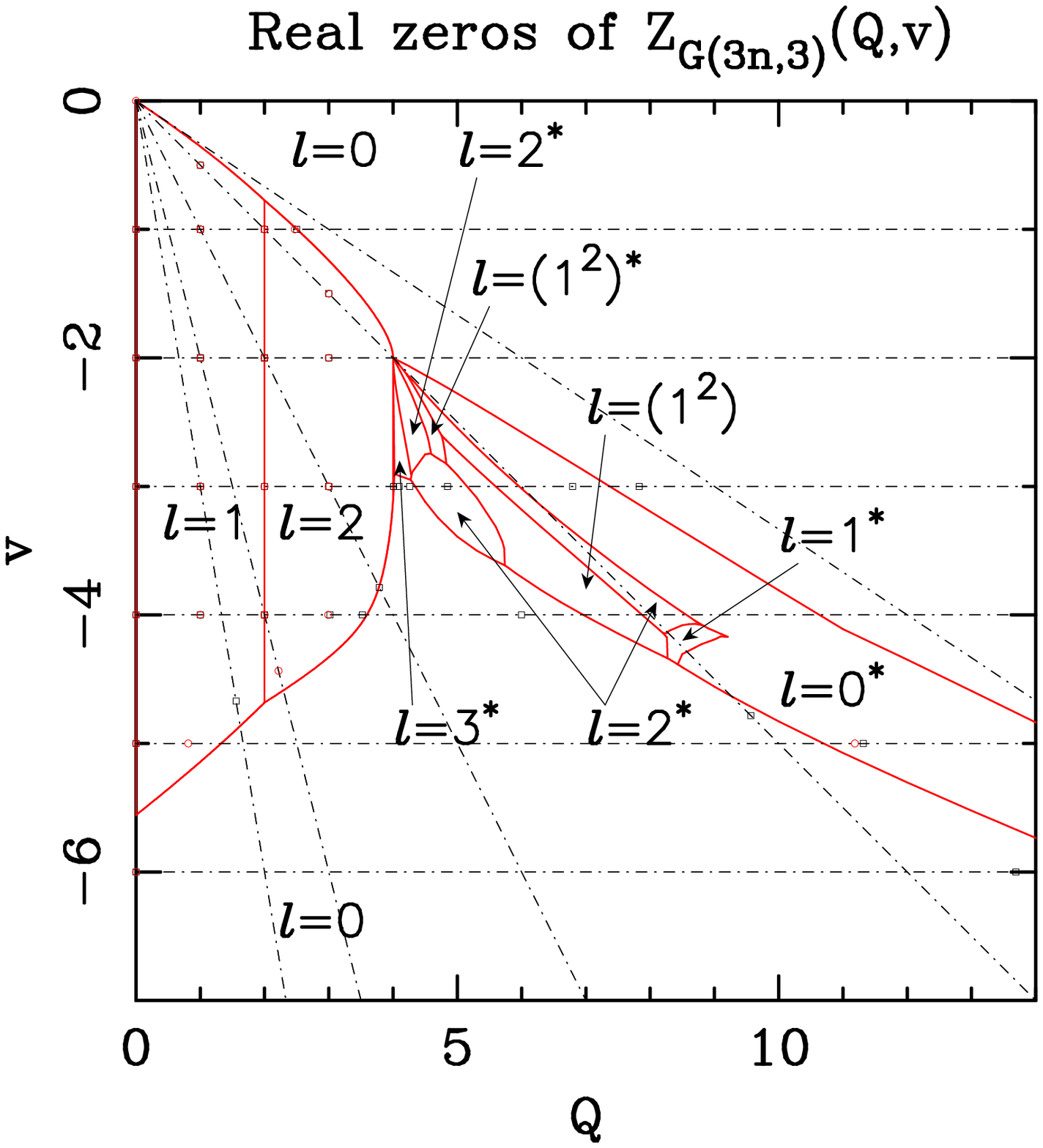} & 
  \includegraphics[width=200pt]{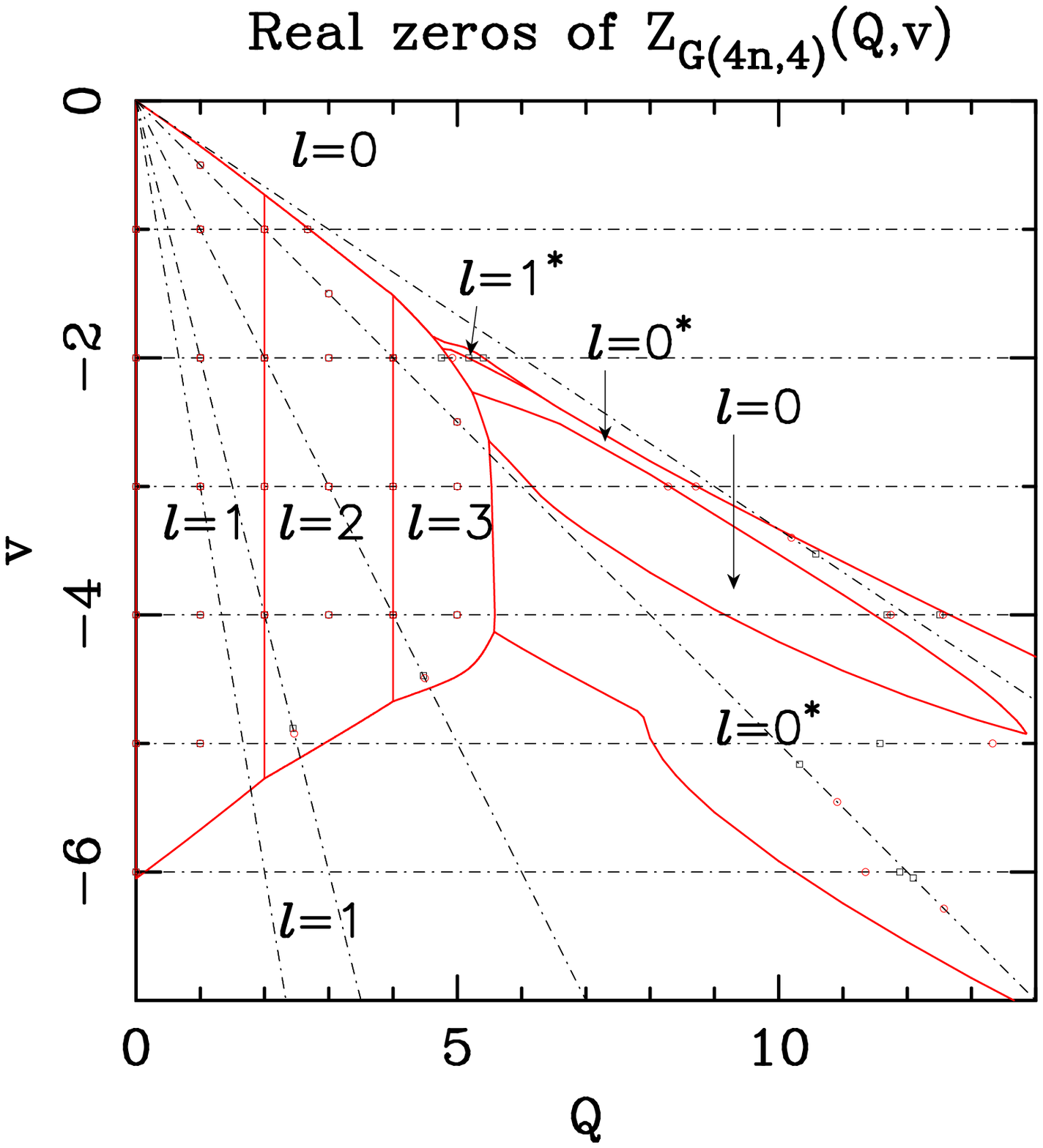} \\
  \qquad (c) & \qquad (d) 
  \end{tabular}
  \caption{
  Real zeros of the Potts-model partition function
  and limiting curves $\mathcal{B}_k$ in the plane $(Q,v)$  
  for the generalized Petersen graphs $G(nk,k)$ with $k=1$ (a), 
  $k=2$ (b), $k=3$ (c), and $k=4$ (d). 
  For each panel, we show the real zeros along several lines 
  (depicted as doted-dashed lines) of the type $Q=-p v$ ($p=1,2,3$),
  $v=-p Q$ ($p=1,2,3$), and $v=-1,-2,-3,-4,-5,-6$. 
  For each value of $k$, the zeros correspond to the generalized Petersen
  graphs $G(9k,k)$ (black $\square$) and $G(10k,k)$ (red $\circ$). 
  Each region is labeled with the sector the dominant eigenvalue belongs to 
  (e.g., $\ell= 1$).
  The asterisk in the sector label (e.g., $\ell = 1^*$) means that there is 
  a pair of complex-conjugate dominant eigenvalues in that region. 
  The label $\ell=(1^2)$ in panel (c) means that the dominant eigenvalue 
  comes from the completely anti-symmetric irreducible representation 
  $(1,1)=(1^2)$ of $S_2$.
  }
\label{Figures_Z1}
\end{figure}

\begin{figure}
  \vspace*{-1cm}
  \centering
  \begin{tabular}{cc}
  \includegraphics[width=200pt]{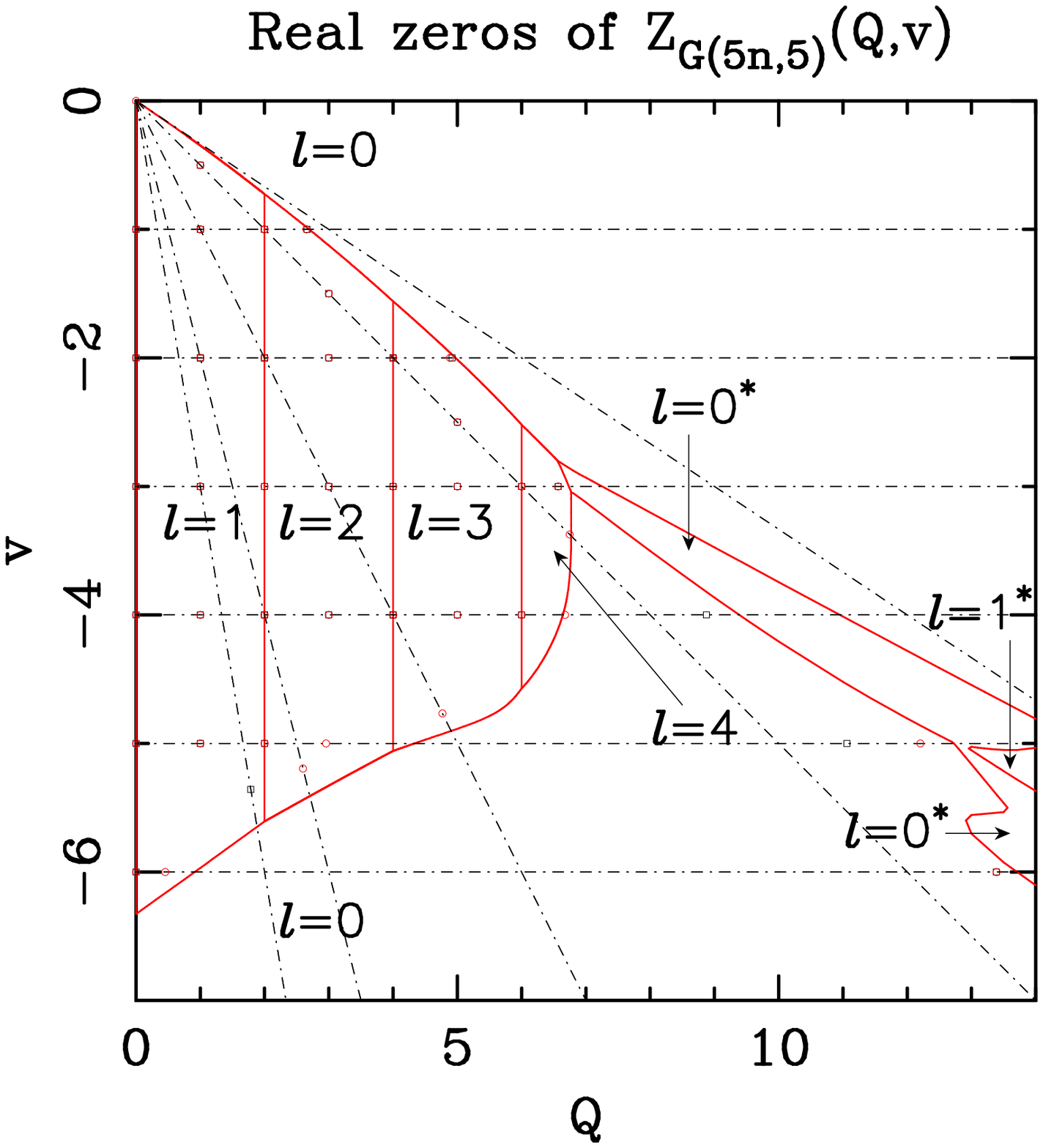} & 
  \includegraphics[width=200pt]{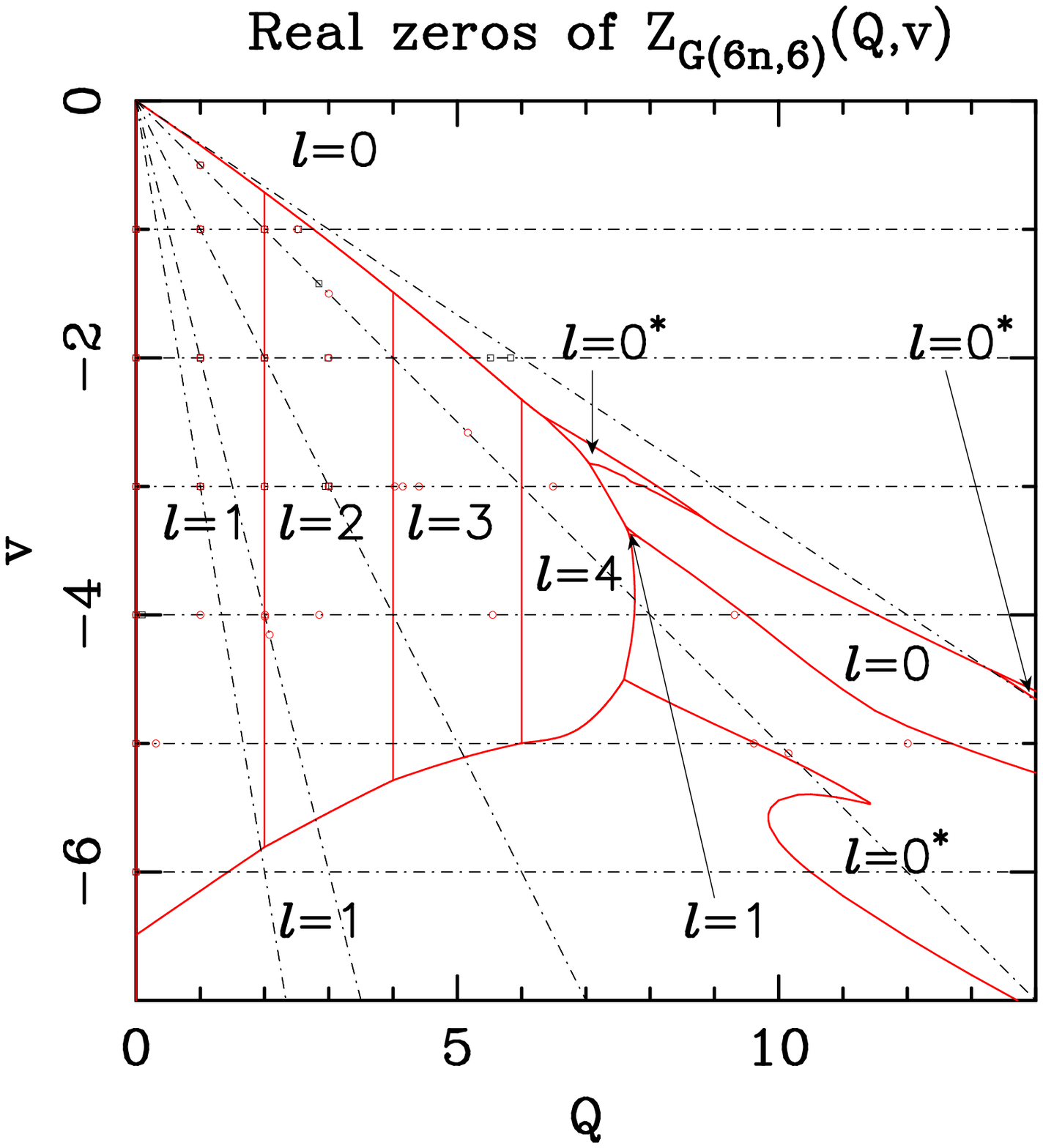} \\
  \qquad (a) & \qquad (b) \\ 
  \multicolumn{2}{c}{%
       \includegraphics[width=200pt]{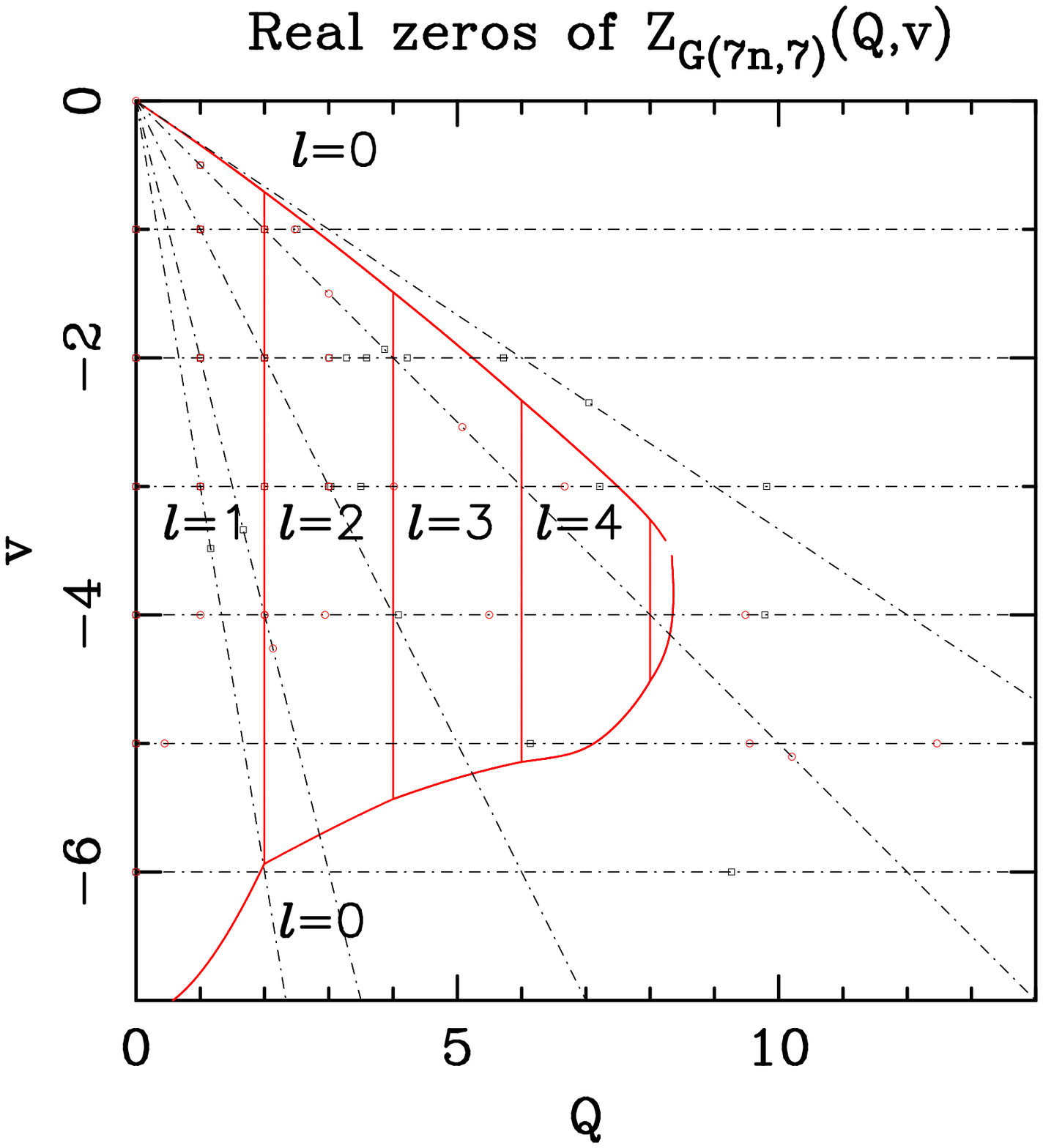}}\\
  \multicolumn{2}{c}{\qquad (c)}
  \end{tabular}
  \caption{
  Real zeros of the Potts-model partition function
  and limiting curves $\mathcal{B}_k$ in the plane $(Q,v)$  
  for the generalized Petersen graphs $G(nk,k)$ with $k=5$ (a), 
  $k=6$ (b), and $k=7$ (c). 
  In (a) the zeros correspond to the generalized Petersen
  graphs $G(45,k)$ (black $\square$) and $G(50,k)$ (red $\circ$).
  For each value of $k$ in (b) and (c), the zeros correspond to the 
  generalized Petersen
  graphs $G(3k,k)$ (black $\square$) and $G(4k,k)$ (red $\circ$). 
  The rest of the notation is as in Figure~\ref{Figures_Z1}.  
  For $k=7$ we could not obtain the more involved structure for large $Q \gtapprox 8.2$.
}
\label{Figures_Z2}
\end{figure}

To study the partition-function zeros in this plane, we took several lines:
$Q= -p v$, $v=-p Q$ (with $p=1,2,3$), and $v=-1,-2,\ldots,-6$. The line 
$v=-Q$ (resp.\/ $v=-1$) corresponds to the flow-polynomial (resp.\/ 
chromatic-polynomial) subspace. Along these 
lines, the two-variable partition function $Z_{G(nk,k)}(Q,v)$ reduces to
a single-variable polynomial with integer coefficients. We then find its 
zeros by using the program {\sc MPSolve} by Bini and Fiorentino
\cite{MPSolve,Bini00}. These lines are depicted as dot-dashed lines 
in Figures~\ref{Figures_Z1} and~\ref{Figures_Z2}, and the zeros are also 
represented as circles and squares in the same figures. Although they
exhibit some systematic features, these zeros are generally too scattered
for any firm conclusions to be drawn.

A much better approach is to compute the limiting curves $\mathcal{B}_k$ 
when $n \to \infty$ in the $(Q,v)$ plane.
We have used the direct--search method \cite{transfer1} to compute 
$\mathcal{B}_k$. But in some areas of the plane, this
method did not work properly. The reason was that there are regions where
the dominant eigenvalue is actually a {\em pair} of complex-conjugate 
eigenvalues belonging to the same sector. 
These regions are labeled by an asterisk in 
Figures~\ref{Figures_Z1} and~\ref{Figures_Z2}. 

%
%

For $k=7,8,9,11$, even though we could not obtain the relevant transfer-matrix
blocks, we were able to numerically compute (parts of) the corresponding 
limiting curves by using a code written in {\sc C}. For $2\le k\le 6$ the 
result of this program coincides with the previous symbolic computation. 

By looking at Figures~\ref{Figures_Z1} and~\ref{Figures_Z2}, we see that for
each fixed value of $k$ (with $1\le k\le 7$) there are two distinct regions 
in the corresponding phase diagram:
\begin{itemize}

\item[(a)] This region is the closest to $Q=0$ and ends approximately at 
           $Q \approx k+1$. It contains vertical lines at even integer 
           values of $Q$, and has isolated limiting points at odd integer
           values of $Q$. It is bounded above and below by simple curves,
           and the dominant eigenvalues are simple ones, hence real valued.
           Close to $Q\gtapprox 0$, the dominant eigenvalue belongs to the 
           sector $\ell=1$, and every time we cross one of these vertical 
           lines the value of $\ell$ increases by one unit. 

\item[(b)] This region starts approximately at $Q\gtapprox k+1$, and it 
           has a more involved structure. 
           For $k=1$ there is a single outward branch starting at 
           $(Q,v)=(4,-2)$ and extending to infinity.
           For $2\le k\le 6$, we find  
           two boundary curves that extend outwards to infinity.
           Inside this region there are sub-regions where the dominant 
           eigenvalue is actually a pair of complex-conjugate eigenvalues
           (in most cases they belong to the sector $\ell=0$). For
           $k=3$, we even find some regions where the dominant eigenvalue 
           comes from the completely anti-symmetric representation 
           $(1,1)\equiv (1^2)$ of the $\ell=2$ sector 
           (see Figure~\ref{Figures_Z1}(c)). 
\end{itemize} 
Above [resp.\ below] both regions the dominant eigenvalue comes from the 
$\ell=0$ [resp.\ $\ell = (k+1) \bmod 2$] sector. 

Region~(a) looks qualitatively similar to the BK phase we have found for
the 2D Potts models on the square and triangular lattices. 
Inside the region enclosed by the limiting curve $\mathcal{B}_k$, we
find that for each integer $p\ge 0$ up to some maximum value $p_\text{max}$, 
there is a ``phase'' corresponding to an eigenvalue belonging to the 
fully symmetric representation $\lambda=(\ell)$ of $S_\ell$ with 
$\ell=p+1$ links. This phase contains all values of $Q$
in the interval $Q\in(2p,2p+2)$. Therefore, there are phase transitions
at $Q=2p$ for $p\ge 0$ in an interval $(v_-(p),v_+(p))$, in which
eigenvalues from the $\ell=p$ and $\ell=p+1$ become equimodular.
As $k$ increases, the maximum value $p_\text{max}$ increases, so we find 
more phases in the BK phase. Even though $Q_2(-1) < 3$ 
is rather small (see Section~\ref{sec.chromatic}), as in the 2D case, 
we now find vertical lines at $Q$-values as large as $Q=8$ for $k=7$ ($p=4$), 
which is the maximum value of $k$ for which we have been able to produce 
the full ``regular part''---region~(a)---of the phase diagram.

We emphasize that the presence of vertical lines of limiting points
extending from $v_-(Q)$ to $v_+(Q)$ provides clear evidence for the
extent of the BK phase. In particular, the perfect verticality means
that the temperature parameter $v$ is RG irrelevant for $v_-(Q) < v <
v_+(Q)$, and the physics is thus controlled by an RG fixed point in
the interior of the BK phase. Ideally one would like to corroborate
the identification---and characterization---of the BK phase by
verifying the algebraic decay of correlation functions inside the BK
phase. In the 2D case this is a rather simple matter, since standard
CFT results relate the strip free energies to critical
exponents. 
In Section~\ref{sec.res2} we pursue a more modest goal, namely to
give numerical evidence that the correlation length $\xi$ is proportional
to $k$ inside the BK phase and tends to a constant outside it. Since $k$ roles
as the effective width in the strip geometry underlying the transfer matrix 
formalism, this implies that the BK phase is characterized by $\xi \to \infty$ 
in the thermodynamic limit $k \to \infty$, and so is critical indeed.

In the phase characterized by $\ell$ links, corresponding to the
interval $Q\in(2(\ell-1),2\ell)$, the amplitude
corresponding to the fully symmetric representation $\lambda=(\ell)$ is
[cf.,~\reff{eigen_amp}] \cite{HR05,JS_flow}
\be
\alpha_{\ell,(\ell)} \;=\; \frac{1}{\ell!} \, (Q-2\ell+1)\,
                           \prod\limits_{i=0}^{\ell-2} (Q-i) \,.
\ee
Therefore, it vanishes at the mid-point $Q=2\ell-1$ of the above interval;
hence, this is an isolated limiting point.

\medskip

\noindent
{\bf Remark}. Notice that the families $G(n,1)$ and $G(2n,2)$ are 
{\em planar}. Indeed, in Figures~\ref{Figures_Z1}(a)--(b), 
we see that there are isolated limiting points
at the non-integer value $Q=B_5$ in agreement with the Beraha conjecture.
This can be explained by the fact that the amplitude for the $\ell=2$ sector
is not the generic one, but $\gamma_2$ for $k=1$, and $\beta_2$ for $k=2$.
(cf.,~\cite[Eq.~(3.28)]{JS_flow}). Indeed, both amplitudes are equal to
$\beta_2=\gamma_2=Q^2 - 3Q + 1$, and $\beta_2(B_5)=0$, as expected.

\bigskip

%
%
\begin{figure}
  \centering
  \includegraphics[width=200pt]{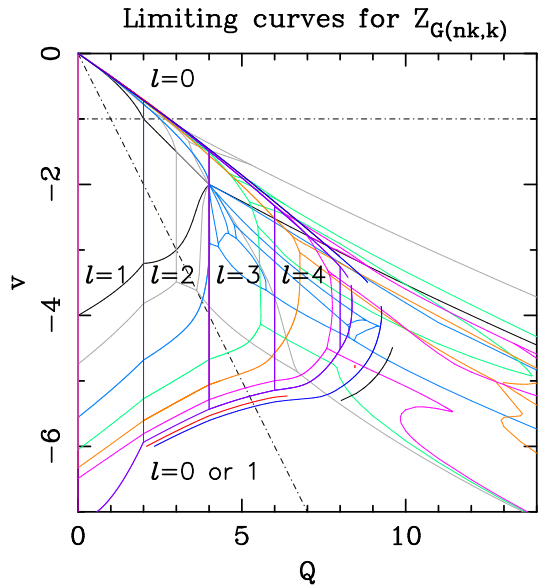}
  \caption{
  Limiting curves $\mathcal{B}_k$ in the plane $(Q,v)$  
  for the generalized Petersen graphs $G(nk,k)$ with $k=1$ (black),
  $k=2$ (gray),
  $k=3$ (light blue), $k=4$ (green), $k=5$ (orange), $k=6$ (pink),
  $k=7$ (violet), $k=8$ (red), $k=9$ (navy blue), and $k=11$ (black).  
  We label some regions with the sector to which the dominant eigenvalue 
  (for large enough $k$) belongs (e.g., $\ell=3$). The horizontal $v=-1$ 
  (resp.\/ inclined $v=-Q$) dot-dashed line corresponds to the chromatic 
  (resp.\/ flow) polynomial subspace.  
  }
\label{Figure_Z_all}
\end{figure}

Figure~\ref{Figure_Z_all} shows all the limiting curves put together.    
If we look at the bottom part of these curves ($Q=0, v\ltapprox -4$), the 
value of $k$ increases as we move outwards. We notice the following empirical
observations: 
\begin{itemize}
 \item As $k$ increases, the BK phase enlarges and more $\ell$ sectors are
       visible in the unphysical region.   

 \item The upper boundary curve converges quickly to a limit. In particular, 
       its slope at $Q=0$ seems to be equal to $-1/3$.\footnote{
          For each value of $k=1,\ldots,6$, we have fitted the data closest to
          the origin to a polynomial Ansatz. For odd values of $k$, we 
          obtained in all cases estimates compatible with a slope equal to 
          $-1/3$ (e.g., $-0.333333(1)$ for $k=1,3$, and $-0.3333(1)$ for 
          $k=5$). For even values of $k$, we obtain estimates that 
          seem to converge to $-1/3$ from below (e.g., $-0.347841(1)$, 
          $-0.339861(1)$, and $-0.3368(1)$ for $k=2,4,6$). 
          A power-law fit to these three data points gives a slope of 
          $\approx -0.329$, not far from the odd-$k$ estimate. Therefore,  
          the numerical data suggest that the slope of the curve $v_+(Q)$
          at $Q=0$ is  exactly $-1/3$, as claimed. 
       } 
       Its crossing with the 
       horizontal line $v=-1$ (depicted as dot-dashed line in 
       Figure~\ref{Figure_Z_all}) defines the value of $Q_2(-1)$. (See
       Section~\ref{sec.chromatic}.)  We can identify this curve with the 
       AF critical curve $v_+(Q)$ for this model. 

 \item The lower boundary curve $v_-(Q)$ converges slowly to some limit. 
       The inclined dot-dashed line in Figure~\ref{Figure_Z_all} corresponds 
       to the flow polynomial. The convergence is not as fast as for 
       $Q_2(-1)$. (See Section~\ref{sec.flow}). 

 \item Using the rough estimates for the limiting curves for $k=9,11$, we see
       that $Q_\text{max} \gtapprox 9.4$, which is far larger than the maximum
       value $Q_\text{max} = 4$ obtainable for 2D Potts models with either 
       cyclic and toroidal boundary conditions.  
        We can obtain a more accurate estimate of $Q_\text{max}$ by looking at
        the line $v =-4$ (see Figures~\ref{Figures_Z1}--\ref{Figures_Z2}).
        This line seems to cross the ``regular'' part of the BK phase at its
        more distant point from the origin, and has the nice property 
        that the eigenvalue that is dominant on the high-$Q$ side of the 
        crossing comes always from the $\ell=0$ sector for all values of
        $k$ we can access numerically: i.e., $3\le k\le 10$.\footnote{
           For $k=11$, Arnoldi's method did not converge, probably due to
                 a complicated eigenvalue structure nearby.  
        } 
        These results are 
        displayed in Table~\ref{table.prelim}. As a prevention against 
        parity effects, we have fitted the odd-- and even--$k$ data subsets 
        separately. We have used several Ans\"atze: i.e., power law 
        and polynomials of different degrees in $1/k$. From the dispersion of 
        the estimates for $Q_\text{max}$, we conclude that our preferred 
        estimate is $Q_\text{max} \gtapprox 12.4(1)$, 
        as the limiting value of the crossings along the line $v=-4$ only 
        provides a lower bound for the true $Q_\text{max}$.
 
%
%
\begin{table}[hbt]
\centering
\begin{tabular}{r|l}
\hline\hline
\multicolumn{1}{c}{$k$} & \multicolumn{1}{|c}{$Q_c(k)$} \\
\hline 
3 &  3.5918146982 \\
4 &  5.5837328676 \\
5 &  6.6418204973 \\
6 &  7.7533996189  \\
7 &  8.3495427730 \\
8 &  8.8740090544 \\
9 &  9.2660844176 \\
10&  9.5702423022 \\
\hline\hline
\end{tabular}
\caption{\label{table.prelim}
Values of $Q_c(k)$ along the line $v=-4$ for the generalized Petersen 
graphs $G(nk,k)$ for $3\le k \le 10$.
}
\end{table}

\end{itemize} 
%

%
%
\subsection{Criticality of the Berker--Kadanoff phase} \label{sec.res2}

%
%
\begin{figure}
  \vspace*{-1cm}
  \centering
  \begin{tabular}{cc}
  \includegraphics[width=200pt]{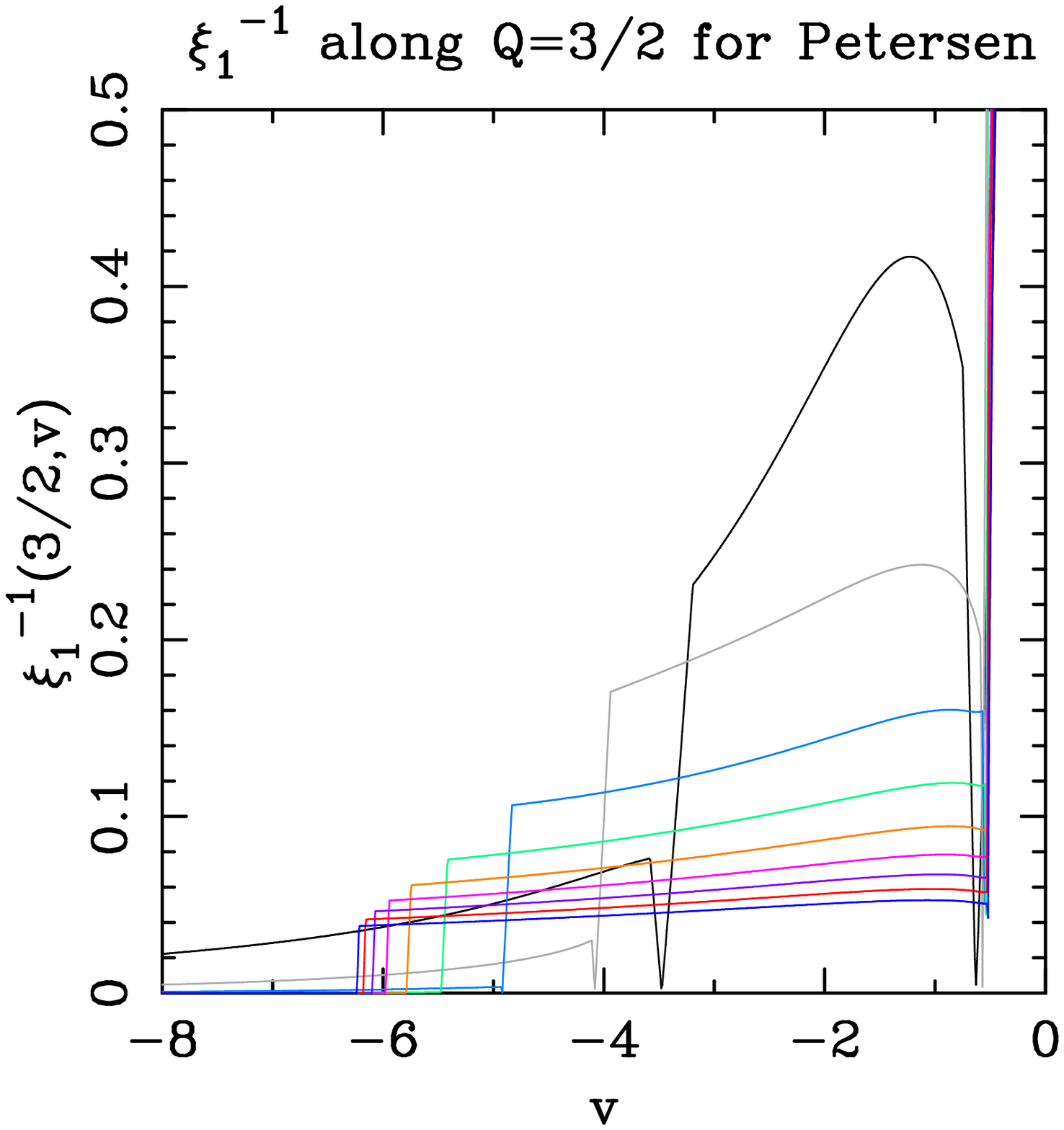} & 
  \includegraphics[width=200pt]{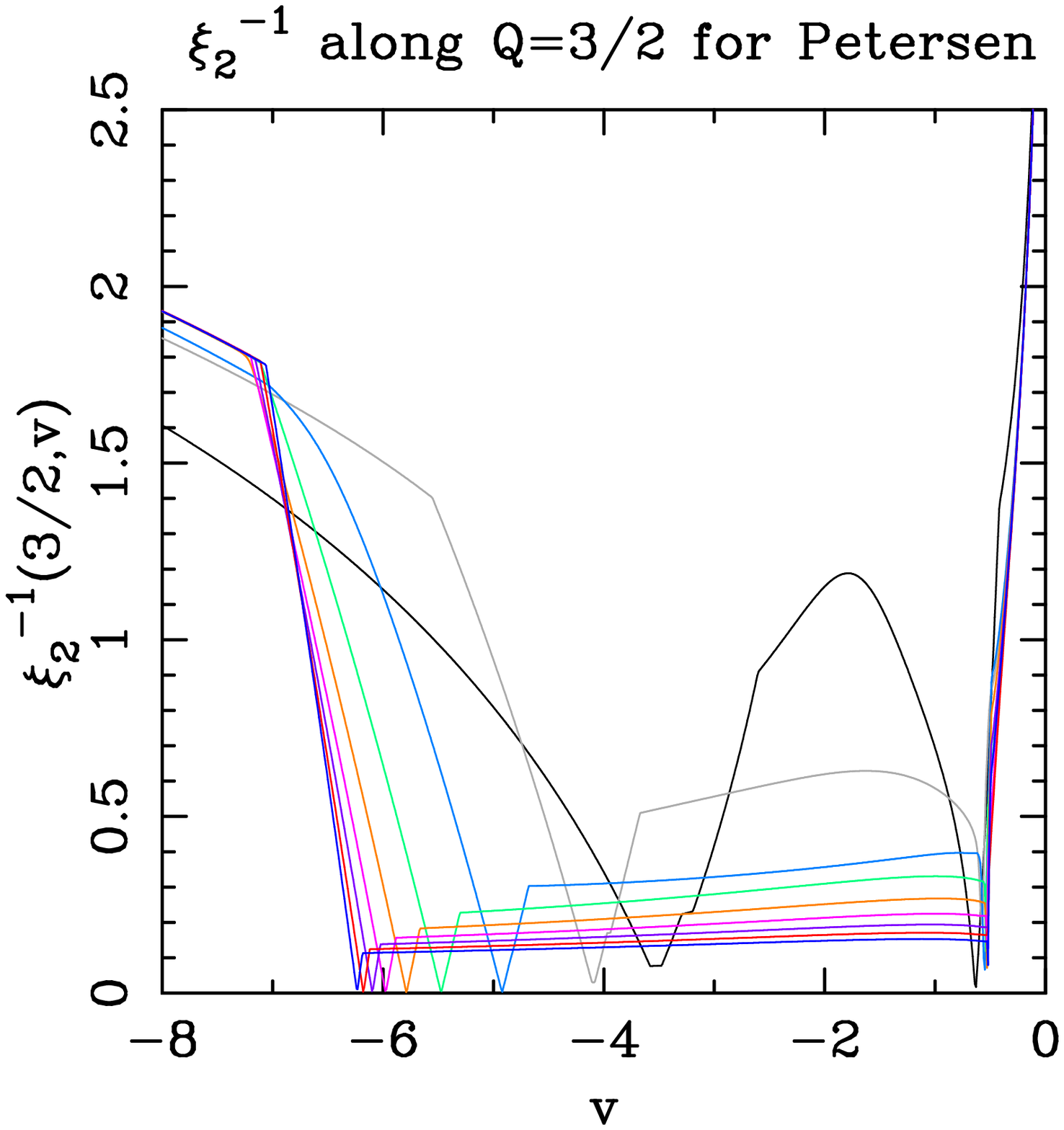} \\
  \qquad (a) & \qquad (b) \\ 
  \includegraphics[width=200pt]{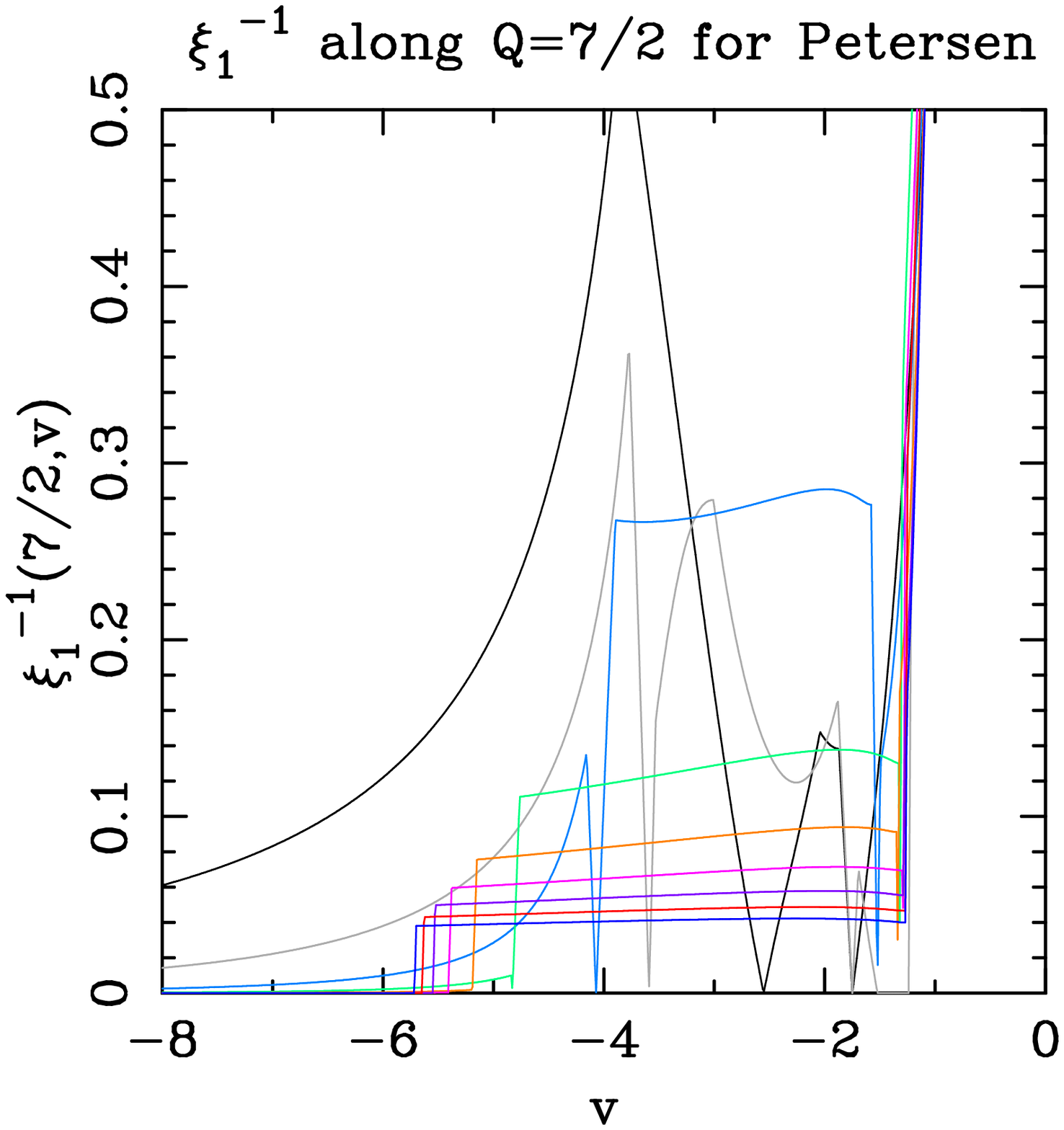} & 
  \includegraphics[width=200pt]{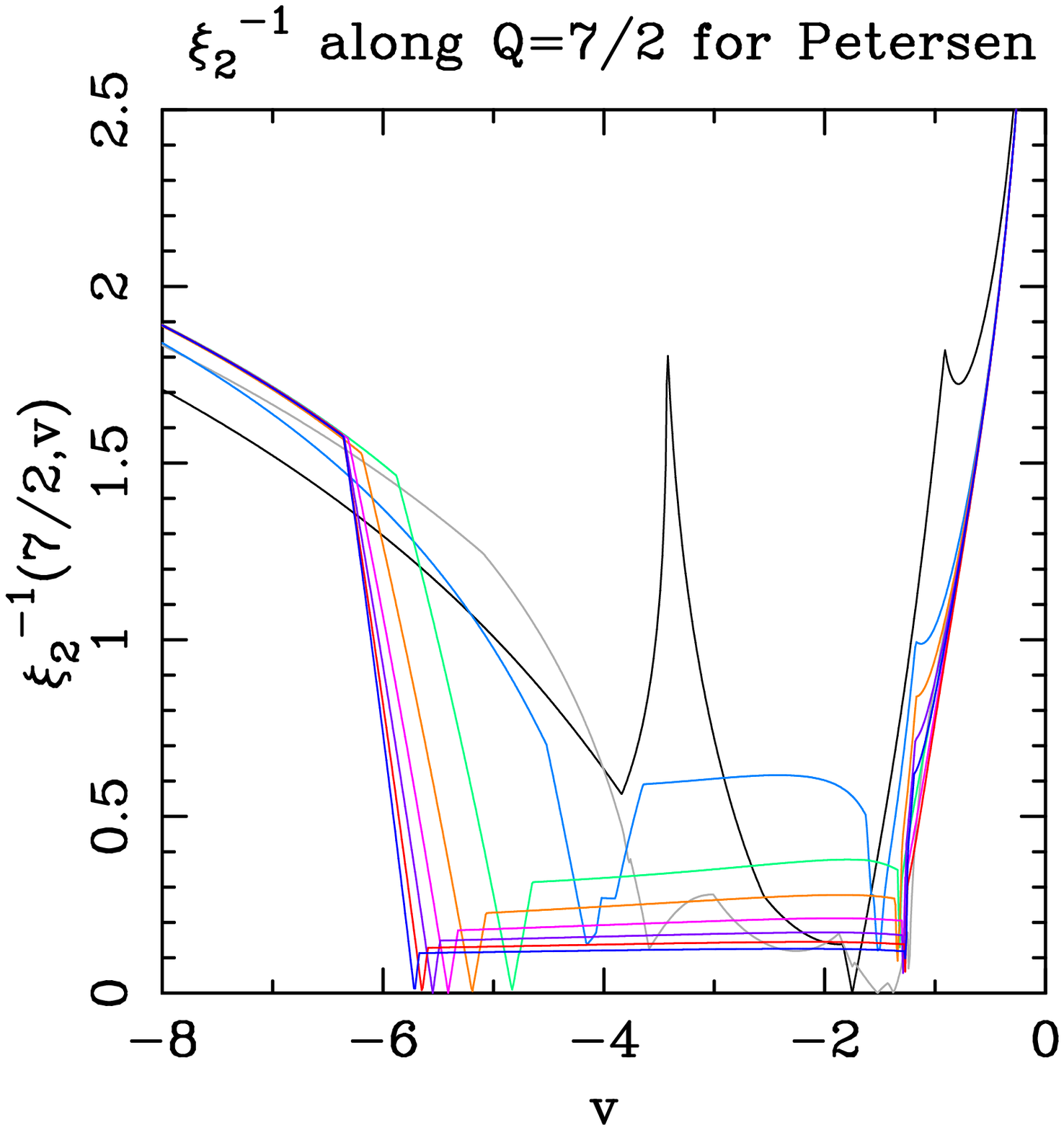} \\
  \qquad (c) & \qquad (d) \\ 
  \end{tabular}
  \caption{
  Inverse correlation lengths $\xi_i^{-1}(Q,v)$ for $Q=3/2$ (a,b),
  and $Q=7/2$ (c,d). The index $i=1$ (resp.\/ $i=2$) corresponds to the 
  the ratio of the first (resp.\/ second) sub-dominant eigenvalue and 
  the dominant one [cf.,~\reff{def_inverse_xi}]. 
  We show the inverse correlation lengths $\xi_i^{-1}$ ($i=1,2$)
  for the Petersen graphs $G(nk,k)$ with $k=1$ (black), $k=2$ (gray), 
  $k=3$ (light blue), $k=4$ (green), $k=5$ (orange), $k=6$ (pink),
  $k=7$ (violet), $k=8$ (red), and $k=9$ (navy blue).
}
\label{Figures_gap}
\end{figure}

The BK phase in 2D Potts models is a critical phase: it corresponds to a 
massless phase with algebraic decaying of correlations. In this section we
want to study whether the BK phase for the non-planar Petersen graphs 
$G(nk,k)$ is critical or not. One way to achieve this goal is to consider 
the inverse correlation length (or mass gap) given by the ratios 
\be
\xi_i^{-1}(Q,v) \;=\; \log \left| \frac{\mu_i(Q,v)}{\mu_*(Q,v)} \right| \,,
\label{def_inverse_xi}
\ee
where $\mu_*$ is the dominant eigenvalue, and $\mu_i$ with $i=1,2,\ldots$ 
is the $i$-th subdominant eigenvalue 
(i.e., $|\mu_*| \ge |\mu_1| \ge |\mu_2| \ge \ldots$). 
In Figure~\ref{Figures_gap} we have shown the values of $\xi_i^{-1}$ 
for $i=1,2$ versus the parameter $v$ at fixed values of $Q=3/2,7/2$.\footnote{
   We have also made computations for $Q=11/2$, but the results are 
   qualitatively very similar to those for $Q=7/2$, so we refrain from 
   displaying the corresponding plots. 
}
Note that for $Q=3/2$ we cross the region of the BK phase dominated by the 
$\ell=1$ sector, while for $Q=7/2$, we cross the nearby region dominated by 
the $\ell=2$ sector. 
We expect to find a disordered phase for $v\in(v_{+}(Q),0]$,
the BK phase for $v\in (v_{-}(Q),v_{+}(Q))$, and another phase for
$v< v_{-}(Q)$. The critical/non-critical nature of the last two phases 
needs to be determined.

%
%
\begin{figure}
  \vspace*{-1cm}
  \centering
  \begin{tabular}{cc}
  \includegraphics[width=200pt]{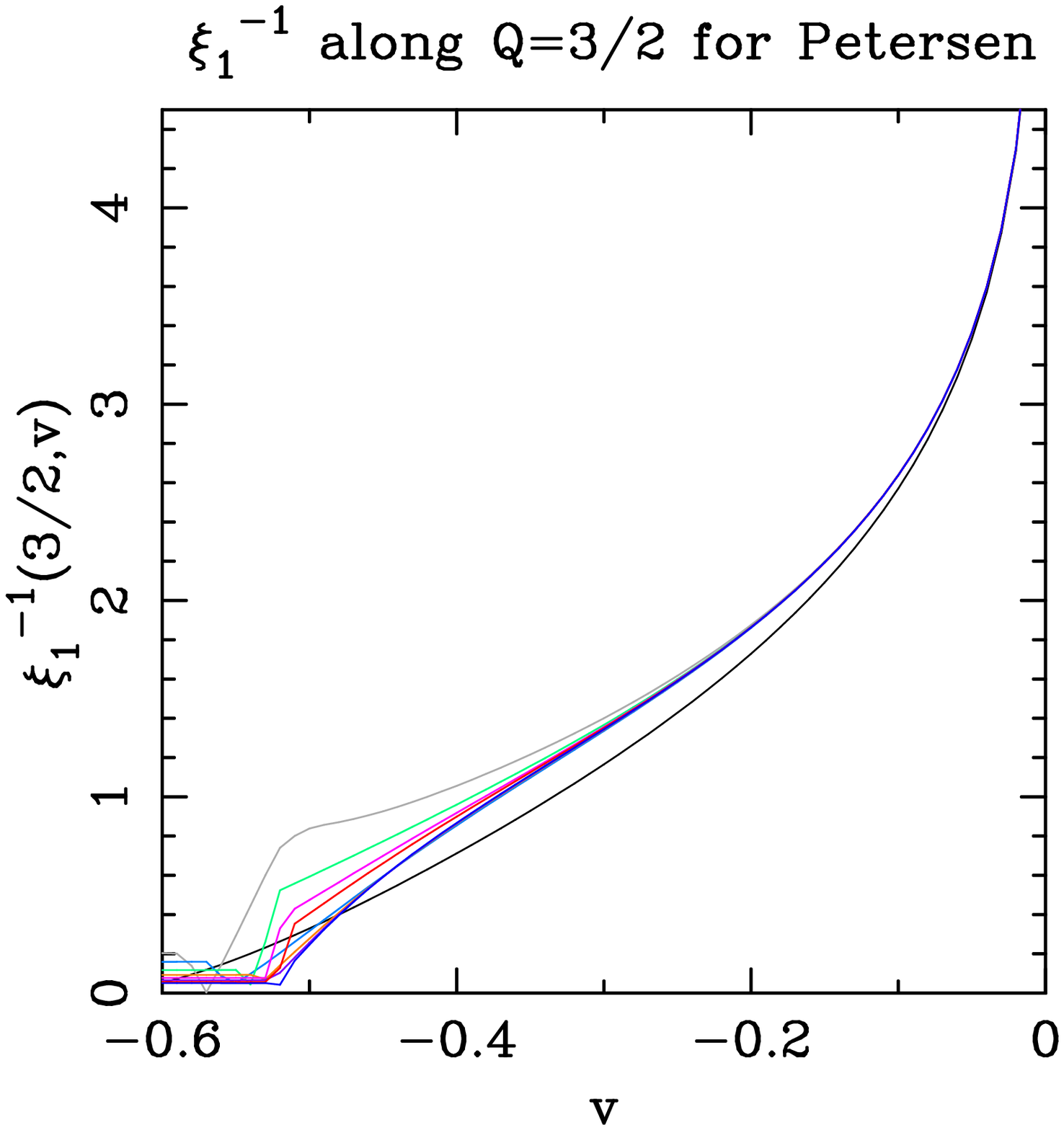} & 
  \includegraphics[width=200pt]{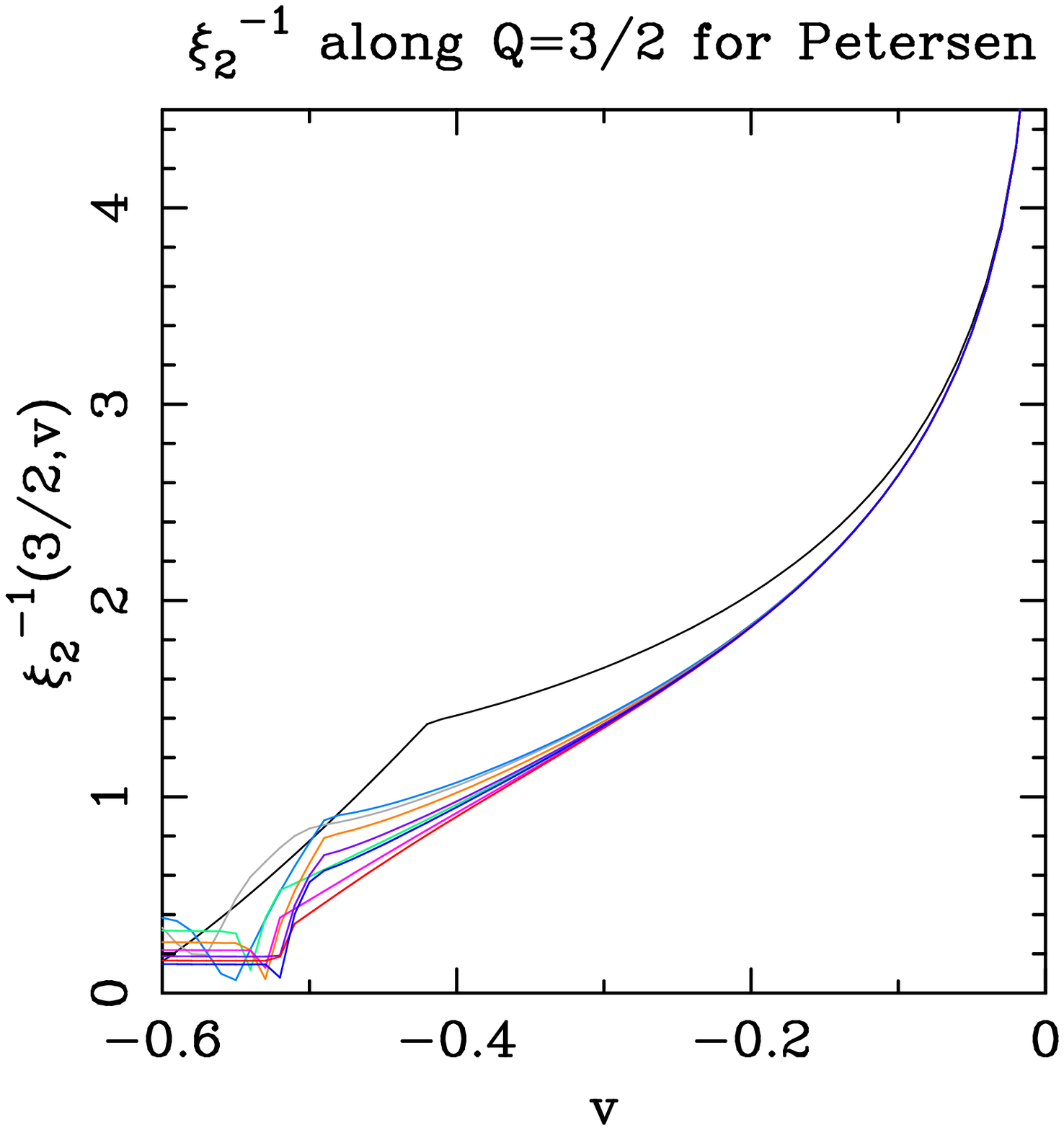} \\
  \qquad (a) & \qquad (b) \\ 
  \end{tabular}
  \caption{
  Zoom-out of Figures~\ref{Figures_gap}(a,b) showing the   
  inverse correlation lengths $\xi_i^{-1}(Q,v)$ for $Q=3/2$ and 
  $i=1$ (a) and $i=2$ (b) for the Petersen graphs $G(nk,k)$.   
  The color code is as in Figure~\ref{Figures_gap}. We show in both  
  panels the curves for $1\le k\le 9$.
}
\label{Figures_gap_region1}
\end{figure}

Let us now start with the first region $v\in(v_{+}(Q),0]$. 
Figure~\ref{Figures_gap} does not give much information about this regime;
so we provide a zoom of this region in Figure~\ref{Figures_gap_region1} 
for $Q=3/2$. Now the behavior of the different curves as $k$ increases is
clear: they tend to some finite limit $\xi^{-1}_{1,\infty}(v) > 0$. This 
limit curve is essentially attained for $k\ge 3$ and $v\gtapprox -0.3$. 
We conclude that in this region there is a {\em finite} gap in the 
thermodynamic limit, and therefore, this is a non-critical phase. 
For $Q=5/2$ the plots are very similar. To summarize, the region 
$v\in(v_{+}(Q),0]$ is a non-critical (disordered) phase. 

%
%
\begin{figure}
  \vspace*{-1cm}
  \centering
  \begin{tabular}{cc}
  \includegraphics[width=200pt]{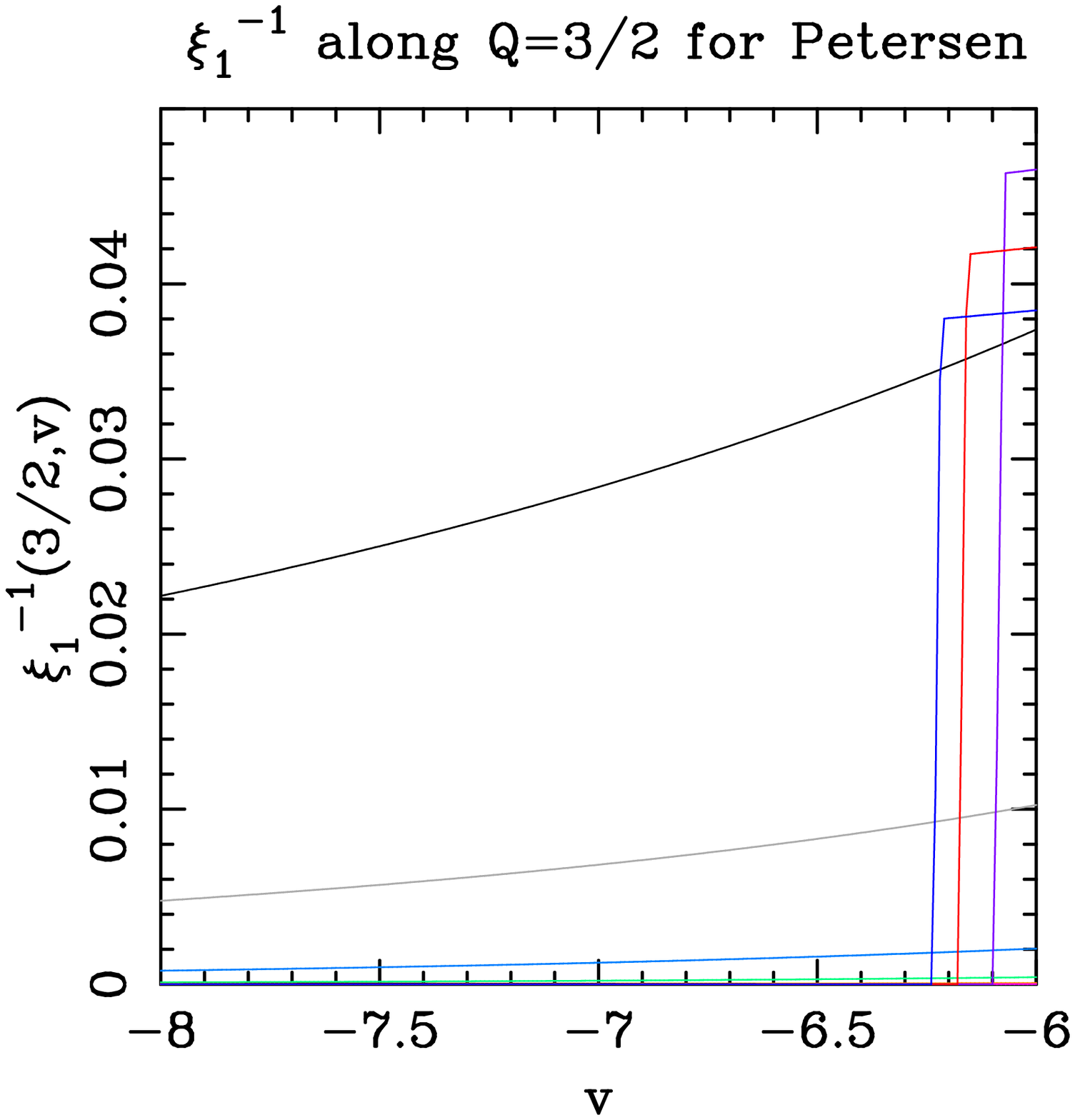} & 
  \includegraphics[width=200pt]{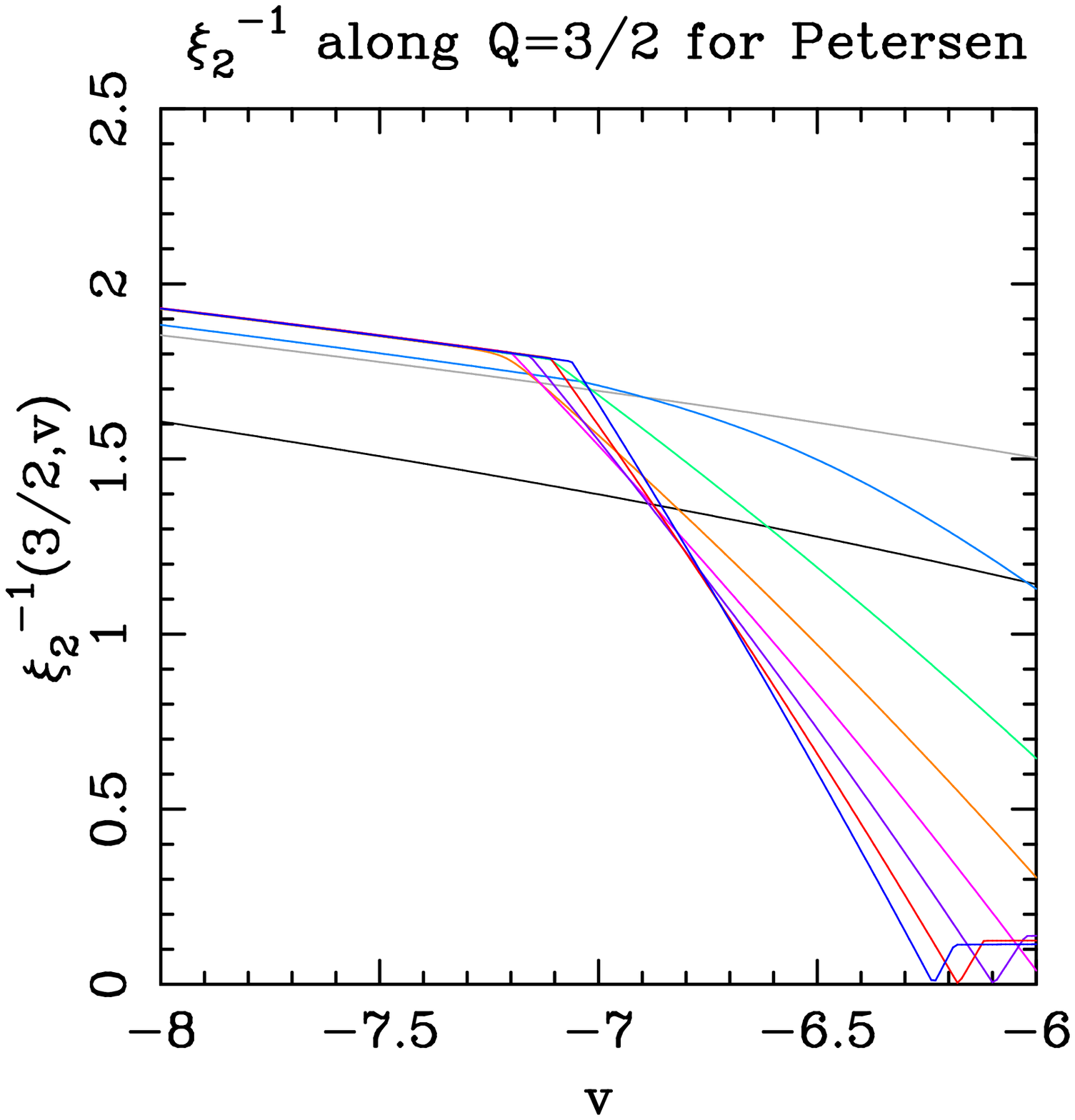} \\
  \qquad (a) & \qquad (b) \\ 
  \end{tabular}
  \caption{
  Zoom-out of Figures~\ref{Figures_gap}(a,b) showing the   
  inverse correlation lengths $\xi_i^{-1}(Q,v)$ for $Q=3/2$ and 
  $i=1$ (a) and $i=2$ (b) for the Petersen graphs $G(nk,k)$.   
  The color code is as in Figure~\ref{Figures_gap}. We show in both
  panels the curves for $1\le k\le 9$.
}
\label{Figures_gap_region3}
\end{figure}

Let us now consider the last region $v\in (-\infty,v_{-}(Q))$.
The most relevant part of this regime is displayed in 
Figure~\ref{Figures_gap_region3} for $Q=3/2$. The plot of $\xi_1^{-1}$ 
displayed in Figure~\ref{Figures_gap_region3}(a) show clearly that this 
inverse correlation length tends to zero as $k$ increases. Therefore, one
could naively conclude that this also a massless phase. There are two 
almost identical eigenvalues, one coming from the $\ell=0$ sector, and the
other from the $\ell=1$ sector. But the plot of $\xi_2^{-1}$ in 
Figure~\ref{Figures_gap_region3}(b) show that the second correlation 
length is {\em finite}. For $k\ge 4$ and $v\gtapprox 7.2$ all 
curves collapse almost perfectly. Therefore, we expect that as $k$ tends 
to infinity, these curves will tend to some finite limit 
$\xi_{2,\infty}^{-1}(v) > 0$. 
The plots for $Q=5/2$ are very similar. Therefore we conclude that the region 
$v\in (-\infty,v_{-}(Q))$ is also a non-critical phase.
The two most dominant eigenvalues for {\em finite} values of $k$ will tend to
a common function, so the mass gap in this phase is given in the 
thermodynamic limit by $\xi_{2,\infty}^{-1}$.

\begin{figure}
  \vspace*{-1cm}
  \centering
  \begin{tabular}{cc}
  \includegraphics[width=200pt]{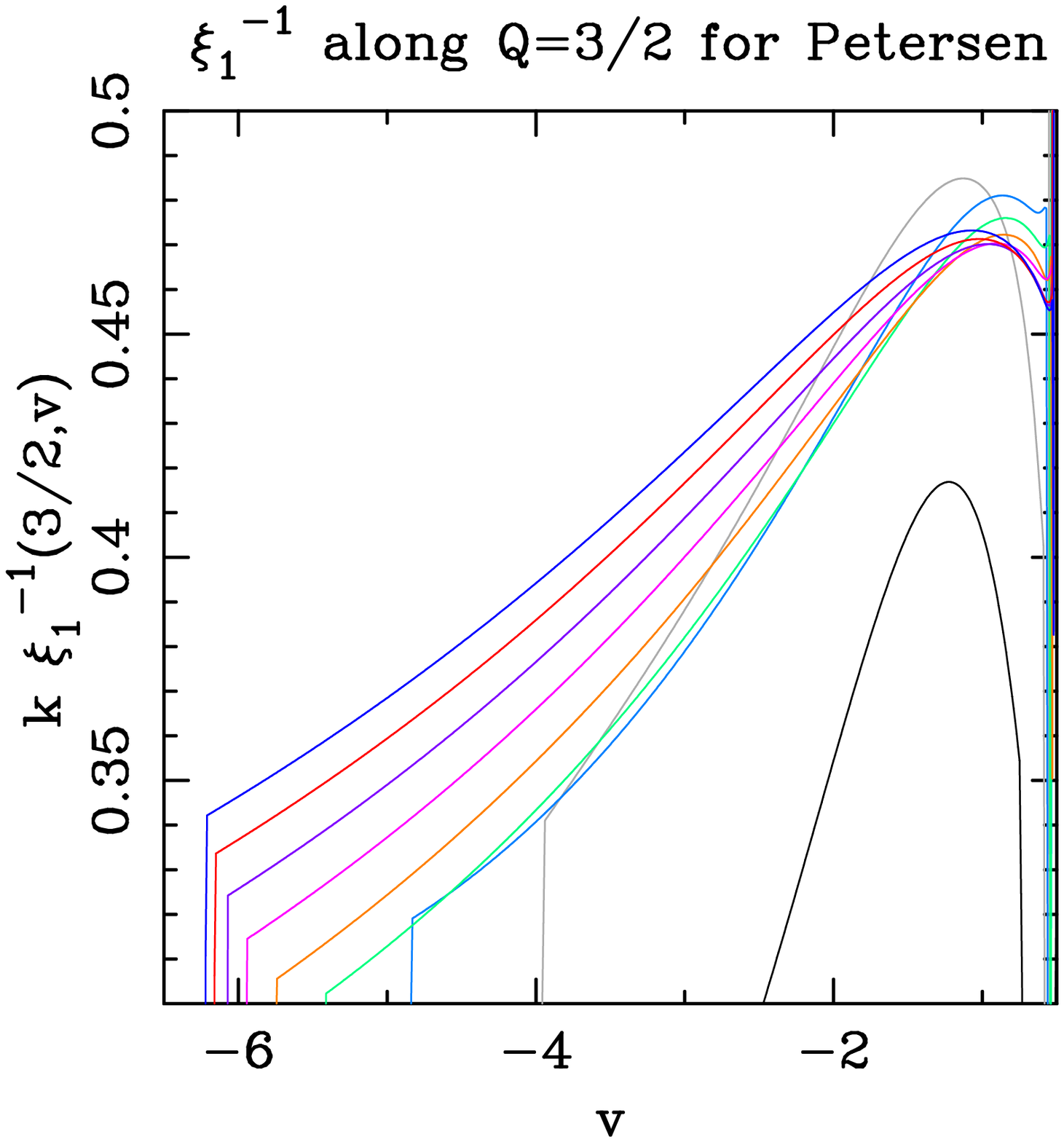} & 
  \includegraphics[width=200pt]{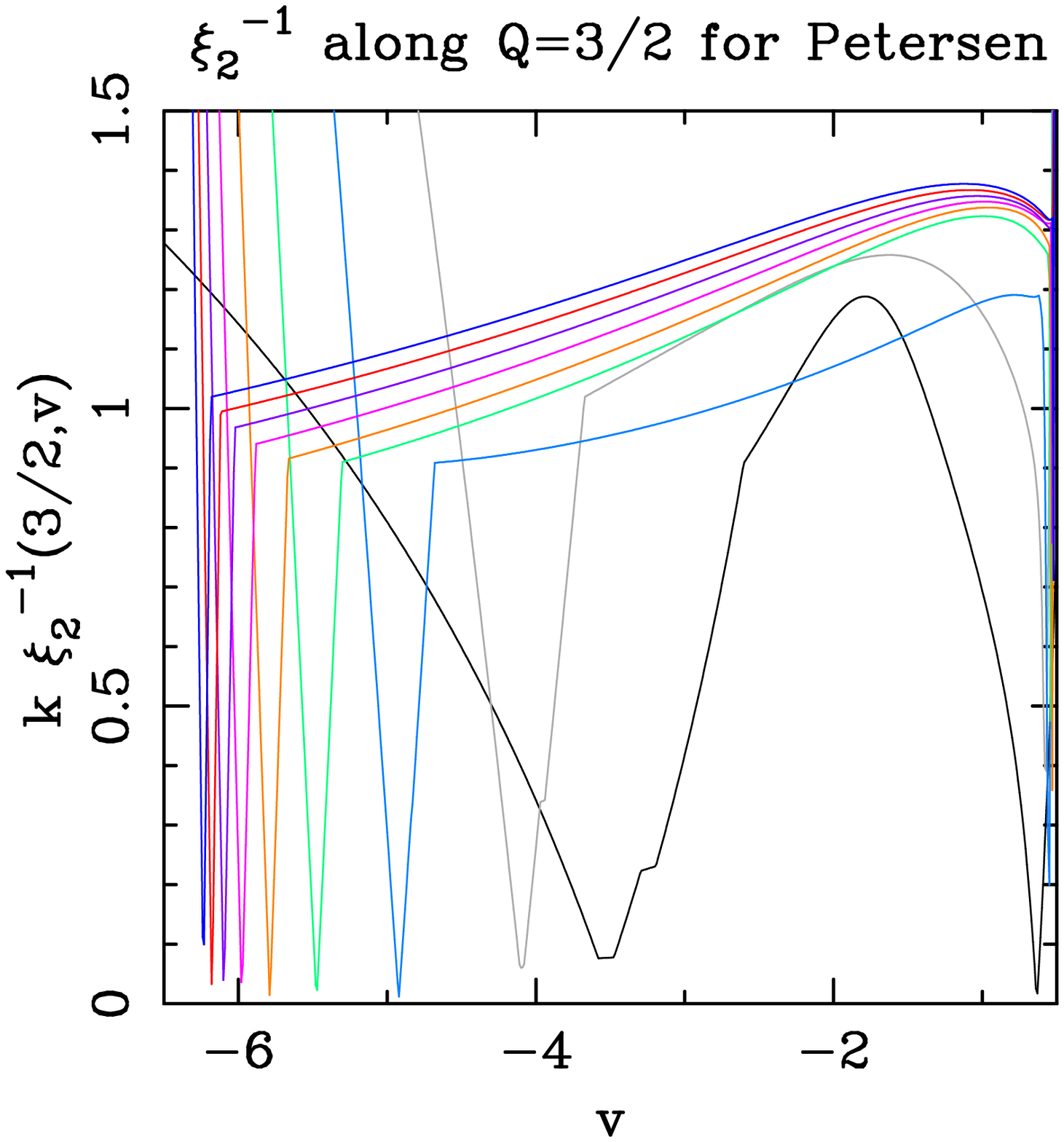} \\
  \qquad (a) & \qquad (b) \\ 
  \includegraphics[width=200pt]{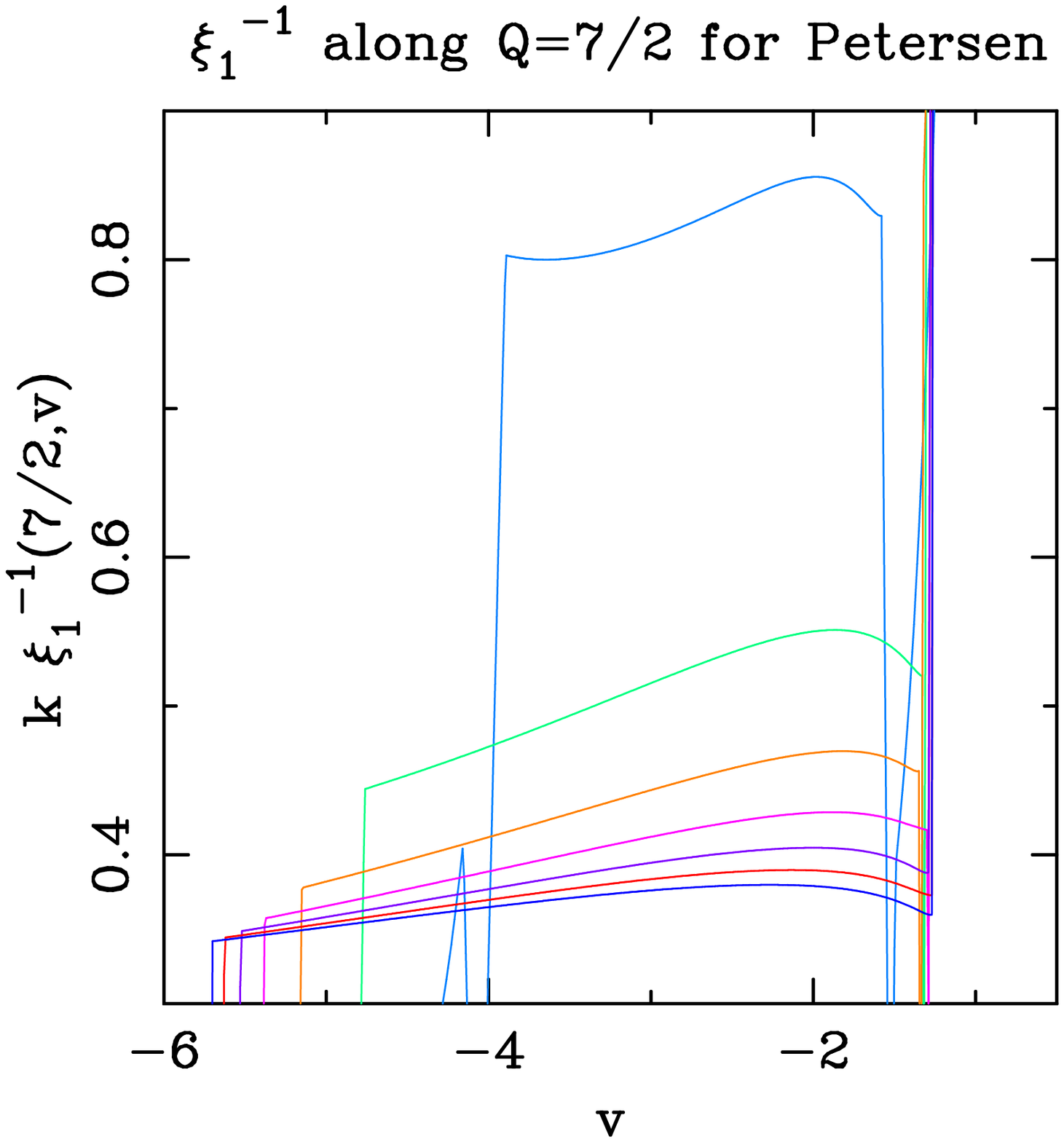} &
  \includegraphics[width=200pt]{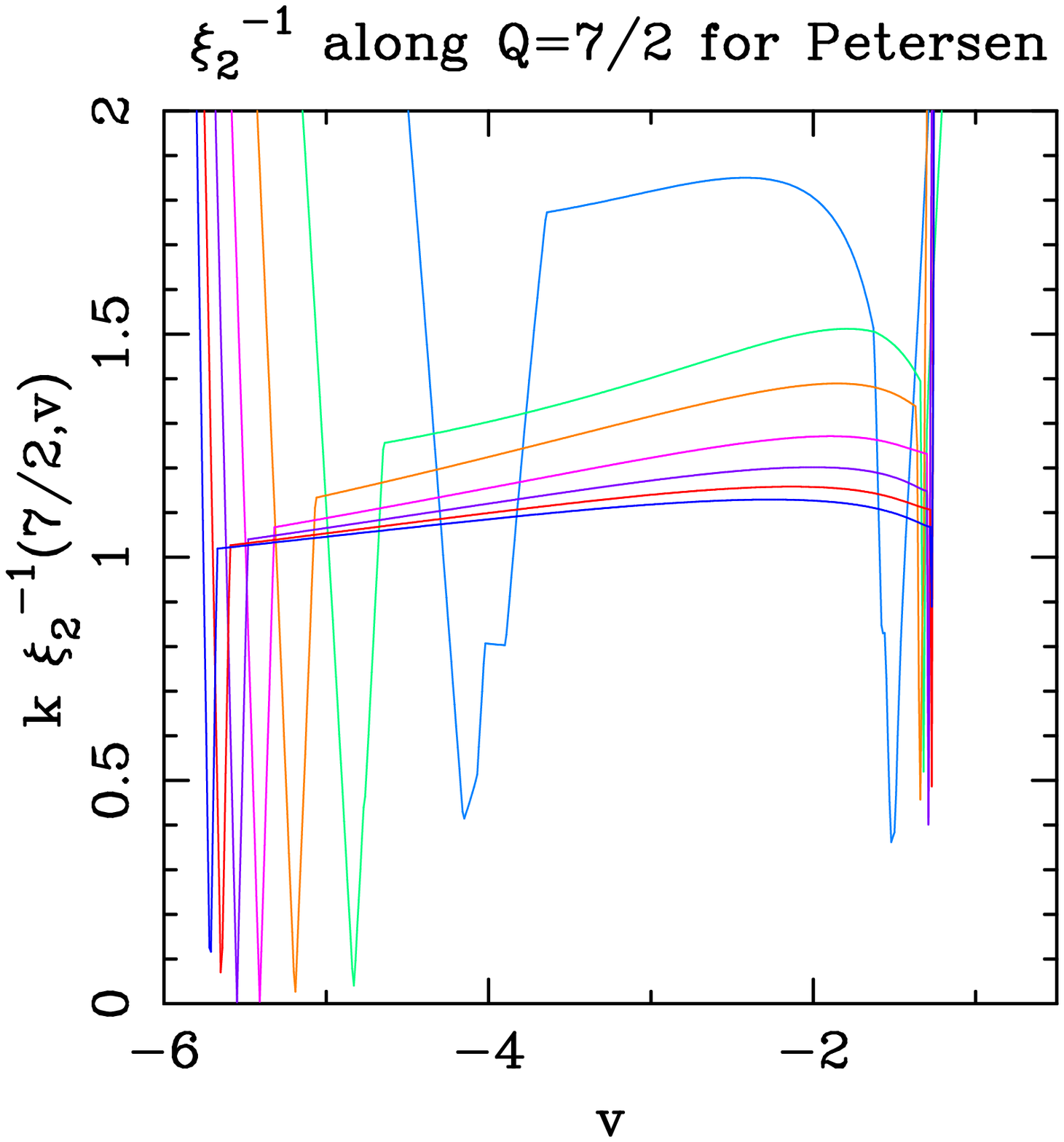} \\
  \qquad (c) & \qquad (d) \\ 
  \end{tabular}
  \caption{
  Zoom-out of Figure~\ref{Figures_gap} showing 
  $k$ times the inverse correlation length,
  $k \xi_i^{-1}(Q,v)$, with $Q=3/2$ (a,b)
  and $Q=7/2$ (c,d) for the Petersen graphs $G(nk,k)$ in the region 
  $v\in( v_{-}(Q),v_{+}(Q))$ for $i=1$ (a,c) and $i=2$ (b,d).  
  For $Q=3/2$, we show in panels (a,b) the curves for 
  $1\le k\le 9$. For $Q=7/2$, we only show in panels (c,d) those 
  curves with $3\le k\le 9$, as for smaller values of $k$ the line 
  $Q=7/2$ crosses the limiting curves $\mathcal{B}_k$ at their non-regular 
  part. The color code is as in Figure~\ref{Figures_gap}.  
}
\label{Figures_gap_region2}
\end{figure}

Finally, we focus on the BK phase $v\in (v_{-}(Q),v_{+}(Q))$. 
Figure~\ref{Figures_gap} shows that in this region the curves 
$\xi_i^{-1}$ are rather flat around some non-zero value for $k\ge 3$, 
and as $k$ increases, this value becomes smaller. This is an indicator that
as $k$ increases the mass gap tends to zero in this phase. Therefore, 
we expect that this is a critical (massless) phase (as for the 2D Potts
model). In 2D models, CFT predicts that the mass gap for a critical theory 
defined on a strip graph of width $L$ and cylindrical boundary conditions 
behaves like  $\xi_i^{-1} = A_i/L + o(L^{-1})$, where $A_i$ is 
a {\em universal} constant, proportional to the critical exponent that
governs the algebraic decay of an appropriate correlation function.
For the non-planar recursive strip graphs studied here, there is no such strong
link to CFT, but it is nevertheless true that $k$ roles as the effective width
of the strip in the transfer matrix formalism \cite{JS_flow}. Critical behavior
will therefore result if the mass gap for the non-planar Petersen 
graphs $G(nk,k)$ behaves asymptotically like $1/k$. To determine whether this
is the case, we have plotted in Figure~\ref{Figures_gap_region2} the quantity 
$k \, \xi_i^{-1}$ versus $v$ for fixed values of $Q=3/2,7/2$.  

%
%

Even though there are obvious finite-size effects, for $Q=3/2$ it 
looks like the curves tend to some limiting curve from below. However, for
$Q=7/2$,\footnote{
  For $Q=11/2$, the finite-$k$ curves seem to tend to some 
  limiting curve from above. 
}
this convergence is from above. This means that there is a change of sign in 
the dominant finite-size-scaling correction term. The best convergence is 
clearly attained for $Q=7/2$ and $i=1$. We have tried to fit the data
for $Q=3/2$ and $7/2$ and $i=1,2$ to an Ansatz 
\be
k \xi_i^{-1}(Q,v) \;=\; g_i(Q,v) + \frac{h_i(Q,v)}{k} \,,
\ee
where the second term intends to mimic the finite-size-scaling corrections.
For the four cases we have considered here, we find reasonable fits 
leading to smooth functions $g_i(Q,v)$ as functions of $v$.  
Therefore, we conjecture that in the BK phase  
\be  
\xi_{i}^{-1}(Q,v) \;=\; \frac{ g_i(Q,v)}{k} + o(k^{-1}) \,. 
\ee

This implies that the BK phase is {\em critical} in the sense that it is
massless. This coincides with the properties of the BK phase found for 
2D Potts models. 

In Section~\ref{sec.res1}, we have shown that the limiting curves 
$\mathcal{B}_k$ contained vertical lines for even integer values of 
$Q=0,2,\ldots$ and $v$ in the interval $v\in (v_{-}(Q),v_{+}(Q))$. 
The perfect verticality of these lines implies that the temperature 
is an irrelevant operator in the RG sense. Moreover, in 
Section~\ref{sec.res2}, we have shown that inside 
the region $v\in (v_{-}(Q),v_{+}(Q))$ for fixed values of $Q$, the
mass gap vanishes in the infinite-$k$ limit. This implies that
this region is critical. Therefore, both conditions that define the BK
phase (as we know from 2D Potts models) are met also for the $Q$-state 
Potts model on the Petersen graphs $G(nk,k)$.

We conclude that the BK phase exists also at least for recursive 
families of non-planar graphs, and a role analogous to the one played by 
the Beraha numbers 
for planar graphs is now played by the non-negative integers for non-planar
graphs. The natural question is to know, in the non-planar case, 
whether there are also cancellations among eigenvalues and vanishing of 
amplitudes when $Q$ is a non-negative integer. This question will be 
answered in the next section.

%
%
\section{What happens for integer values of $\bm{Q}$?} \label{sec.integers}

In this section we will consider the partition function for the
generalized Petersen graphs $Z_{G(nk,k)}(Q,v)$ when $Q$ takes a
non-negative integer value. 

Let us fix $k\ge 3$. For each $0\le \ell\le k$, let us denote by 
$\mathcal{E}_\ell$ the set of {\em non-trivial} eigenvalues 
$\mu_{k,\ell,\lambda,s}$ associated to the sector of $\ell$ links.
Moreover, let $\mathcal{E}_{k+1}$ be the one-element set 
containing only the trivial eigenvalue $\mu_{k,k+1}=v^{2k}$. In the same way, 
let $\mathcal{E}_{\ell,\lambda}$ denote the
set of {\em non-trivial} eigenvalues $\mu_{k,\ell,\lambda,s}$ corresponding to
the irreducible representation $\lambda$ of the group $S_\ell$.

When $Q$ is a non-negative integer, some of the non-trivial eigenvalues of
${\sf T}_{k+1,\ell,\lambda}^{(nt)}$ (for $0\le \ell \le k-1$) or 
${\sf T}_{k+1,k}^{(nt)}$ (for $\ell=k$) may reduce to the trivial eigenvalue 
$\mu_{k,k+1}=v^{2k}$. These eigenvalues are excluded by definition 
from the sets $\mathcal{E}_\ell$ and $\mathcal{E}_{\ell,\lambda}$ defined 
above for $0\le\ell\le k$.

Finally, let us denote by $\chi_{k,\ell,\lambda}$ the 
{\em net non-trivial}
contribution to the partition function  $Z_{G(nk,k)}$ 
[cf.,~\reff{def_Z_petersen3}] of the sector with $\ell$ links and the 
irreducible representation $\lambda$ of $S_\ell$:
\be
\chi_{k,\ell,\lambda} \;=\; 
\left.\sum_{s}\right.^\prime  \mu_{k,\ell,\lambda,s}^n \,.
\label{def_chi}
\ee
The prime on the sum in the r.h.s.\/ of \reff{def_chi} means that we 
exclude, for a particular non-negative integer value of $Q$, 
both 1) those non-trivial eigenvalues of ${\sf T}_{k+1,\ell,\lambda}^{(nt)}$ 
that become identical to the trivial one $v^{2k}$, and 2) those eigenvalues 
in $\mathcal{E}_{\ell,\lambda}$ that are exactly canceled by the contribution 
of other sectors. 
For $\ell=k$, we use the eigenvalues $\mu_{k,k}$ of the reduced matrix 
${\sf T}_{k+1,k}^{(nt)}$ instead, and $\lambda$ plays no role. 
For $\ell=0,1$, the sub-index $\lambda$ is also superfluous and can 
be dropped.

We can obtain the eigenvalues of each transfer matrix
${\sf T}_{k+1,\ell,\lambda}$ by computing its characteristic
polynomial symbolically using {\sc Mathematica}. This procedure can
be carried out for $k=2,3,4$; but for $k=5$, this is not feasible due to the
large dimension of some of the transfer matrices. We have observed that
the characteristic polynomials obtained in this way cannot be
factorized over the integers. (This is not true in general when $Q$
takes integer values, as shown below.) Indeed, if we find that two
transfer matrices have characteristic polynomials with a common factor, then
we can deduce that the eigenvalues coming from the common factor appear
in both matrices. On the contrary, if the characteristic polynomials of two
transfer matrices have no common factor, then they do not contain any common
eigenvalue. For $k=5$ we have numerically computed the eigenvalues to high
precision (i.e., 100 digits) using {\sc Mathematica}. For these cases, we
could check the conjectures obtained from the smaller values for $k$.

We summarize our findings for the smallest integer values by using the
exact results for $k=3,4,5$ (which correspond to non-planar generalized
Petersen graphs).

%
%
\subsection{$\bm{Q=0}$}

This case is trivial, as $Z_G(0,v) = 0$ for any graph with at least one 
vertex. Therefore, there should be a massive cancellation of eigenvalues 
and amplitudes, so that the net sum is always zero.

First of all, it is clear from \reff{eigen_amp}, that only the fully
anti-symmetric representations of $S_\ell$ give a non-zero contribution.
In fact, if we represent by $(1^\ell)$, the fully anti-symmetric 
representation of $S_\ell$, then
\be
\alpha_{\ell,(1^\ell)}(0) \;=\; (-1)^\ell\,.
\ee
The two non-generic amplitudes are also \cite{JS_flow}:
\be
\beta_k(0) \;=\; \gamma_{k}(0) \;=\; (-1)^k\,.
\ee

We find that $\mathcal{E}_0 \subset \mathcal{E}_1$: i.e., the set of 
$\ell=0$ eigenvalues  is a proper subset of those of the next sector. 
For the higher sectors (except the last one), we find that
$\mathcal{E}_{\ell}\setminus \mathcal{E}_{\ell-1} \subset \mathcal{E}_{\ell+1}$.
In words, the non-trivial eigenvalues of the sector with $\ell$ links
that do not  belong to the sector with $\ell-1$ links, are a proper
subset of the set of non-trivial eigenvalues of the sector
with $\ell+1$ links. The last step corresponds to $\ell=k$, and it reads
$\mathcal{E}_k\setminus \mathcal{E}_{k-1} = \emptyset$. (Recall that
$\mathcal{E}_\ell$ for $0\le \ell\le k$ contains non-trivial eigenvalues,  
while $\mathcal{E}_{k+1}$ contains the trivial eigenvalue  $v^{2k}$.)

When $Q=0$, some of the eigenvalues $\mu_{k,\ell,\lambda}$ in 
\reff{def_Z_petersen3} reduce to the trivial one $\mu_{k,k+1}=v^{2k}$.
These new trivial eigenvalues appear with some multiplicity $n_{k,\ell}>0$
in sectors $1\le \ell \le k$ (resp.\/ $2\le \ell\le k$) 
for odd (resp.\/ even) values of $k$ 
(e.g., when $k=4$, $n_{4,2}=n_{4,3}=3$, and $n_{4,4}=1$). 
But their net contribution always cancels the last term on the r.h.s.\/ of 
Eq.~\reff{def_Z_petersen3}:
\be
\sum\limits_{\ell=1}^k n_{k,\ell} \, (-1)^\ell + \gamma_{k+1}(0) \;=\; 0 \,.
\ee

The above inclusion-exclusion patterns, in addition to the values taken by
the amplitudes, imply that {\em all} eigenvalues cancel out exactly, giving
the expected result $Z_{G(nk,k)}(0,v)=0$.

%
%
\subsection{$\bm{Q=1}$}

This case is also trivial, as for any graph $G=(V,E)$, $Z_G(1,v) = (1+v)^{|E|}$.
This comes from the FK representation \reff{def_Z_Potts_FK}
of the $Q$-state Potts model. Recall that for the generalized Petersen graph
$G(nk,k)$, $|E|=3nk$.

The generic amplitudes $\alpha_{\ell,\lambda}(1)$ are non-zero only for
the following cases:
\begin{itemize}
\item $\ell=0$ with $\alpha_0 = 1$.
\item $\ell=2$ and the representation $(2)$ with $\alpha_{2,(2)}=-1$.
\item For each $\ell\ge 3$, the representation $(2,1^{\ell-2})$, 
      with $\alpha_{\ell,(2,1^{\ell-2})}=(-1)^{\ell-1}$.
\end{itemize}
The non-generic amplitudes \cite{JS_flow} take the values
\be
\beta_k      \;=\; (k-1)(-1)^{k-1} \,, \qquad
\gamma_{k+1} \;=\; (-1)^k \,.
\ee

The pattern for the non-trivial eigenvalues is as follows:
For $\ell=0$, we have the decomposition 
$\mathcal{E}_0 = \{ \mu_0 \} \cup \mathcal{E}''_0$, with $\mu_0 = (1+v)^{3k}$.  
Then an inclusion-exclusion pattern similar to the $Q=0$ case occurs:
$\mathcal{E}''_0\subset \mathcal{E}_2$,
$\mathcal{E}_{\ell}\setminus \mathcal{E}_{\ell-1} \subset \mathcal{E}_{\ell+1}$
for $3\le \ell\le k-1$, and
$\mathcal{E}_k\setminus \mathcal{E}_{k-1} = \emptyset$.

There is a slight difference with the $Q=0$ case: the non-zero amplitude for
$\ell=k$ is $(k-1)(-1)^{k-1}$, so that the corresponding (non-trivial)
eigenvalues in $\mathcal{E}_k$ contribute to the partition function
with this amplitude. But these eigenvalues also appear in the preceding
sector $\ell=k-1$ with a degeneracy exactly equal to $k-1$; therefore,
these two contributions cancel out in the partition function.

The trivial eigenvalue $v^{2k}$ cancels out in all cases $3\le k \le 5$. 
Unfortunately, we have been unable to find a simple pattern for the 
multiplicities and the amplitudes in this case. 

Therefore, all eigenvalues cancel out exactly, except the simple eigenvalue
$\mu_0 = (1+v)^{3k}$ coming from the $\ell=0$ sector. This eigenvalue gives
the right value of the partition function when $Q=1$. Using \reff{def_chi},
we can write the partition function as
\be
Z_{G(nk,k)}(1,v) \;=\; \chi_{k,0} \;=\; (1+v)^{3nk} \,.
\ee

%
%
\subsection{$\bm{Q=2}$}

This case corresponds to the Ising model on the family of graphs $G(nk,k)$.
The generic amplitudes $\alpha_{\ell,\lambda}(2)$ \reff{eigen_amp} 
are non-zero only for the following cases:
\begin{itemize}
\item $\ell=0$  with $\alpha_0 = 1$.
\item $\ell=1$  with $\alpha_1 = 1$.
\item $\ell=2$ and the representation $(2)$ with $\alpha_{2,(2)}=-1$.
\item $\ell=3$ and the representation $(3)$ with $\alpha_{3,(3)}=-1$.
\item For each $\ell\ge 4$ and the representations $(3,1^{\ell-3})$ and
      $(2^2,1^{\ell-4})$, with
      $\alpha_{\ell,(3,1^{\ell-3})}=\alpha_{\ell,(2^2,1^{\ell-4})}=(-1)^\ell$.
\end{itemize}
The non-generic amplitudes take the values \cite{JS_flow}:
\be
\beta_k      \;=\; (k^2-3k+1)(-1)^k \,, \qquad
\gamma_{k+1} \;=\; -1 \,.
\ee

The pattern for the non-trivial eigenvalues is more involved than for
the previous cases. In the sector $\ell=0$, we find two types of
non-trivial eigenvalues: $\mathcal{E}_0 = \mathcal{E}'_0 \cup \mathcal{E}''_0$.
The first sub-set contains $|\mathcal{E}'_0|=2^k$ eigenvalues, and these 
eigenvalues do contribute to the partition function $Z_{G(nk,k)}(2,v)$. 
The second subset is contained in $\mathcal{E}_3$ 
($\mathcal{E}''_0\subset \mathcal{E}_3$); but the amplitudes of these two
sectors are equal in absolute value with distinct signs, so they 
give a zero net contribution to $Z_{G(nk,k)}(2,v)$. 
In the sector $\ell=1$, a similar phenomenon occurs:
$\mathcal{E}_1 = \mathcal{E}'_1 \cup \mathcal{E}''_1$, with
$|\mathcal{E}'_1|=2^k$ contributing eigenvalues, and the rest cancel out
with part of the $\ell=2$ sector: $\mathcal{E}''_1\subset \mathcal{E}_2$.
The non-trivial eigenvalues in $\mathcal{E}_2\setminus \mathcal{E}''_1$
and $\mathcal{E}_3\setminus \mathcal{E}''_0$ also appear in the higher
sectors $\ell\ge 4$, and their net contribution is always zero.
We also find for $k=4,5$ and sector $\ell=k$, $2k+1$ 
distinct eigenvalues with amplitude $\beta_k$. The same $2k+1$ non-trivial 
eigenvalues also appear at lower sectors with multiplicities that 
make the their net contribution equal to zero.

The trivial eigenvalue $v^{2k}$ cancels out in all cases
$3\le k \le 5$. But its pattern does not look simple, and probably has
parity effects.

Therefore, all eigenvalues cancel out exactly, except the $2^{k+1}$
eigenvalues in the set $\mathcal{E}'_0 \cup \mathcal{E}'_1$. This is
expected, as on each layer of $G(nk,k)$  there are exactly
$k+1$ vertices, and hence the corresponding transfer matrix in the
{\em spin} representation has dimension $2^{k+1}$. Therefore, the
partition function reads:
\be
Z_{G(nk,k)}(2,v) \;=\; \chi_{k,0} + \chi_{k,1} \,, 
\ee
where each $\chi_{k,\ell}$ contains $2^k$ non-zero eigenvalues.  
This formula looks like the expression found for the RSOS representation 
of the 2D Ising model \cite{RSOS}. 

%
%
\subsection{$\bm{Q=3}$}

The situation is now more involved, but the result is rather simple:
\be
Z_{G(nk,k)}(3,v) \;=\; \chi_{k,0} + 2\chi_{k,1} + \chi_{k,2,(1^2)}\,,
\ee
where there are $\lfloor 3^k/2 \rfloor$+1, $3^k$, and 
$\lfloor 3^k/2 \rfloor$ non-zero eigenvalues in the above 
``characters'', respectively. In total, there are $2\times 3^k$
contributing eigenvalues in the set 
$\mathcal{E}'_0 \cup \mathcal{E}'_1 \cup \mathcal{E}'_2$, where we have 
split each eigenvalue set 
$\mathcal{E}_k = \mathcal{E}'_k \cup \mathcal{E}''_k$, as in the case
$Q=2$. The final formula resembles the formula found for the RSOS 
representation of the 3--state Potts model in 2D \cite{RSOS}.

We also find for $k=4,5$ the following patterns for the other eigenvalues 
(whose net contribution to $Z_{G(nk,k)}(3,v)$ cancels out): 
in the sector $\ell=1$,
$\mathcal{E}_1\setminus \mathcal{E}'_1 \subset \mathcal{E}_{3,(3)}$.
In the sector $\ell=2$, the only non-vanishing amplitude is
$\alpha_{2,(1^2)}=1$, while for sector $\ell=3$, there are two
non-zero amplitudes $\alpha_{3,(3)}=-2$ and $\alpha_{3,(2,1)}=-1$.
We then find that 
$\mathcal{E}_2\setminus \mathcal{E}'_2 \subset \mathcal{E}_{3,(2,1)}$.
Let us note that $Q=3$ is inside the phase characterized by
$\ell=2$ links; but from this sector most eigenvalues cancel out.

The trivial eigenvalue $v^{2k}$ cancels out in all cases
$3\le k \le 5$. But its pattern does not look simple, and probably has
again parity effects.

%
%
\subsection{$\bm{Q=4}$}

The analysis of the eigenvalues for $Q=4$ is more complicated (in fact,
for $k=3$ we do not find any cancellation at all!). For $k=4,5$ we
do find cancellations; in particular, all eigenvalues associated to
negative generic amplitudes $\alpha_{\ell,\lambda}$ cancel out, so that
we conjecture the following formula:
\begin{eqnarray}
Z_{G(nk,k)}(4,v) &=& \chi_{k,0} + 3\chi_{k,1} + 2\chi_{k,2,(2)} 
                  + 3\chi_{k,2,(1^2)} + \chi_{k,3,(1^3)} \nonumber\\
                 & & \qquad + A v^{2kn} I[\text{$k$ even}]\,,
\end{eqnarray}
where $I[A]$ is the indicator function for the event $A$ ($I[A]=1$ if $A$ is
true, and zero, otherwise). We cannot determine the value of the constant
$A$ (or even if it is really a constant), as we only have one data
point ($k=5$).

The non-trivial eigenvalues associated to $\ell=3$ and $\lambda=(3)$ (with
$\alpha_{3,(3)}=-2$) cancel out with those appearing in $\mathcal{E}_{2,(2)}$:
i.e., $\mathcal{E}_{3,(3)}\subset \mathcal{E}_{2,(2)}$. For $k=5$, we
find $11$ distinct eigenvalues for $\ell=k$. These eigenvalues appear also
for $(\ell,\lambda)= (1,(1)), (4,(3,1)), (4,(2,1^2))$; and their net
contribution is zero.
The other 305 eigenvalues that appear in $\mathcal{E}_{4,(3,1)}$
(resp.\/ $\mathcal{E}_{4,(2,1^2)}$) cancel with those appearing in
$\mathcal{E}_{2,(1^2)}$ (resp.\/ $\mathcal{E}_{3,(1^3)}$). In summary,
all eigenvalues associated to negative amplitudes cancel out (except the
trivial one for even values of $k$), and there are only five non-trivial
contributions to the partition function. (Indeed, this is in sharp contrast
with the infinite sum obtained using the RSOS representation for the 4-state
Potts model on a planar graph \cite{RSOS}).

%
%
\subsection{Conjectured behavior for integer $\bm{Q}$}

The above results for $Q=0,1,2,3,4$ suggest the following conjecture 
for the partition function of the generalized Petersen graphs 
$Z_{G(nk,k)}(Q,v)$ \reff{def_Z_petersen2} when $Q$ takes a non-negative
integer value.

\begin{conjecture}
Let us fix $Q$ to a non-negative integer value $N$. Then the partition function 
of the generalized Petersen graphs \reff{def_Z_petersen2} reduces for 
any $k\ge N$ to:
\be
Z_{G(nk,k)}(N,v) \;=\; \sum\limits_{\ell=0}^{N-2} \left[ 
   \sum\limits_{\begin{scarray}
            \lambda\in S_\ell \\ 
            \alpha_{\ell,\lambda}(N)>0 
         \end{scarray}}
   \alpha_{\ell,\lambda} \, 
   \chi_{k,\ell,\lambda} \right] \;+\; \chi_{k,N-1,(1)^{N-1}} +  
   \rho_k(N) v^{2nk} \,,
\ee
where we have defined the ``characters'' $\chi_{k,\ell,\lambda}$ in 
\reff{def_chi}, $\rho_k$ is the number of the trivial eigenvalue 
contributes to the partition function, and the second sum is over all 
irreducible representations $\lambda\in S_\ell$ with a positive value of
the amplitude $\alpha_{\ell,\lambda}$ at $Q=N$. 
\end{conjecture}

\bigskip

\noindent
{\bf Remarks.} 1. Notice that $\alpha_{N-1,(1)^{N-1}} = 1$.

2. The function $\rho_k(N)$ is simple for small values of $0\le N\le 4$:
\be
\rho_k(N) \;=\; \begin{cases}
                 0                      & \quad \text{if $0\le N\le 3$} \\
                 6\, I[\text{$k$ even}] & \quad \text{if $N=4$} \end{cases}
\ee

3. For any $k\ge N$, we find the following sum rule:
\be
\sum\limits_{\ell=0}^{N-1} \alpha_{\ell,\lambda} \widetilde{N}_k(\ell,\lambda)
+ \rho_k(N) \;=\; N^{k+1}\,,
\label{sumrule}
\ee
where $\widetilde{N}_k(\ell,\lambda)$ is the number of contributing non-trivial
eigenvalues to the partition function $Z_{G(nk,k)}(N,v)$ and coming from the
sector with $\ell$ links and the representation $\lambda\in S_\ell$. 
Note that \reff{sumrule} is compatible with the fact that, for $Q$ integer,
the transfer matrix can alternatively be represented in terms of Potts spins.
In this representation the partition function with periodic longitudinal
boundary conditions is an ordinary matrix trace, and no eigenvalue
cancellations can take place. Therefore, the number of eigenvalues
(counted with multiplicities) is $Q^{k+1}$ indeed.

%
%
\section{Flow polynomial for the generalized Petersen graphs} 
\label{sec.flow} 

\begin{figure}
  \vspace*{-1cm}
  \centering
  \begin{tabular}{cc}
  \includegraphics[width=200pt]{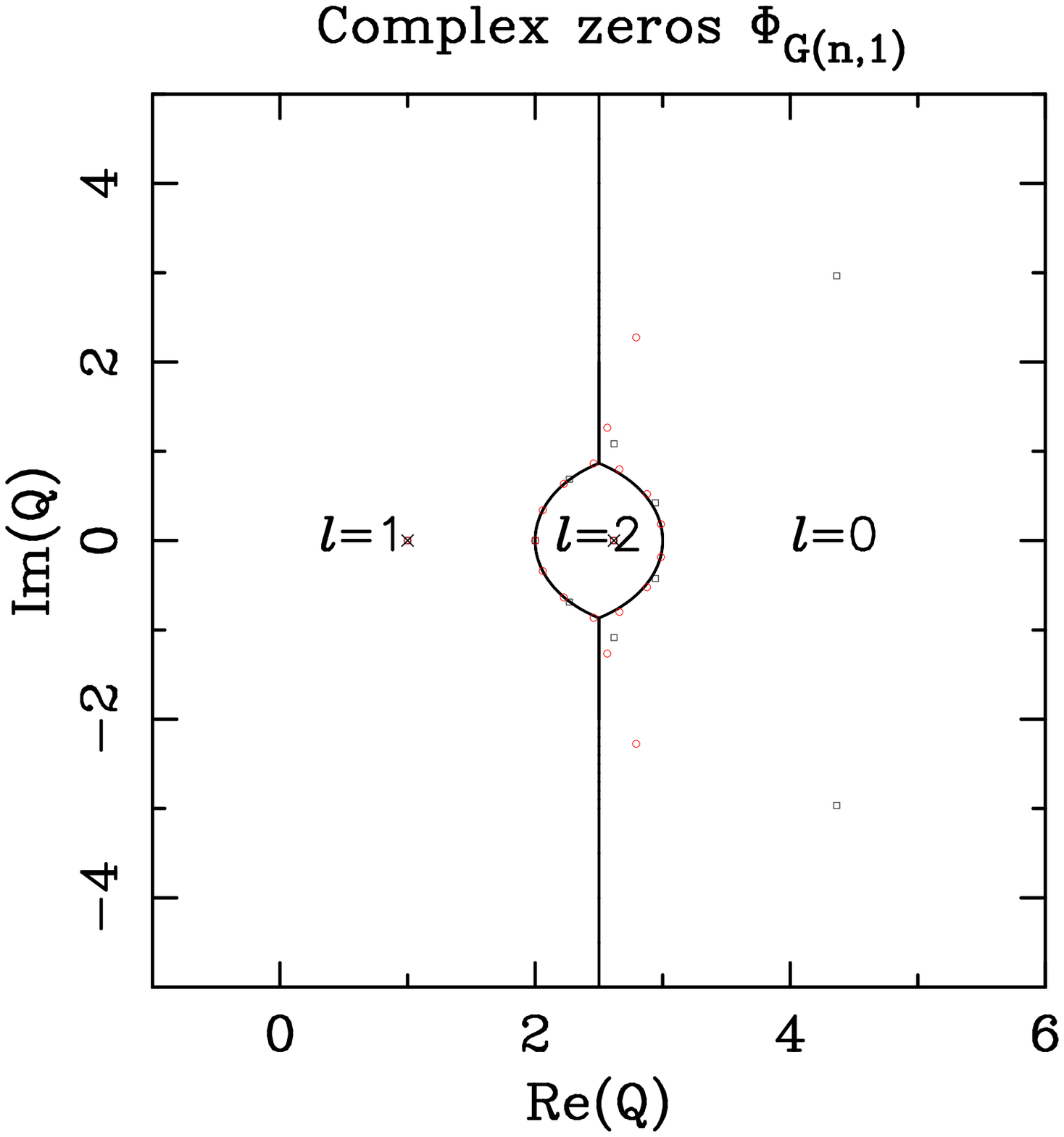} & 
  \includegraphics[width=200pt]{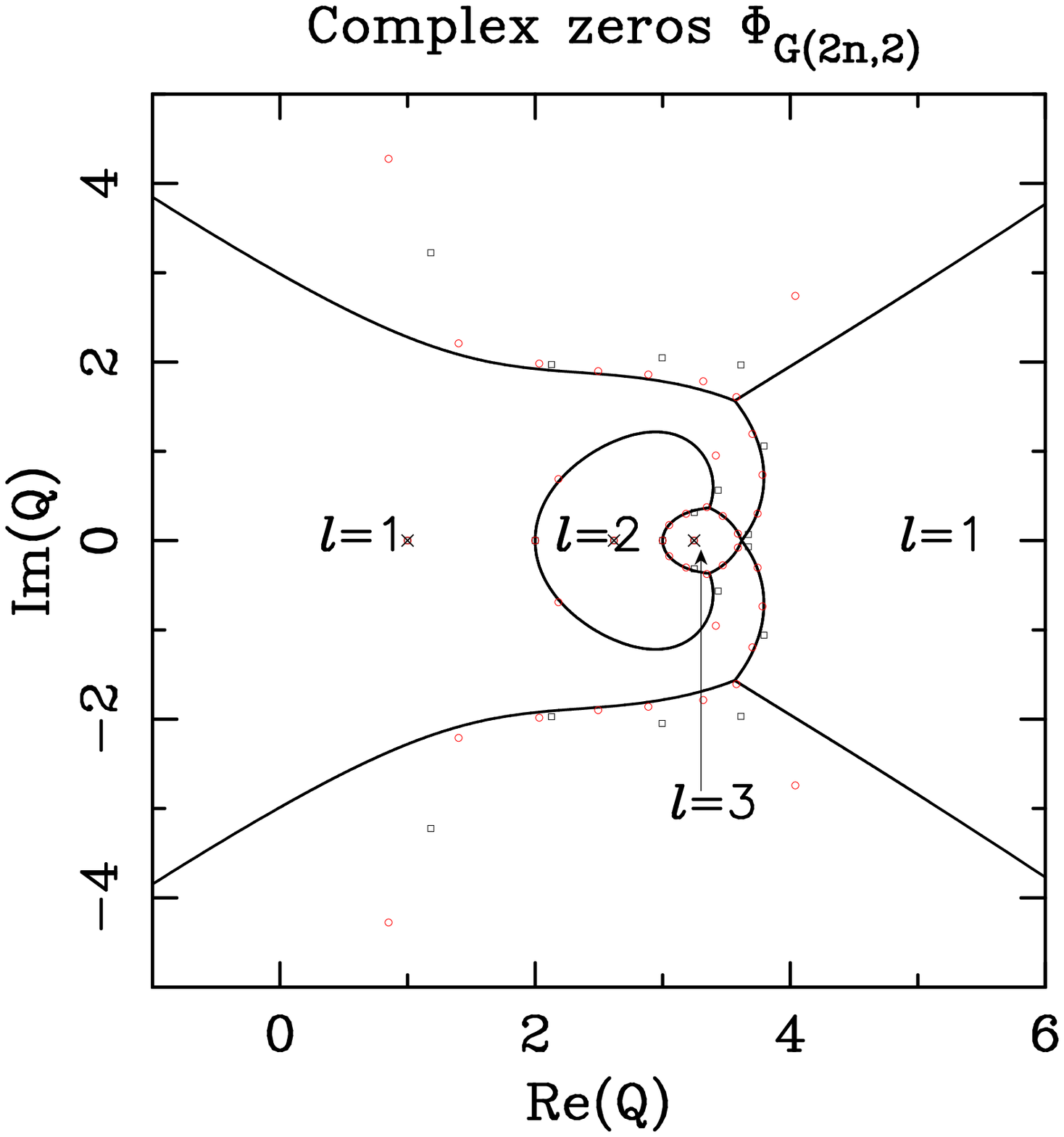} \\
  \qquad (a) & \qquad (b) \\[2mm]
  \includegraphics[width=200pt]{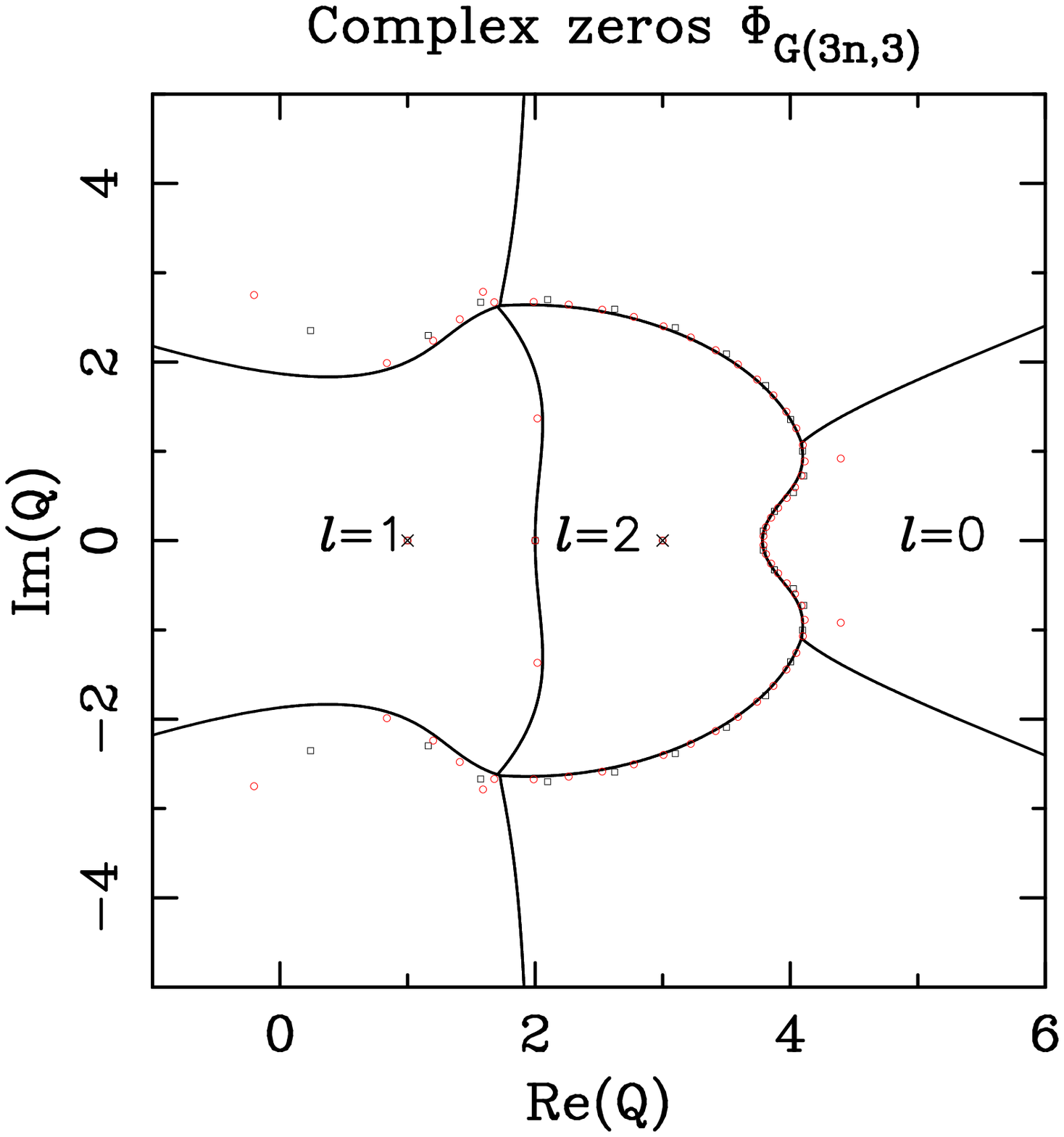} & 
  \includegraphics[width=200pt]{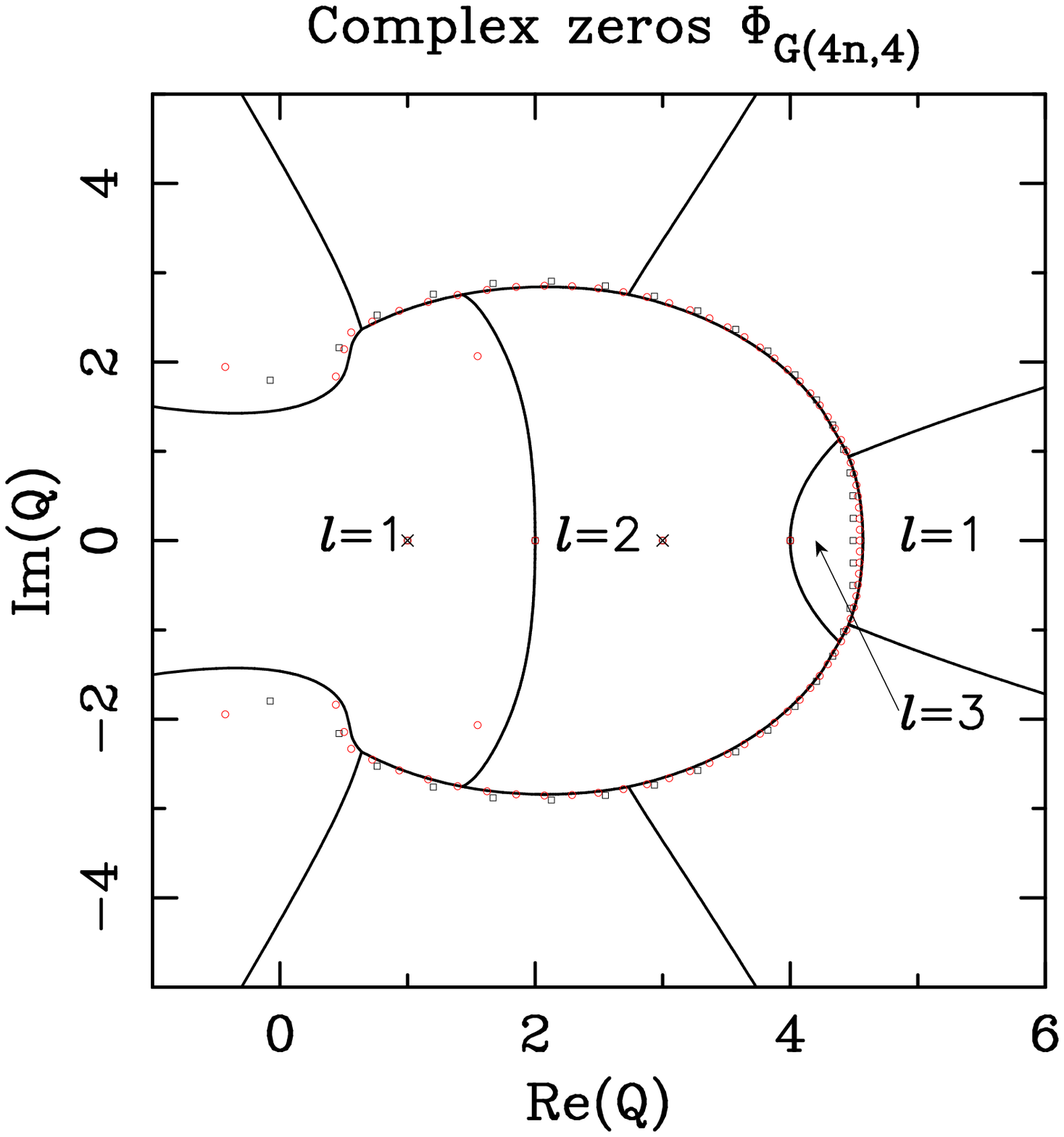} \\
  \qquad (c) & \qquad (d) 
  \end{tabular}
  \caption{
  Complex zeros of the flow polynomial and limiting curves
  $\mathcal{B}_k$ in the complex $Q$--plane
  for the generalized Petersen graphs $G(nk,k)$ with $k=1$ (a),
  $k=2$ (b), $k=3$ (c), and $k=4$ (d).
  For each value of $k$, the zeros correspond to the generalized Petersen
  graphs $G(10k,k)$ (black $\square$) and $G(20k,k)$ (red $\circ$).
  The labels (e.g., $\ell= 1$) show the sector the dominant eigenvalue 
  belongs to. We only label the regions that have a non-empty intersection with
  the real axis.  
  }
\label{Figures_flow1}
\end{figure}

\begin{figure}
  \vspace*{-1cm}
  \centering
  \begin{tabular}{cc}
  \includegraphics[width=200pt]{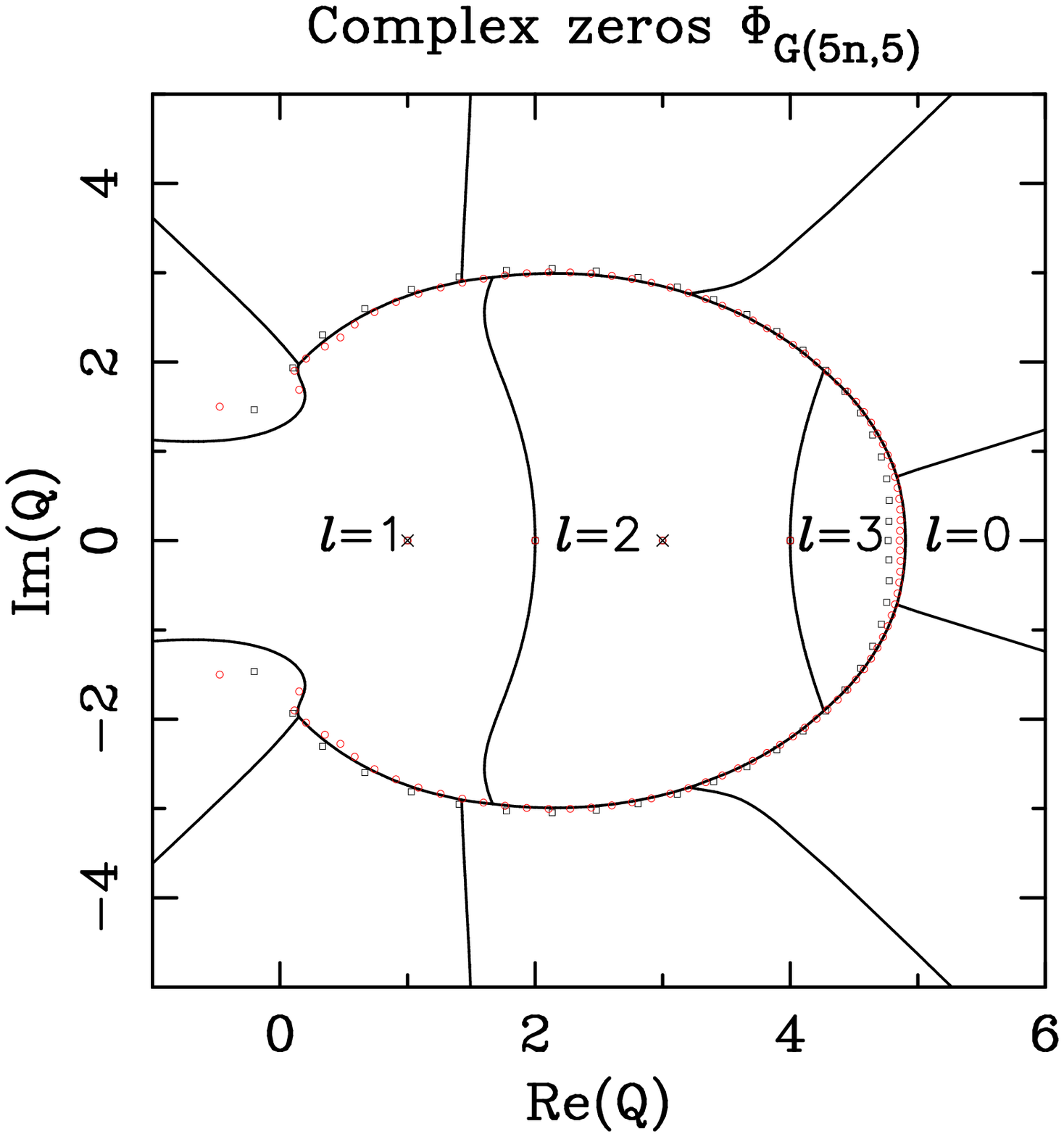} & 
  \includegraphics[width=200pt]{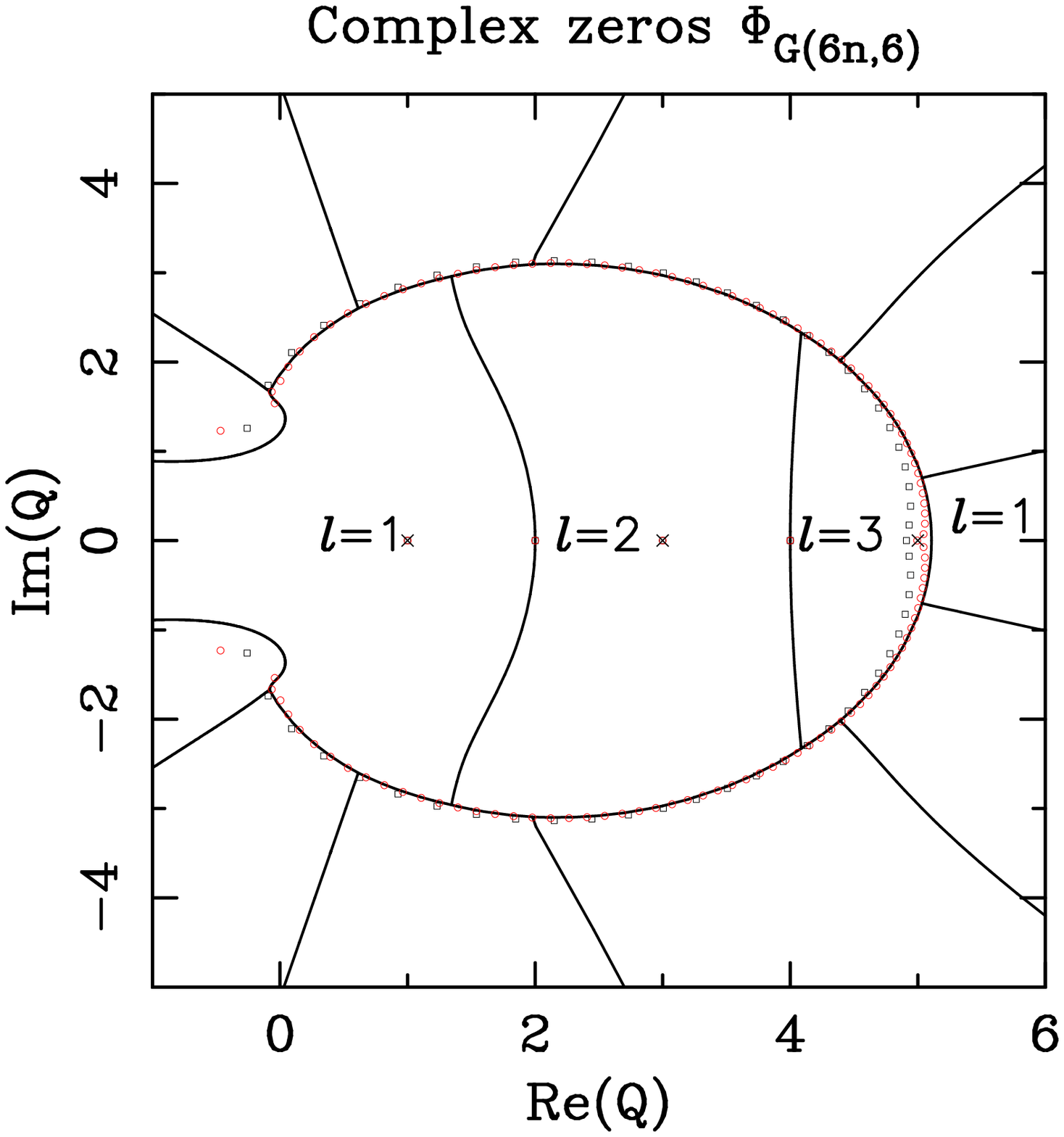} \\
  \qquad (a) & \qquad (b) \\[2mm]
  \multicolumn{2}{c}{%
       \includegraphics[width=200pt]{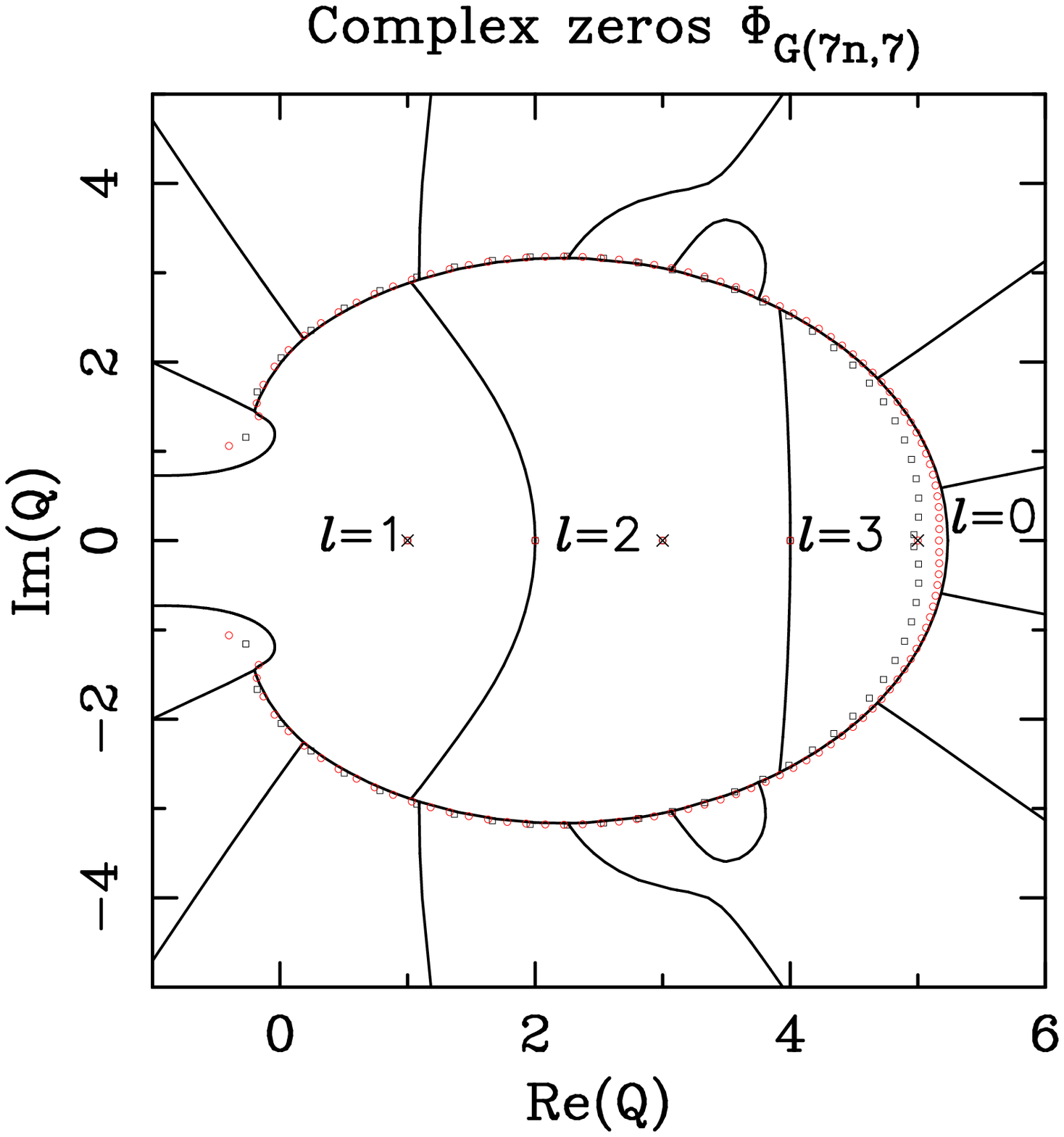}}\\ 
  \multicolumn{2}{c}{\qquad (c)} 
  \end{tabular}
  \caption{
  Complex zeros of the flow polynomial and limiting curves            
  $\mathcal{B}_k$ in the complex $Q$--plane
  for the generalized Petersen graphs $G(nk,k)$ with $k=5$ (a),
  $k=6$ (b), and $k=7$ (c).
  The black squares ($\square$) correspond to the zeros of
  $G(50,5)$ (a), $G(60,6)$ (b), and $G(63,7)$ (c).
  The red circles ($\circ$) correspond to the zeros of
  $G(100,5)$ (a), $G(120,6)$ (b), and $G(119,7)$ (c).
  Labels are as in Figure~\ref{Figures_flow1}.
  }
\label{Figures_flow2}
\end{figure}

In Ref.~\cite{JS_flow} we focused on the particular case of the flow
polynomial of the generalized Petersen graphs $G(nk,k)$ for $1\le k\le 7$.
Using the eigenvalues found there, our present goal is to compute the 
accumulation sets of flow zeros in the limit $n\to \infty$. In particular
we consider the limiting curves $\mathcal{B}_k$ for $1\le k\le 7$.
We have computed these curves for $k=1,2,\ldots,7$ using the
direct-search method, as described in \cite{transfer1}.  The results
are shown in Figures~\ref{Figures_flow1} and~\ref{Figures_flow2} in
the complex $Q$-plane. 

We have noticed several empirical patterns in
Figures~\ref{Figures_flow1} and~\ref{Figures_flow2}. The first one is 
related to the number of outward branches of the limiting curve $\mathcal{B}_k$,
and the dominant eigenvalue in each asymptotic sector:%
\footnote{This is exactly the same empirical behavior found for
     the limiting curves $\mathcal{B}_m$ associated to the chromatic 
     zeros of the family $S_{m,n}$ when $n\to\infty$ 
     \cite[Conjecture~7.1]{transfer6}.
     The graph $S_{m,n}$ can be regarded as a square-lattice grid with $m$ 
     columns, $n$ rows, free boundary conditions in the longitudinal
     direction, and special boundary conditions in the transverse direction:
     we introduce two extra vertices such that every vertex on the leftmost
     column of the grid is connected to one of them, and every vertex on the
     rightmost column is connected to the other one.
}

\begin{conjecture} \label{conj.flow1}
Fix $k\ge 1$. Then, the limiting curve $\mathcal{B}_k$ has exactly
$2k$ outward branches extending to $Q=\infty$, with asymptotic angles
$\arg Q = \theta_{k,j}$ where
\be
\theta_{k,j} \;=\; \left( j - \frac{1}{2} \right) \frac{\pi}{k} \,,
\qquad \text{for $j=1,\ldots,2k$.}
\ee
Moreover, the dominant eigenvalue is $\mu_{k+1,0,\star}$ in the
asymptotic regions
\be
\theta_{k,j} \;\le\; \arg Q  \;\le \; \theta_{k,j+1} \quad \text{for} \quad
\begin{cases}
   \; j = 1,3, \ldots, 2k-1 & \text{if $k$ is even} \\
   \; j = 2,4, \ldots, 2k   & \text{if $k$ is odd}
\end{cases}
\ee
while in the other asymptotic sectors the dominant eigenvalue is
$\mu_{k+1,1,(1),\star}$.
\end{conjecture}

\begin{figure}
  \centering
  \includegraphics[width=200pt]{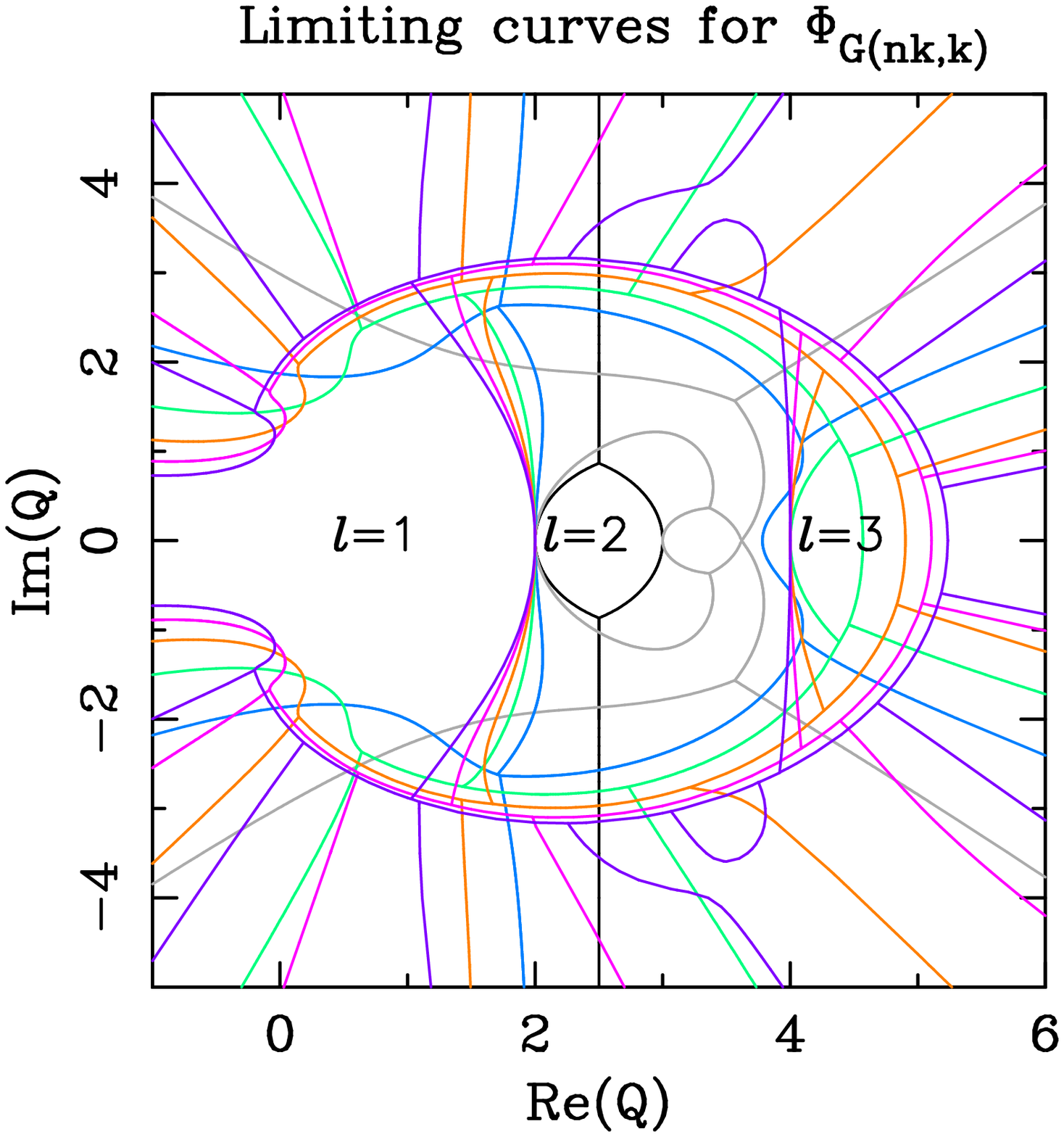}
  \caption{
  Limiting curves $\mathcal{B}_k$ for the flow polynomial in the
  complex $Q$-plane for the generalized Petersen graphs $G(nk,k)$ with
  $k=1$ (black), $k=2$ (gray), $k=3$ (light blue), $k=4$ (green),
  $k=5$ (orange), $k=6$ (pink), and $k=7$ (violet).
  Labels are as in Figure~\ref{Figures_flow1}.
  }
\label{Figure_flow_all}
\end{figure}

We may also conjecture that, as $k\to \infty$, the limiting curves 
$\mathcal{B}_k$ (without the outward branches) converge to some curve 
$\mathcal{B}_\infty$.
This is what Figure~\ref{Figure_flow_all} suggests. Indeed, outside this
curve, the outward branches seem to get denser as $k$ increases. Therefore,
we conjecture that the set of accumulation points of the flow--polynomial
zeros for the generalized Petersen graphs $G(nk,k)$ is {\em dense} in
the whole complex $Q$-plane, except in the interior of the curve
$\mathcal{B}_\infty$. If this conjecture is true, it implies in particular
that flow roots of {\em non-planar cubic graphs} are dense in the complex 
plane, except possibly in some bounded region near the origin. 
Note that a similar result holds (by duality) for the flow roots of 
planar graphs \cite{Sokal04}, but without the restriction to cubic graphs.

%
%

%
%

%
%

If we restrict to the cases with $k\ge 4$ (see
Figures~\ref{Figures_flow1}(d) and~\ref{Figures_flow2}), we notice some
regularities about the isolated limiting points and the dominant
eigenvalues in the non-asymptotic region of the complex $Q$-plane. If we
denote by $Q_c^{(\Phi)}$ the largest real value of $Q$ where $\mathcal{B}_k$
crosses the real $Q$-axis, then we conjecture that:

\begin{conjecture} \label{conj.flow2}
Fix $k\ge 4$. Then, the dominant eigenvalue in the regions that
intersect the real $Q$-axis is
\be
\mu_\star \;=\; \begin{cases}
            \; \mu_{k+1,1,(1),\star} & \; \text{for $\Re Q \in (-\infty,2]$}\\
            \; \mu_{k+1,2,(2),\star} & \; \text{for $\Re Q \in [2,4]$}\\
            \; \mu_{k+1,3,(3),\star} & \; 
                                   \text{for $\Re Q \in [4,Q_c^{(\Phi)}(k)]$}\\
            \; \mu_{k+1,1,(1),\star} & \;
\text{for $\Re Q \in [Q_c^{(\Phi)}(k),\infty)$ and even $k$} \\
            \; \mu_{k+1,0,\star} & \;
\text{for $\Re Q \in [Q_c^{(\Phi)}(k),\infty)$ and odd $k$}
 \end{cases}
\ee
Therefore, $Q=1$, $Q=3$ and $Q=5$ (only when $Q^{(\Phi)}_c(k)>5$) are isolated
limiting points; and $Q=2$, $Q=4$, and $Q=Q^{(\Phi)}_c(k)$ are non-isolated
limiting points.
\end{conjecture}

Notice that the above conjecture is also valid for $k=3$, except that in
this case, there is no region where the eigenvalue $\mu_{k+1,3,(3),\star}$
becomes dominant, since $Q_c^{(\Phi)}(3)<4$. It is worth noting that the dominant
eigenvalue comes from the completely symmetric irreducible representation
of $S_\ell$ for $\ell=1,2,3$.

Assuming that the dominant eigenvalue on the real $Q$-axis invariably
comes from the completely symmetric representation, it is possible to
extend the numerical determination of $Q_c^{(\Phi)}(k)$ to higher values of
$k$. The relevant transfer matrices can be numerically diagonalized by a
standard iterative scheme that uses their decomposition as a product of 
sparse matrices, as in \reff{def_T}. The resulting values of $Q_c^{(\Phi)}(k)$ 
are shown in Table~\ref{table.QcFlow}.

%
%
\begin{table}[hbt]
\centering
\begin{tabular}{r|l}
\hline\hline
\multicolumn{1}{c}{$k$} & \multicolumn{1}{|c}{$Q_c^{(\Phi)}(k)$} \\
\hline 
1 &  3 \\
2 &  3.6180339887 \\
3 &  3.7818423129 \\
4 &  4.5697435537 \\
5 &  4.9029018077 \\
6 &  5.1079785012 \\
7 &  5.2352605291 \\
8 &  5.3246966903 \\
9 &  5.3886186958 \\
10&  5.4364766073 \\
11&  5.4729804532 \\
\hline\hline
\end{tabular}
\caption{\label{table.QcFlow}
Values of $Q_c^{(\Phi)}(k)$ for the generalized Petersen graphs $G(nk,k)$ for
$1\le k \le 11$.
}
\end{table}

The values of $Q_c^{(\Phi)}(k)$ are seen to be well fitted by a power-law
Ansatz: $Q_c^{(\Phi)}(k) = Q_c^{(\Phi)} + A k^{-\Delta}$. To prevent 
possible parity effects, we performed the fits for odd and even values of $k$
separately. Both fits were consistent with each other and give a common
limit $Q_c^{(\Phi)} = 5.69(1)$. In fact, we tried a fit of the full data 
with an Ansatz of the form
\be
Q_c^{(\Phi)}(k) \;=\; Q_c^{(\Phi)}  + \begin{cases}
            A  \, k^{-B}  & \quad \text{if $k$ is even} \\  
            A' \, k^{-B'} & \quad \text{if $k$ is odd} 
            \end{cases}
\label{Ansatz.fulldata}
\ee
We found that 1) $Q_c^{(\Phi)} = 5.69(1)$, 2) that the differences between
$(A,B)$ and $(A',B')$ were very small, so the parity effects are 
almost eliminated by fixing a common limit for the odd and even data sub-sets,
and 3) the fits were rather stable (the fits for data $5\le k\le 9$, 
$6\le k\le 10$, and $7\le k\le 11$, gave almost the same results).  
Finally, we also tried to extrapolate our numerical data using the 
Bulirsch-Stoer algorithm for the odd-- and even--$k$ subsets, and for the 
full data set. The convergence is rather slow, and our best estimate for
the limit is $Q_c^{(\Phi)} = 5.685(10)$, which agrees with the previous 
estimates. We conclude that: 
\be
 \label{Qc.flow.Petersen}
 Q_c^{(\Phi)} \;\equiv\; \lim_{k\to\infty} Q_c^{(\Phi)}(k) 
              \;=\; 5.69 \pm 0.01 \,.
\ee
Formalizing slightly these experimental results, we conjecture that
the exists a real number $Q_c^{(\Phi)} \approx 5.69$ that 
is an accumulation
point of real flow zeros for the graph family $G(nk,k)$, in the limit
$\lim_{n \to \infty} \lim_{k \to \infty}$, and that no generalized
Petersen graph has a real flow zero above $Q_c^{(\Phi)}$. The hope that no
graph family fares better than the generalized Petersen graph is our
main motivation---and evidence---for stating Conjecture~1.9 of 
Ref.~\cite{JS_flow}:
\begin{conjecture} \label{conj.JS}
For any bridgeless graph $G$, $\Phi_G(Q) > 0$ for $Q \in [6,\infty)$.
\end{conjecture}
This conjecture has a great interest in Combinatorics, especially in 
connection to Tutte's five--flow conjecture \cite{Tutte54,Tutte_book,Jaeger}:  
\begin{conjecture} \label{conj.tutte}
For any bridgeless graph $G$, $\Phi_G(5) > 0$.
\end{conjecture}
This conjecture seems at least as difficult to prove or disprove as the 
famous four-color conjecture (which eventually turned out to be a theorem 
\cite{AH77}). 

%
%
\section{Chromatic polynomial for the generalized Petersen graphs} 
\label{sec.chromatic} 

We have also symbolically computed the chromatic polynomial $P_{G(nk,k)}(Q)$ 
for the generalized Petersen graphs $G(nk,k)$ with $1\le k\le 6$ by 
specializing the full partition functions $Z_{G(nk,k)}(Q,v)$ of 
Sections~\ref{sec.tm} and~\ref{sec.num.res} to the case $v=-1$. 
Again, we have checked our symbolic results against exact computations 
made by using {\sc Maple} for the smallest values of $n\ge 1$.

%
%
\begin{figure}
  \vspace*{-1cm}
  \centering
  \begin{tabular}{cc}
  \includegraphics[width=200pt]{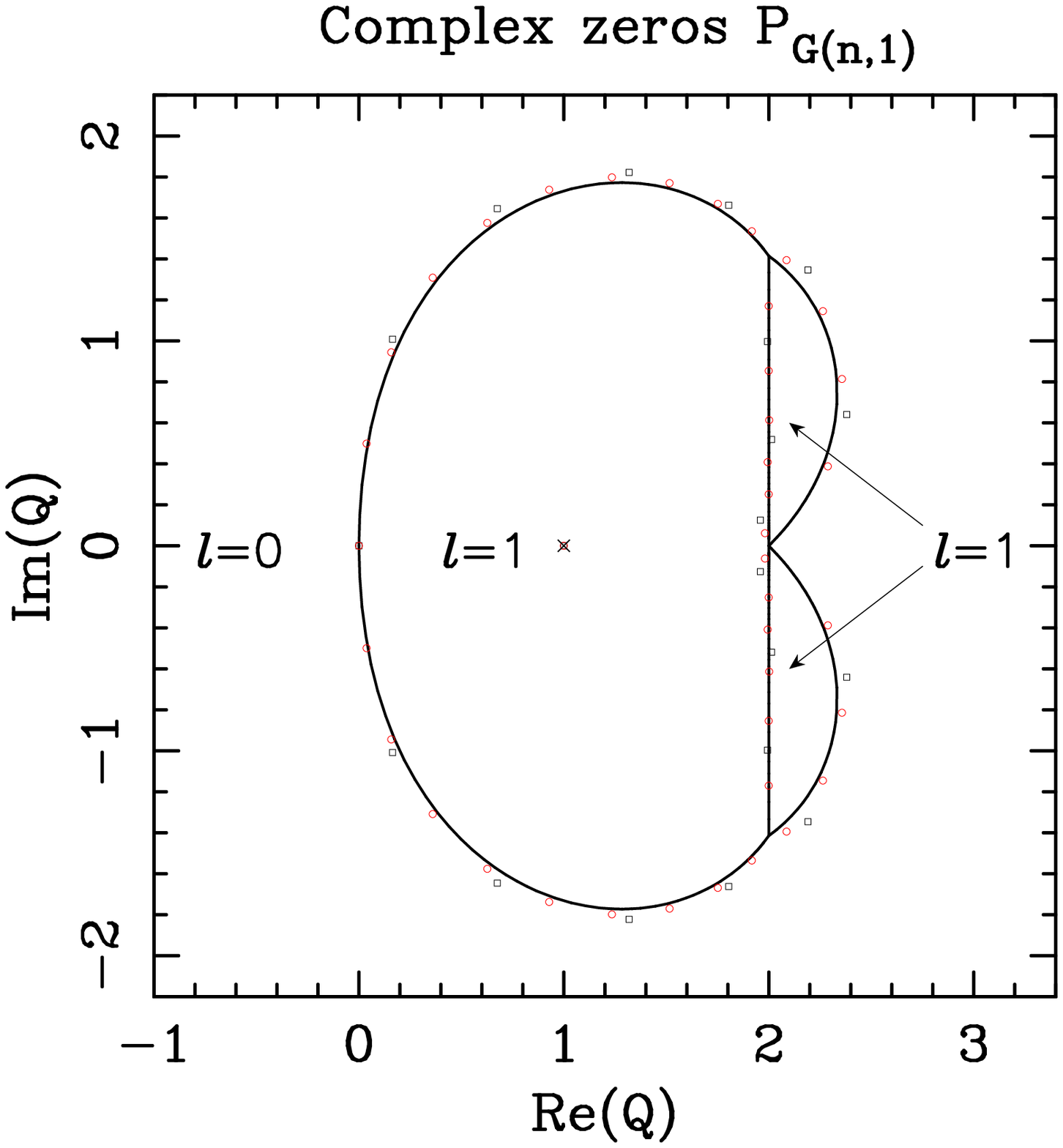} & 
  \includegraphics[width=200pt]{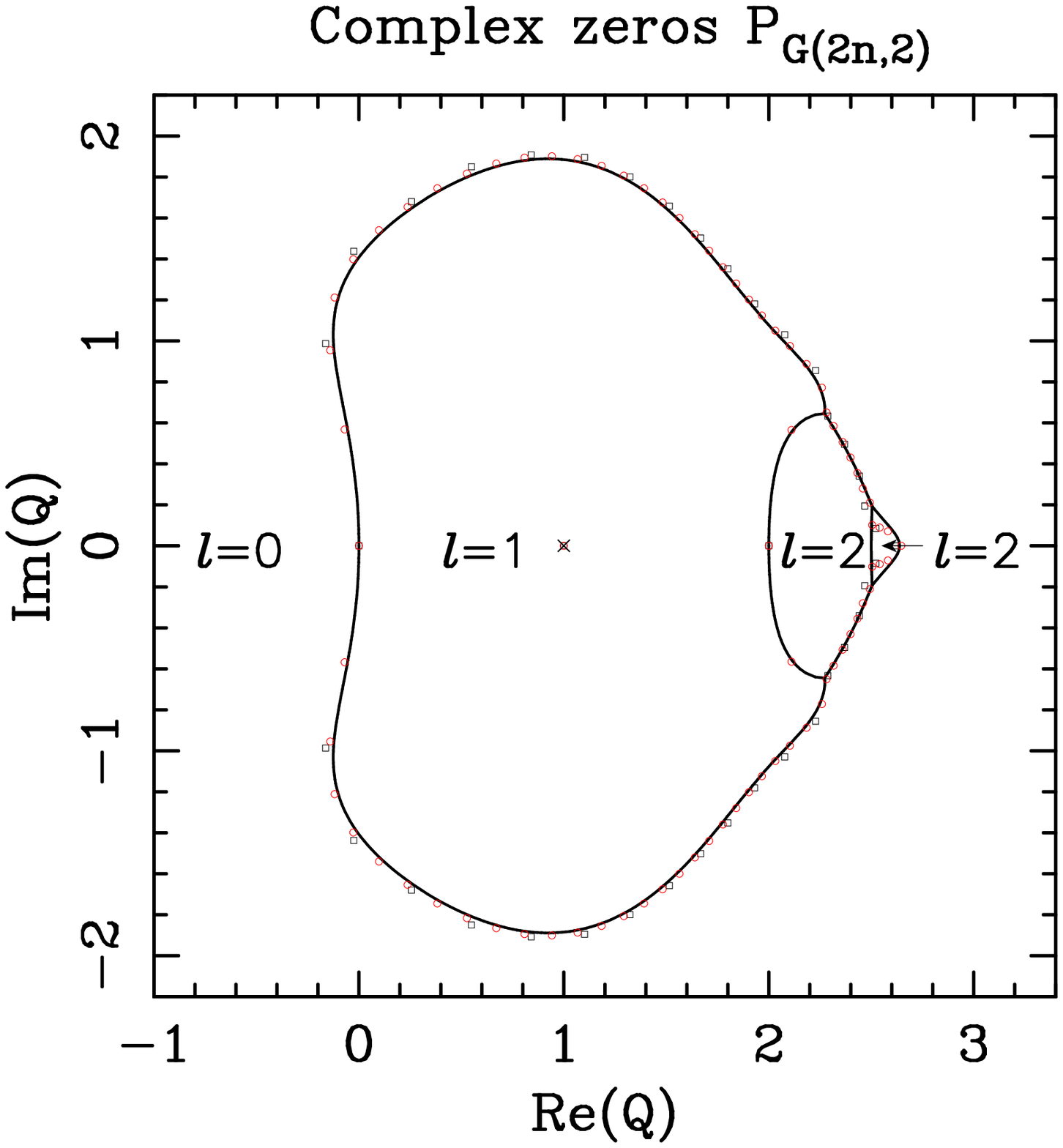} \\
  \qquad (a) & \qquad (b) \\[2mm]
  \includegraphics[width=200pt]{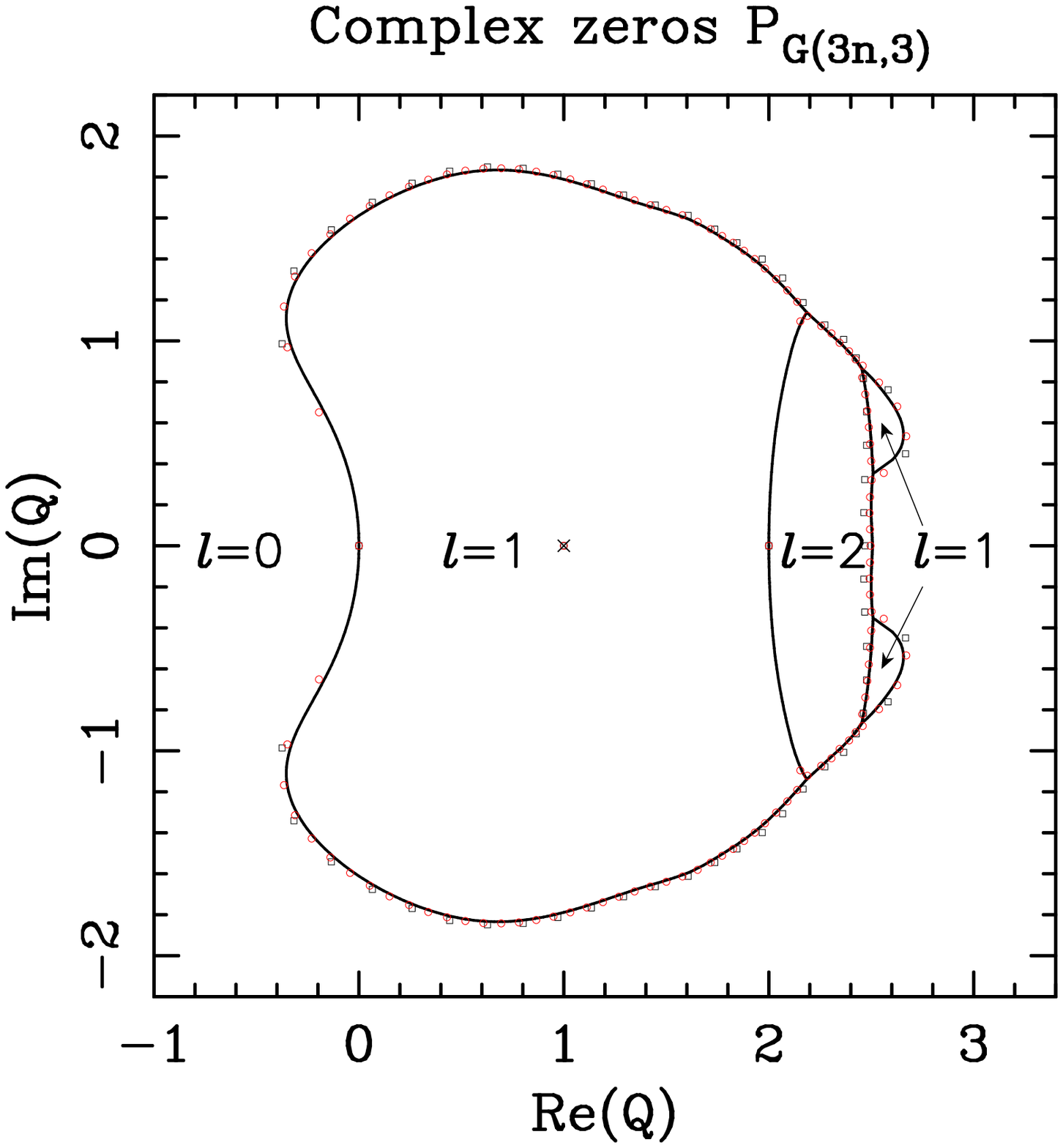} & 
  \includegraphics[width=200pt]{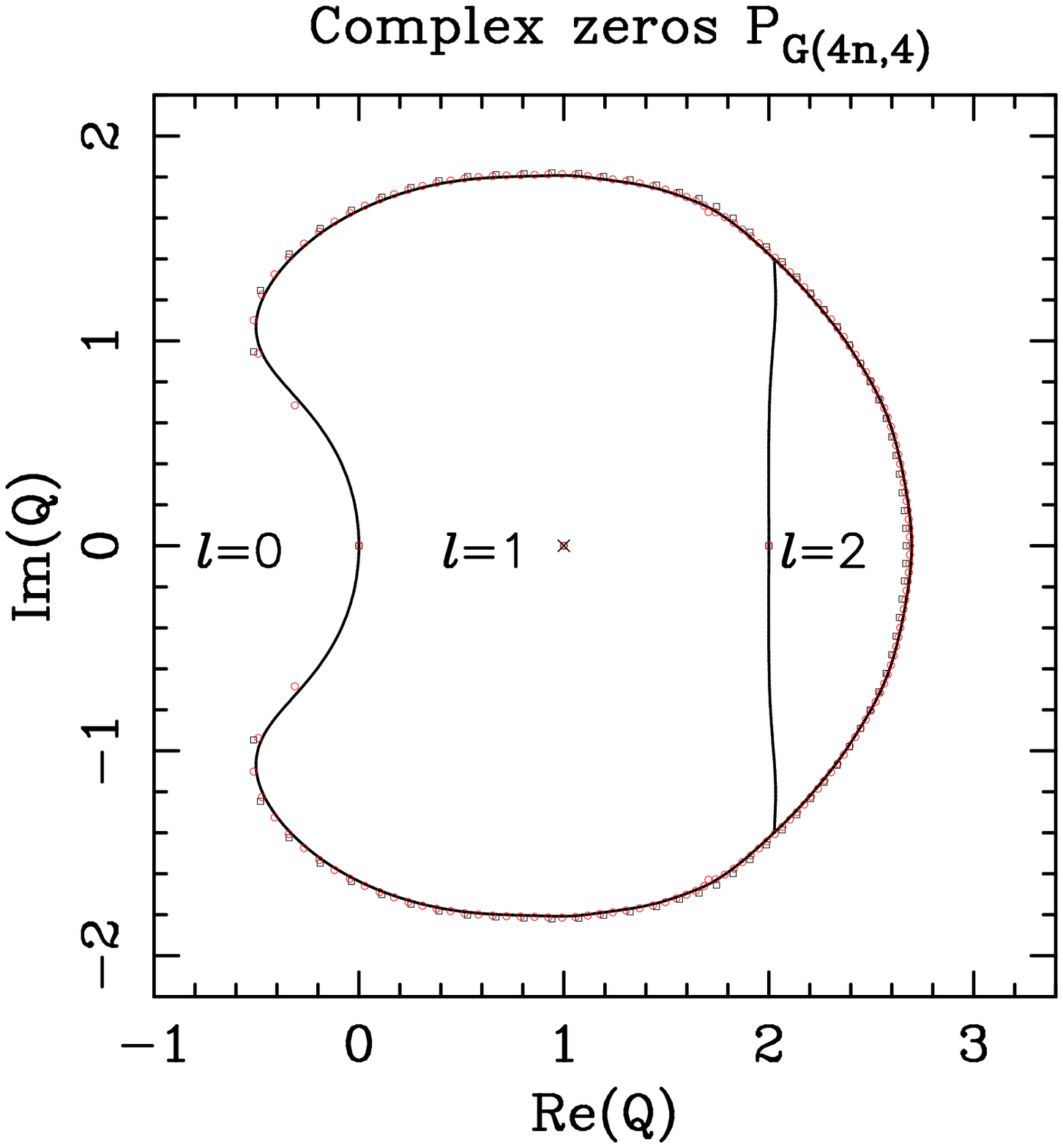} \\
  \qquad (c) & \qquad (d) 
  \end{tabular}
  \caption{
  Complex zeros of the chromatic polynomial and limiting curves 
  $\mathcal{B}_k$ in the complex $Q$--plane 
  for the generalized Petersen graphs $G(nk,k)$ with $k=1$ (a), 
  $k=2$ (b), $k=3$ (c), and $k=4$ (d). 
  For each value of $k$, the zeros correspond to the generalized Petersen
  graphs $G(10k,k)$ (black $\square$) and $G(20k,k)$ (red $\circ$). 
  Each region is labeled with the sector the dominant eigenvalue belongs to 
  (e.g., $\ell= 1$).
  }
\label{Figures_P1}
\end{figure}

%
%
\begin{figure}
  \vspace*{-1cm}
  \centering
  \begin{tabular}{cc}
  \includegraphics[width=200pt]{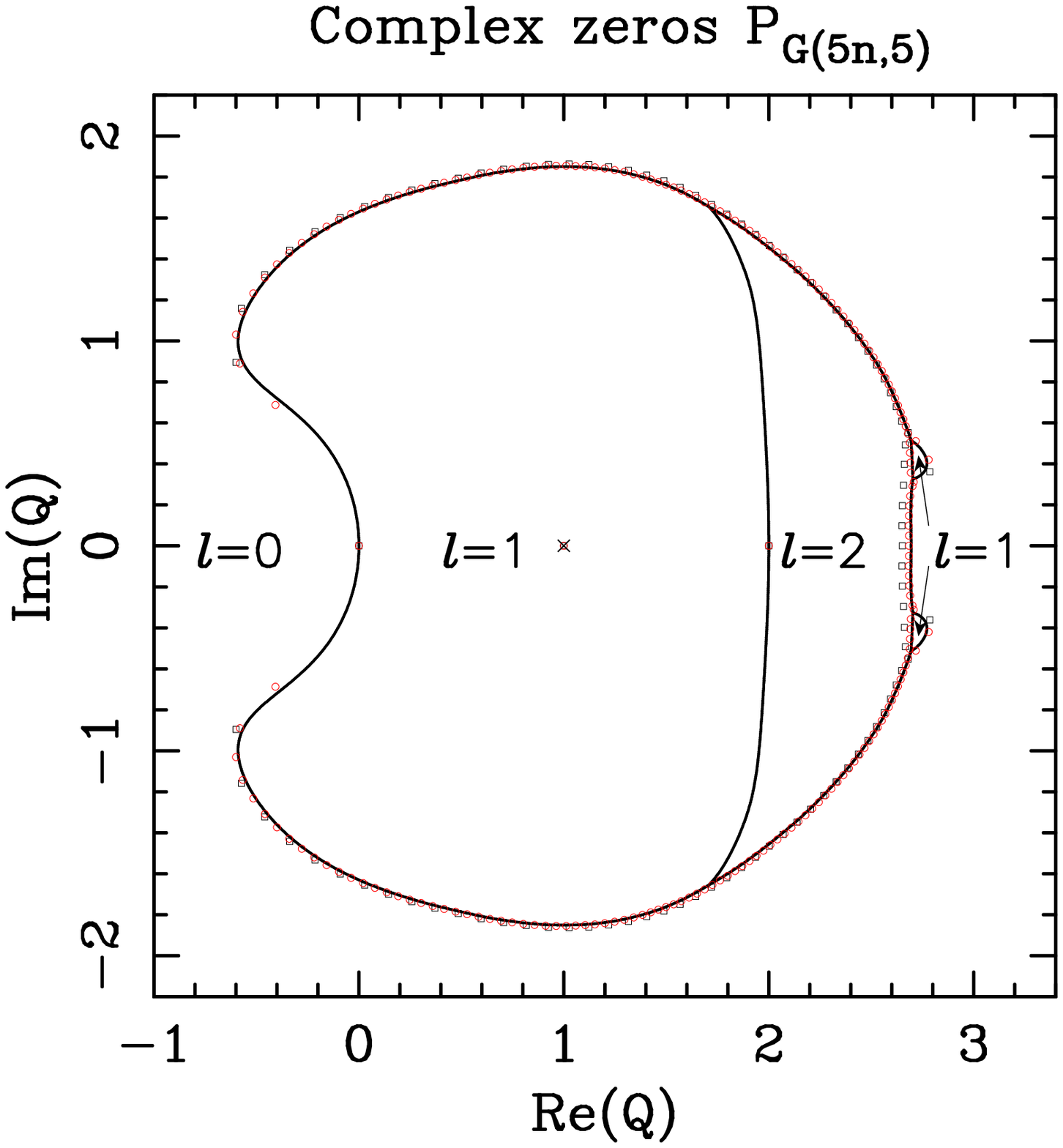} & 
  \includegraphics[width=200pt]{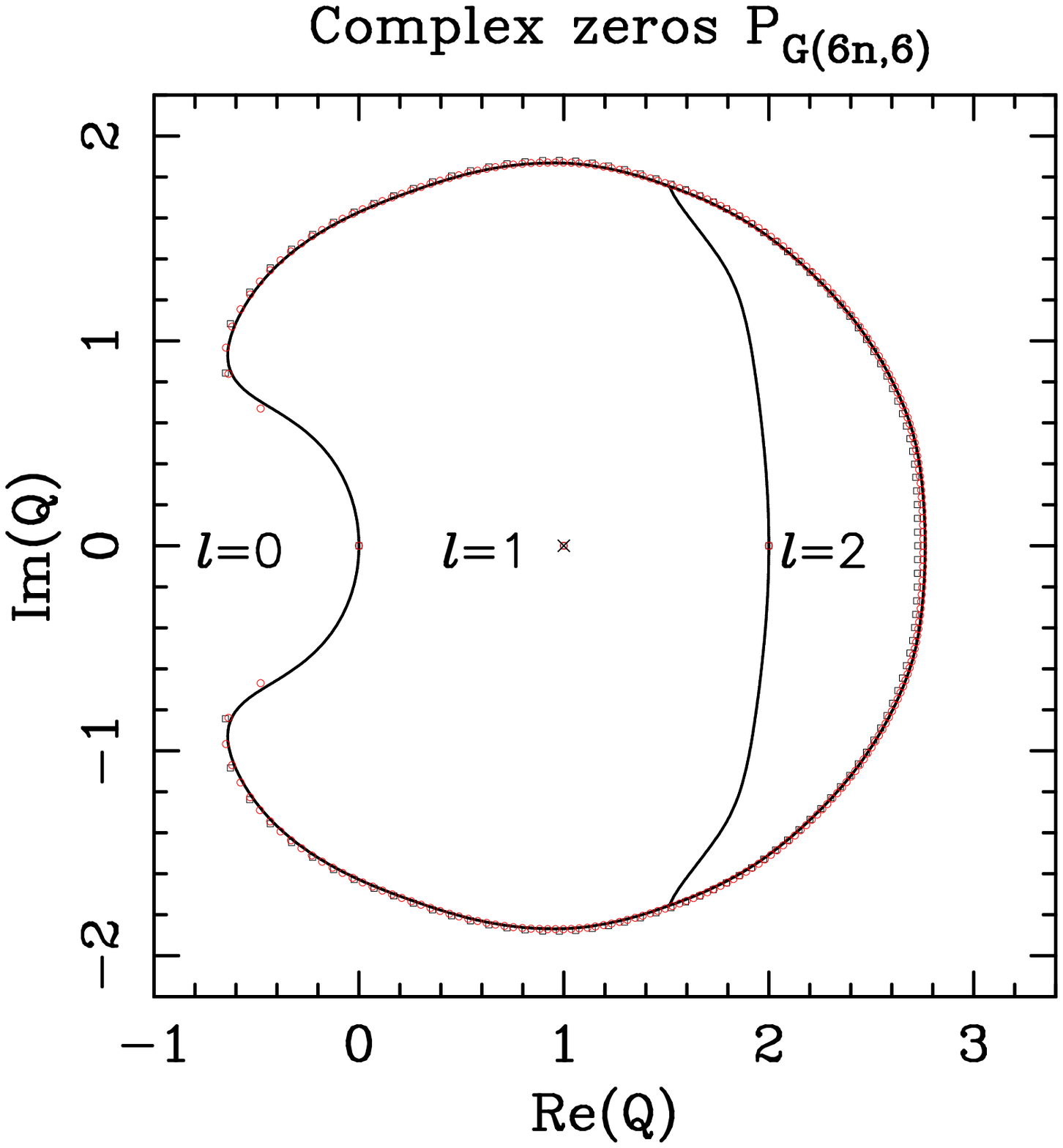} \\
  \qquad (a) & \qquad (b) \\[2mm]
  \includegraphics[width=200pt]{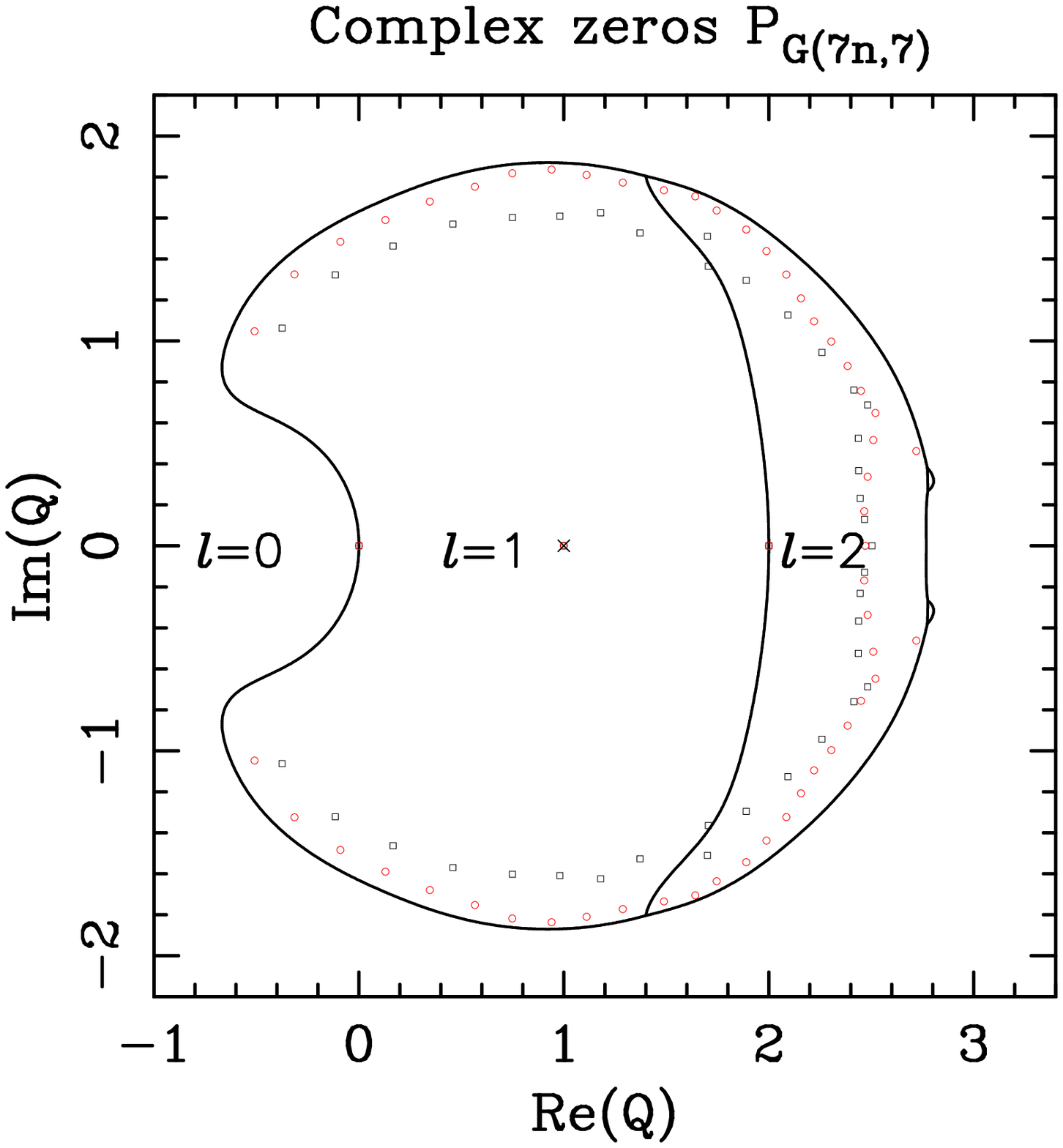} &   
  \includegraphics[width=200pt]{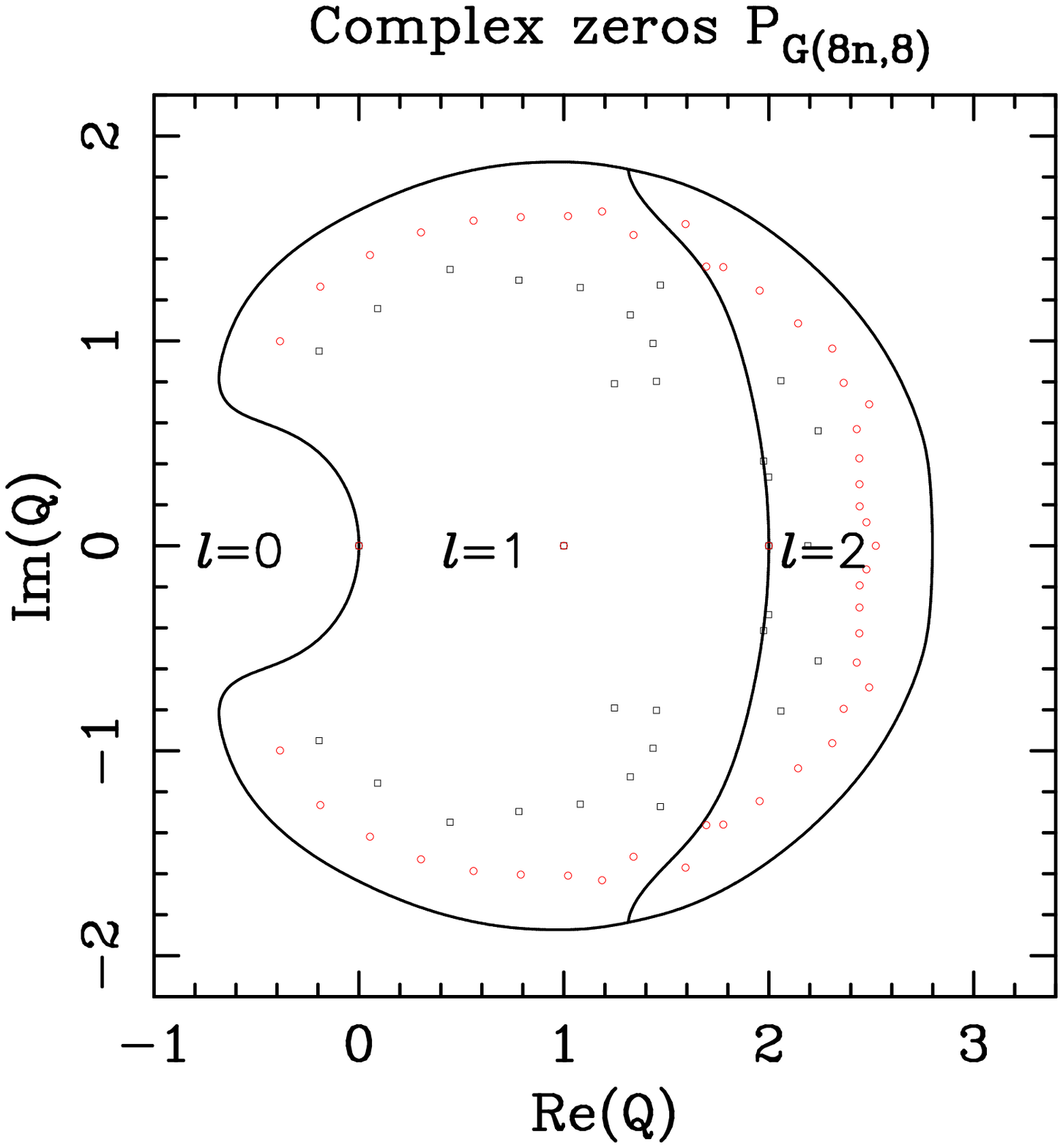} \\   
  \qquad (c) & \qquad (d) 
  \end{tabular}
  \caption{
  Complex zeros of the chromatic polynomial and limiting curves 
  $\mathcal{B}_k$ in the complex $Q$--plane 
  for the generalized Petersen graphs $G(nk,k)$ with $k=5$ (a), 
  $k=6$ (b), $k=7$ (c), and $k=8$ (d). 
  The black squares ($\square$) correspond to the zeros of 
  $G(50,5)$ (a), $G(60,6)$ (b), $G(21,7)$ (c), and $G(18,8)$ (d). 
  The red circles ($\circ$) correspond to the zeros of 
  $G(100,5)$ (a), $G(120,6)$ (b), $G(28,7)$ (c), and $G(24,8)$ (d). 
  The labels are as in Figure~\ref{Figures_P1}.
  }
\label{Figures_P2}
\end{figure}

%
%
\begin{figure}
  \centering
  \includegraphics[width=200pt]{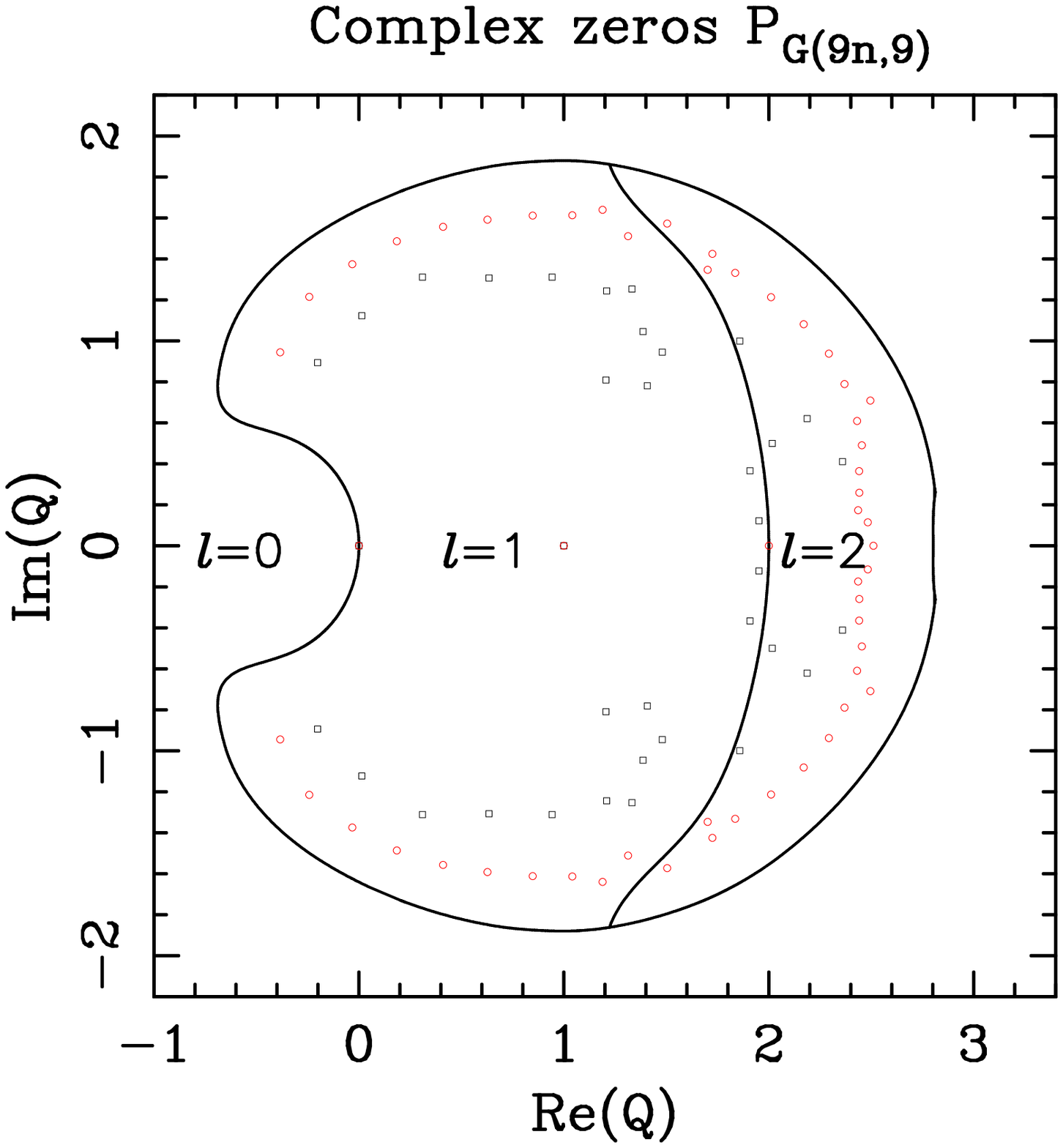}
  \caption{
  Complex zeros of the chromatic polynomial and limiting curves
  $\mathcal{B}_k$ in the complex $Q$--plane
  for the generalized Petersen graphs $G(9n,9)$. 
  The black squares ($\square$) correspond to the zeros of 
  $G(18,9)$, and the red circles ($\circ$) correspond to the zeros of 
  $G(27,9)$. The labels are as in Figure~\ref{Figures_P1}. 
  There are two tiny complex-conjugate oval-like regions (as for 
  the other odd values of $k$) around $(Q,v) \approx (2.80,0.36)$;
  see Figure~\ref{Figure_P_all}(b).
  }
\label{Figures_P3}
\end{figure}

Again, in addition to finding the chromatic roots for finite Petersen graphs
$G(nk,k)$, we are interested in their accumulation sets as $n\to \infty$.
We have obtained in this way, for each value of $k$, the corresponding 
limiting curve $\mathcal{B}_k$ of non-isolated limiting points.    
We have used the direct search method \cite{transfer1} to obtain this
curves with high precision arithmetic. These curves are shown in 
Figures~\ref{Figures_P1}--\ref{Figures_P3}.
In the cases $k=7,8,9$, the chromatic
zeros were obtained by using {\sc Maple}. The corresponding limiting 
curves $\mathcal{B}_k$ were computed by numerically diagonalizing the 
relevant transfer matrices by the Arnoldi algorithm and the use of 
their decomposition as a product of sparse matrices, as in \reff{def_T}. 
This latter procedure was implemented in {\sc C}.

Contrary to what happens for the flow--polynomial case, the limiting curves
are bounded in the complex $Q$-plane (i.e., there are no outward branches
going to infinity). This is expected because a theorem by Sokal 
\cite[Corollary~5.3 and Proposition~5.4]{Sokal01} states that if $G$
is a loopless finite undirected graph of maximum degree $\Delta$, then 
all the chromatic zeros lie in the disk $|Q| < K \Delta$, with
$\Delta \approx 7.963906$. For the generalized Petersen graphs $\Delta=3$, 
and the bound given by this theorem is far from sharp for this family of 
graphs.  If we restrict to the cases with $k\ge 3$ (see
Figures~\ref{Figures_P1}(c)--(d), \ref{Figures_P2}, and~\ref{Figures_P3}), 
we notice some regularities about the isolated limiting points and the dominant
eigenvalues in the complex $Q$-plane. If we
denote by $Q_c^{(P)}$ the largest real value of $Q$ where $\mathcal{B}_k$
crosses the real $Q$-axis, then we conjecture that:

\begin{conjecture} \label{conj.chromatic1}
Fix $k\ge 3$. Then, the dominant eigenvalue in the regions that
intersect the real $Q$-axis is
\be
\mu_\star \;=\; \begin{cases}
            \; \mu_{k+1,0,\star} & \; \text{for $\Re Q \in (-\infty,0]$}\\
            \; \mu_{k+1,1,(1),\star} & \; \text{for $\Re Q \in [0,2]$}\\
            \; \mu_{k+1,2,(2),\star} & \;
                                \text{for $\Re Q \in [2,Q_c^{(P)}(k)]$}\\
            \; \mu_{k+1,0,\star} & \; 
                          \text{for $\Re Q \in [Q_c^{(P)}(k),\infty)$} 
 \end{cases}
\ee
Therefore, $Q=1$ is the only isolated limiting point; and $Q=0$, $Q=2$ and 
$Q=Q^{(P)}_c(k)$ are non-isolated limiting points.
\end{conjecture}
%

%
%
\begin{figure}[ht]
  \centering
  \begin{tabular}{cc}
  \includegraphics[width=200pt]{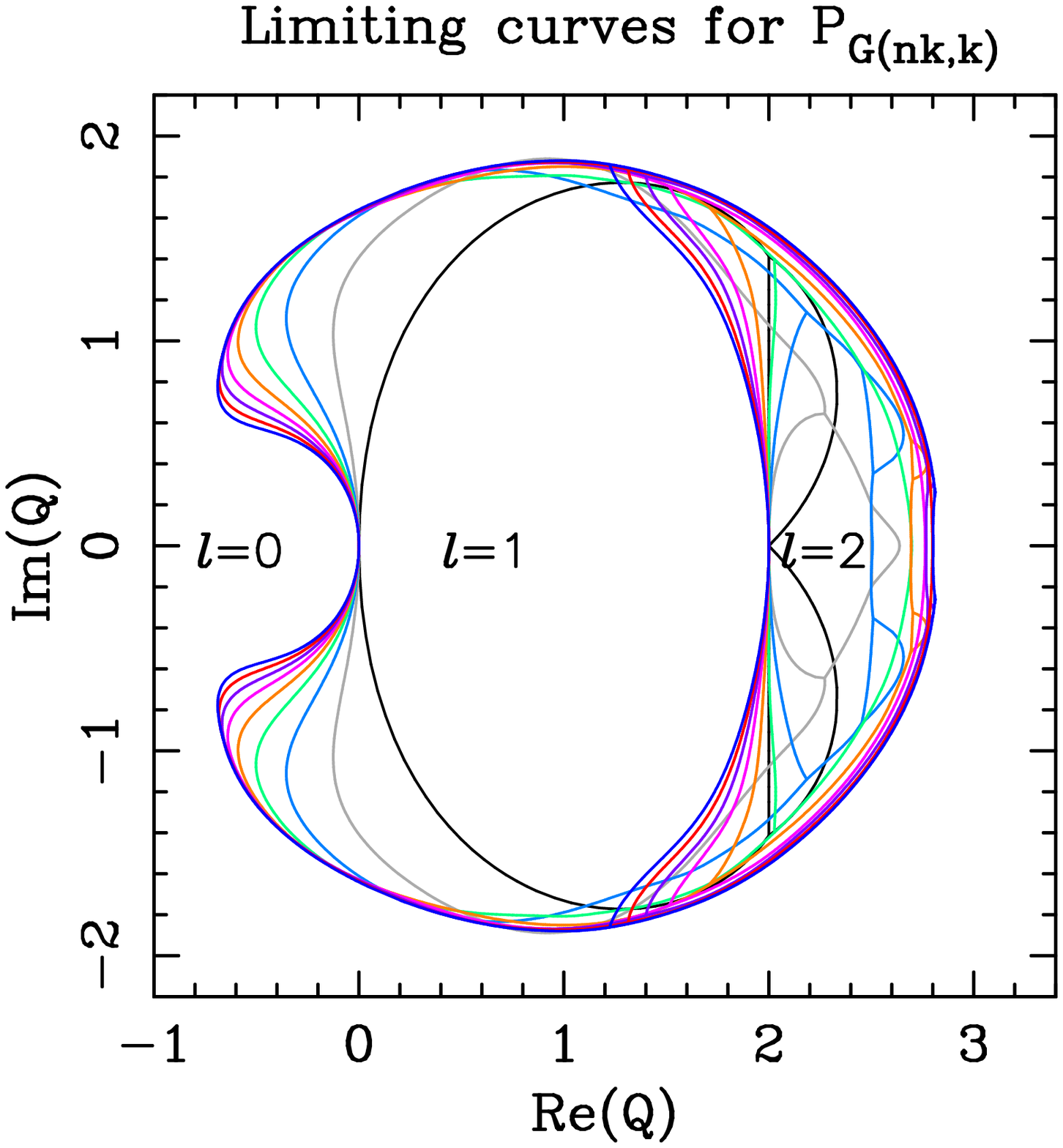} & 
  \includegraphics[width=200pt]{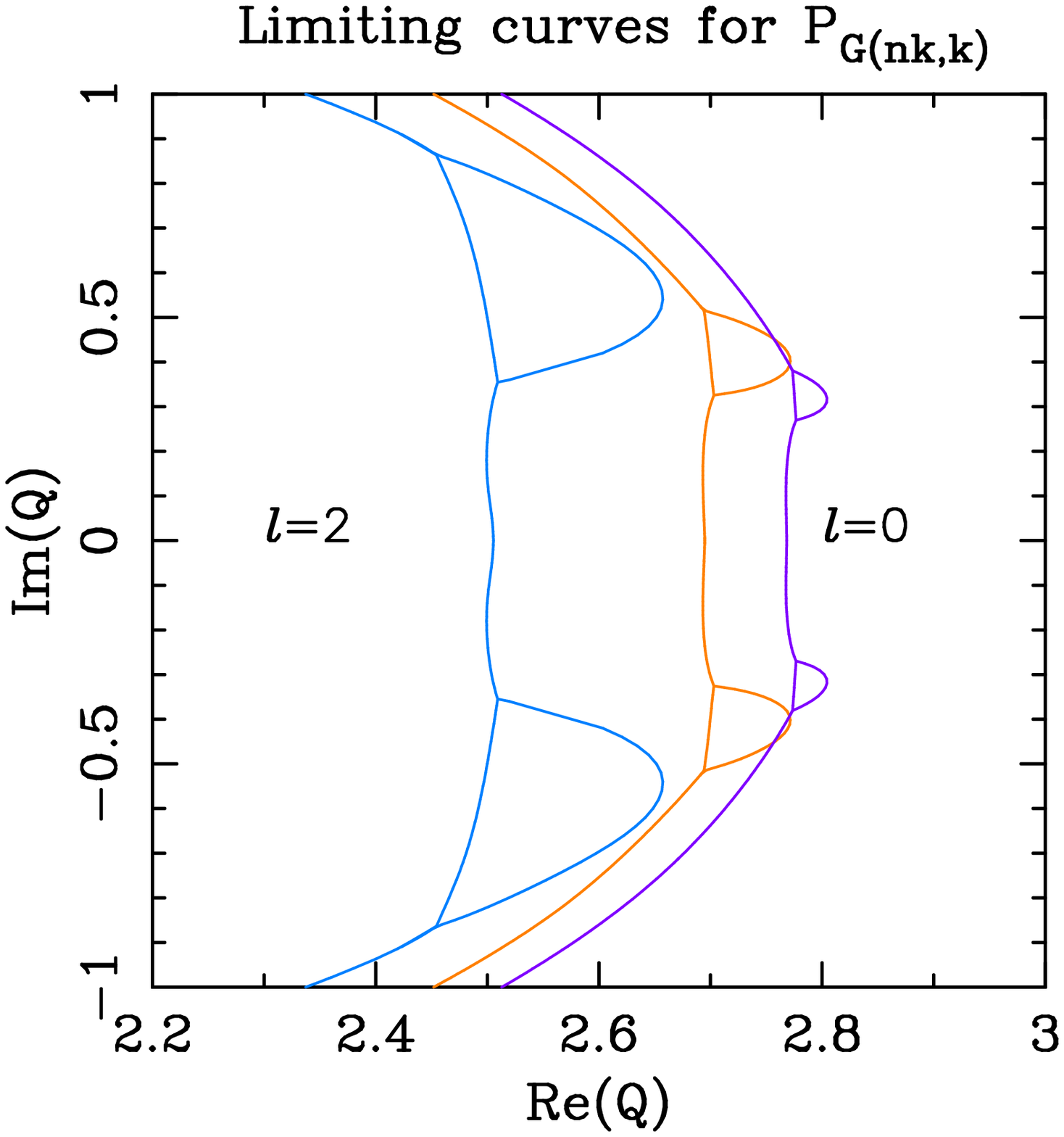} \\
  \qquad (a) & \qquad (b) \\[2mm]
  \end{tabular}
  \caption{
  (a) Limiting curves $\mathcal{B}_k$ for the chromatic polynomial in the 
  complex $Q$-plane for the generalized Petersen graphs $G(nk,k)$ with 
  $k=1$ (black), $k=2$ (gray), $k=3$ (light blue), $k=4$ (green), 
  $k=5$ (orange), $k=6$ (pink), $k=7$ (violet), $k=8$ (red), and $k=9$ 
  (navy blue).  
  (b) Zoom for odd values of $k=3,5,7$,
  where we find two small complex-conjugate closed regions. 
  Inside them, the dominant eigenvalue comes from the $\ell=1$ sector. 
  Notice the change of scale with respect to panel (a). We do not show the
  curve for $k=1$, as then the small regions for $k=7$ would be difficult to 
  see. The labels are as in Figure~\ref{Figures_P1}.
  }
\label{Figure_P_all}
\end{figure}

%
%
\begin{table}[h]
\centering
\begin{tabular}{r|l}
\hline\hline
\multicolumn{1}{c}{$k$} & \multicolumn{1}{|c}{$Q_c^{(P)}(k)$} \\
\hline
 1 &  2 \\
 2 &  2.6383423072 \\
 3 &  2.5054257523 \\
 4 &  2.6971127211 \\
 5 &  2.6947536196 \\
 6 &  2.7631358388 \\
 7 &  2.7682556521 \\
 8 &  2.7983688222 \\
 9 &  2.8023368562 \\
10 &  2.8176338815 \\
11 &  2.8203525818 \\
\hline\hline
\end{tabular}
\caption{\label{table.QcChromatic}
Values of $Q_c^{(P)}(k)$ for the generalized Petersen graphs $G(nk,k)$ for
$1 \le k \le 11$.
}
\end{table}

Note that, contrary to the flow--polynomial case, the integer $Q=0$ is 
a (non-isolated) limiting point. The similarities of the limiting curves
can be appreciated from Figure~\ref{Figure_P_all}(a), where we show together the
limiting curves $\mathcal{B}_k$ for $1\le k\le 9$. The limiting curves 
seem to converge to some infinite-width curve $\mathcal{B}_\infty$, and the
picture is qualitatively very similar to those for the square--lattice
chromatic polynomial with both cyclic \cite{JScyclic} and toroidal 
\cite{JStorus} boundary conditions. For odd values of $k$, the limiting 
curve $\mathcal{B}_k$ displays two complex-conjugate regions on its right-most
part. Inside these regions, the dominant eigenvalue comes from
the $\ell=1$ sector. These regions are shown for $k=3,5,7$ in 
Figure~\ref{Figure_P_all}(b). It is clear that, as $k$ increases, they 
become smaller and closer to the real $Q$-axis. Therefore, we conjecture 
that they completely disappear in the limit $k\to\infty$.

Assuming that the dominant eigenvalue on the real $Q$-axis invariably
comes from the completely symmetric representation, it is possible to
extend the numerical determination of $Q_c^{(P)}(k)$ to higher values of
$k$. The resulting values of $Q_c^{(P)}(k)$ are shown in 
Table~\ref{table.QcChromatic}.

The values of $Q_c^{(P)}(k)$ show parity effects, so we fitted the data
for $k$ even and the data for $k$ odd separately. Using the same
techniques as for $Q_c^{(\Phi)}$, we obtain in this case  
\be
 \label{Qc.chromatic.Petersen}
 Q_c^{(P)} \;\equiv\; \lim_{k\to\infty} Q_c^{(P)}(k)
           \;=\; 2.861 \pm 0.003 \,.
\ee
The power-law fit to the odd-$k$ data gave a constant term 
$\approx 2.86(1)$, while for the even-$k$ data the result was $\approx 2.86(8)$.
If we used the full data set with the Ansatz \reff{Ansatz.fulldata}, we
obtained $\approx 2.861$ for $7\le k\le 11$, and $\approx 2.864$ 
for $6\le k\le 10$. The error bar was taken as the difference between 
the last two estimates.

%
%
\section{Discussion} \label{sec.final} 

In this concluding section we would like to make an unified picture
of the results presented in this paper, and to propose a number of
conjectures that should stimulate further research. We have seen that
the $Q$-state Potts model with temperature parameter $v$ comprises
both the flow polynomial ($v=-Q$), and the chromatic polynomial
($v=-1$) as special cases. It is therefore useful to compare our
results on the partition function and its flow-- and chromatic--polynomial
specializations for the particular class of non-planar graphs $G(nk,k)$ 
with what is known for the same objects evaluated on certain classes of 
planar graphs. Our graphs are taken to consist of $n$
identical layers of width $L$ vertices and we impose periodic boundary
conditions in the $n$-direction.

Before starting with the main discussion it is worth recalling the 
BKW theorem \cite{BKW75,BKW78,BK79,BKW80,Sokal04} in a simple way: 
Let $D$ be a domain (connected open set) $D$ in the complex
plane, and let $\alpha_1,\ldots,\alpha_M,\mu_1,\ldots,\mu_M$ ($M \ge 2$)
be analytic functions on $D$, none of which is identically zero.
Then, for each integer $n \ge 0$, define the functions 
\be
   f_n(z)   \;=\;   \sum\limits_{k=1}^M \alpha_k(z) \, \mu_k(z)^n
   \;.
\ee
Each function $f_n$ has a set of complex zeros, and we are interested 
in the accumulation set of these zeros in the limit $n\to\infty$. Then, 
under a mild  ``no-degenerate-dominance'' condition, these limiting zeros
can be characterized as follows: a point $z$ lies in this limiting set 
if and only if either
\begin{itemize}
 \item[(a)] There is a unique dominant eigenvalue $\mu_k$ at $z$ (i.e., 
            $|\mu_k| > |\mu_i|$ for all $i\neq k$), and $\alpha_k(z) =0$;
 \item[(b)] There are two or more dominant eigenvalues $\mu_k$ at $z$.
\end{itemize}
 
Let us now consider first the planar chromatic case. The general situation 
is expected to be the following:

\begin{conjecture} \label{conj.chrom.1}
  Consider a family $G_{n,L}$ of planar connected graphs, consisting of $n$
  identical layers of width $L$ vertices, with periodic boundary
  conditions in the $n$-direction. Let $\mathcal{A}_L$ (resp.\
  $\mathcal{B}_L$) denote the set of limiting points of chromatic
  zeros for $G_{n,L}$, in the limit $n \to \infty$, that satisfies
  condition (a) (resp.\ condition (b)) of the BKW Theorem.
  Further let $B_p$ be the $p$-th Beraha number \reff{def_Bn},
  and let
  $$
  B^{\rm even} \;=\; \left\lbrace Q = B_p \, \colon \, 
                      p \in 2 \N \right\rbrace \,, 
  \qquad
  B^{\rm odd} = \left\lbrace Q = B_p  \, \colon \, 
                      p \in 2 \N + 1 \right\rbrace
  $$
  denote, respectively, the set of even and odd Beraha numbers $B_p$. Then:
  \begin{enumerate}
   \item There exists, for each $L$, a number $Q_c^{(P)}(L) \in (0,4)$ so
         that
   \begin{enumerate}
    \item $\mathcal{A}_L \cap \R = B^{\rm odd} \cap [0,Q_c^{(P)}(L))$
    \item $\mathcal{B}_L \cap \R = B^{\rm even}\cap [0,Q_c^{(P)}(L)) 
                                               \cup \{Q_c^{(P)}(L)\}$
   \end{enumerate}
    \item For any $p \in N$, the dominant eigenvalue of the transfer matrix
          for
          $$
          Q \;\in\; (B_{2p},B_{2(p+1)}) \cap [0,Q_c^{(P)}(L))
          $$
          comes from the sector with $p$ marked clusters.
  \end{enumerate}
\end{conjecture}
Compelling evidence for Conjecture~\ref{conj.chrom.1} comes from the
results of Ref.~\cite{JScyclic}, where it was shown that 
Conjecture~\ref{conj.chrom.1} is correct for cyclic strips of the
square and triangular lattices with $L \le 8$. The corresponding
values of $Q_c^{(P)}(L)$ can be found in \cite{JScyclic}.\footnote{
    In Ref.~\cite{JScyclic} $Q_c^{(P)}(L)$ is simply denoted by 
    $Q_c(L)$, as there is no danger of confusion with the similar 
    quantity $Q_c^{(\Phi)}$ associated to the flow polynomial. 
}

The natural step suggested by these finite-$L$ results is to go to the 
infinite--volume limit, $\lim_{L \to \infty}$. We expect that for each family
of planar graphs $G_{n,L}$, defined as above, the limit 
$Q_c^{(P)} = \lim_{L \to \infty} Q_c^{(P)}(L)$ exists and satisfies 
$0 < Q_c^{(P)} < 4$ \cite{Saleur90,Saleur91,JScyclic}.
In particular for cyclic strip graphs of the square lattice, we have that
$Q_c^{(P)}=3$ \cite{Baxter82,AFPotts}. For cyclic strips of the triangular
lattice, we expect that either $Q_c^{(P)} \approx 3.8196$ 
\cite{Baxter86,Baxter87}, or $Q_c^{(P)}=B_{12}=2 + \sqrt{3}$ \cite{JSStri}.  

The deep meaning of these results is that the special role of the
Beraha numbers in the production of real chromatic zeros is brought
out whenever the chromatic line ($v=-1$) intersects the 
BK phase in the $(Q,v)$ plane of the Potts model \cite{Saleur90,Saleur91}. 
For each graph family, and for each $Q \in [0,4)$, it can be argued 
\cite{JSboundary} that the extent of the BK
phase is a finite non-empty interval $v_-(Q) < v < v_+(Q)$ with the
properties $v_+(0) = 0$ and $\lim_{Q \to 4} v_+(Q) = \lim_{Q \to 4} v_-(Q)$.  
In particular, other choices of the temperature curve $v(Q)$
are possible and might in some cases lead to higher values of $Q_c$
than those quoted above.\footnote{
  We use $Q_c$ in this case to note that the limit $n\to \infty$ is taken 
  along a curve $v(Q)$. Indeed, when $v(Q)=-1$, $Q_c = Q_c^{(P)}$; and when
  $v(Q) = -Q$, $Q_c = Q_c^{(\Phi)}$.
}

The bound on $Q_c$ given above  
is even sharp: i.e., there exists a family of planar graphs $G_{n,L}$, defined 
as above, and a curve $v=f(Q)$ in the $(Q,v)$ plane, so that the limiting set
of zeros for the Potts-model partition function $Z_G(Q,v)$ satisfies
Conjecture~\ref{conj.chrom.1} and $Q_c=4$.   
Indeed, this is the case \cite{Saleur90,Saleur91,JSboundary} for
cyclic strips of the square lattice with the choice \cite{Baxter73}
$v^2 = Q$ for $0 \le Q \le 4$ and $-2 \le v \le 0$.  Another example
is provided by cyclic strips of the triangular lattice with the choice
\cite{Baxter78} $v^3 + 3 v^2 = Q$ for $0 \le Q \le 4$ and $-2 \le v
\le 0$.

As the amplitudes do not depend on $v$, we can extend 
Conjecture~\ref{conj.chrom.1} to the zeros of the full partition function
$Z_{G(n,L)}(Q,v)$ in the $(Q,v)$ plane:

\begin{conjecture} \label{conj.chrom.2}
  Consider a family $G_{n,L}$ of planar connected graphs, consisting of $n$
  identical layers of width $L$ vertices, with periodic boundary
  conditions in the $n$-direction. Let the sub-sets of Beraha numbers
  $B^{\rm even}$ and $B^{\rm odd}$ be as in Conjecture~\ref{conj.chrom.1}.
  Then:
  \begin{enumerate}
   \item There are two curves $v_\pm(Q)$ such that $v_-(Q) \le v_+(Q)$, 
         $v_+(0) = 0$, and a number $Q_0 \le 4$, such that 
         $\lim_{Q \to Q_0} v_+(Q) = \lim_{Q \to Q_0} v_-(Q)$.
  \end{enumerate}
  Furthermore, consider any smooth curve $v(Q)$ in between these two 
  curves: $v_-(Q) \le v(Q) \le v_+(Q)$ with $Q\in [0,Q_0]$. 
  Let us denote its
  graph in the $(Q,v)$ plane as $\mathcal{C}$.  
  Let $\mathcal{A}_L$ (resp.\ $\mathcal{B}_L$) denote the set of 
  limiting points of partition-function zeros for $G_{n,L}$, in the limit 
  $n \to \infty$, that satisfies
  condition (a) (resp.\ condition (b)) of the BKW Theorem. Then:
  \begin{enumerate}
   \setcounter{enumi}{1}
   \item For each curve $v(Q)$ with the above properties, there exists, 
         for each $L$, a number $Q_c(L) \in (0,Q_0)$ so that
   \begin{enumerate}
    \item $\Re(\mathcal{A}_L \cap \mathcal{C})= B^{\rm odd} \cap [0,Q_c(L))$
    \item $\Re(\mathcal{B}_L \cap \mathcal{C})= B^{\rm even}\cap [0,Q_c(L)) 
                                                         \cup \{Q_c(L)\}$
   \end{enumerate}
    \item For any $p \in N$, the dominant eigenvalue of the transfer matrix
          for
          $$
          \Re Q \;\in\; (B_{2p},B_{2(p+1)}) \cap [0,Q_c(L)) \, 
          $$
          and $\Im Q$ in between the curves $v_\pm(Q)$, 
          comes from the sector with $p$ marked clusters.
  \end{enumerate}
\end{conjecture}

Let us finally mention that Conjectures~\ref{conj.chrom.1} 
and~\ref{conj.chrom.2} enjoys strong support from the special representation 
theory of the quantum group at roots of unity \cite{Saleur90,Saleur91}.

\bigskip

\noindent
{\bf Remark.} A natural question is to determine the value of the number 
$Q_0$ defined in the previous conjecture for planar strip graphs. From 
the exact results for the free energy of the $Q$-state Potts antiferromagnet 
on the square, triangular, and hexagonal lattices
\cite{Baxter73,Baxter78,Baxter82,Baxter_book,Baxter86,Baxter87}, one
would be tempted to claim that $Q_0=4$. This value is also supported 
by the numerical results for the $(4,8^2)$ and $(3,12^2)$ lattices, obtained
using the Jacobsen--Scullard method \cite{Scullard13}. However, the
same method \cite{Scullard12,Scullard13} suggest that $Q_0< 4$ for the
kagome lattice. Therefore, we prefer to be conservative and not make any 
firm conjecture about the value of $Q_0$ for general planar lattices. 

\bigskip

Turning now to the partition-function zeros of non-planar graphs,
Figures~\ref{Figure_Z_all}, \ref{Figure_flow_all}, and~\ref{Figure_P_all},
Conjectures~\ref{conj.flow2} and~\ref{conj.chromatic1}, and the observations
made in Section~\ref{sec.num.res}, makes us confident that the statements of 
Conjecture~\ref{conj.chrom.1} hold
true also in the partition-function case for non-planar graphs, 
provided that one replaces even/odd Beraha
numbers by even/odd integers. On a fundamental level, the replacement
of Beraha numbers by integers stands out most clearly by comparing the
eigenvalue amplitudes in the two cases [see, e.g.,
\cite[Eqs.~(2.28)--(2.29)]{JScyclic} and Eq.~(\ref{eigen_amp})].
To be precise:

\begin{conjecture} \label{conj.petersen.1}
  Consider a family $G_{n,L}$ of non-planar connected graphs, consisting of $n$
  identical layers of width $L$ vertices, with periodic boundary
  conditions in the $n$-direction. Then:
  \begin{enumerate}
   \item There are two curves $v_\pm(Q)$ such that $v_-(Q) \le v_+(Q)$, 
         $v_+(0) = 0$, and there is a value $Q_0$ (possibly $Q_0=\infty$), 
         such that $\lim_{Q \to Q_0} v_+(Q) = \lim_{Q \to Q_0} v_-(Q)$.
  \end{enumerate}
  Furthermore, consider any smooth curve $v(Q)$ in between these two 
  curves: $v_-(Q) \le v(Q) \le v_+(Q)$ with $Q\in [0,Q_0]$. Let us denote its
  graph in the $(Q,v)$ plane as $\mathcal{C}$.  
  Let $\mathcal{A}_L$ (resp.\ $\mathcal{B}_L$) denote the set of 
  limiting points of partition-function zeros for $G_{n,L}$, in the limit 
  $n \to \infty$, that satisfies
  condition (a) (resp.\ condition (b)) of the BKW Theorem. Then:
  \begin{enumerate}
   \setcounter{enumi}{1}
   \item For each curve $v(Q)$ with the above properties, there exists, 
         for each $L$, a number $Q_c(L) \in (0,Q_0)$ so that
   \begin{enumerate}
    \item $\Re(\mathcal{A}_L \cap \mathcal{C})= (2\N-1) \cap [0,Q_c(L))$
    \item $\Re(\mathcal{B}_L \cap \mathcal{C})=  2\N    \cap [0,Q_c(L)) 
                                                        \cup \{Q_c(L)\}$. 
          For some curves $v(Q)$ the value $0$ might not be present 
          in this set.
   \end{enumerate}
    \item For any $p \in N$, the dominant eigenvalue of the transfer matrix
          for
          $$
          \Re Q \;\in\; (2(p-1),2p) \cap [0,Q_c(L)) \, 
          $$
          and $\Im Q$ in between the curves $v_\pm(Q)$, 
          comes from the fully symmetric irreducible representation $(p)$
          of the sector with $p$ marked clusters.
  \end{enumerate}
\end{conjecture}

This conjecture has been validated in this paper for $G_{n,L} =
G(nk,k)$, the generalized Petersen graphs, with $L=k+1$, and $k\le 7$ 
for the flow polynomial $v(Q)=-Q$, and the chromatic polynomial $v(Q)=-1$.  
We have also presented numerical evidence that the limits $Q_c$ exist 
along these two curves $v(Q)$. In addition, we have directly checked the
consistency of this picture (at least for $k\le 7$) by considering directly 
the partition-function zeros along several lines $v(Q)=-pQ$, $v(Q)=-Q/p$, and
$v(Q)=-p$ with $p$ a positive integer (see Section~\ref{sec.num.res}).  

\medskip

\noindent
{\bf Remark.} 
The set $\Re(\mathcal{B}_L \cap \mathcal{C})$ does not contain the value $0$
for the flow polynomial; but it contains this value for the 
chromatic-polynomial case. The difference might be due to the fact that 
when $Q<0$ we enter in the ferromagnetic regime in the former case, while 
we stay in the antiferromagnetic or unphysical regime in the latter case.  

\bigskip

To conclude, we recall that for the Potts model on planar graphs
$G_{n,L}$, when $Q = \big( 2 \cos(\pi/p) \big)^2$ is a Beraha number
the representation theory of the underlying quantum algebra
\cite{Pasquier} leads to massive degeneracies in the spectrum of the
transfer matrix. In particular, depending on $p$, complete sectors of
eigenvalues are contained in other sectors in an inclusion-exclusion
fashion, completely independent of the value of $v$. We have seen in
Section~\ref{sec.integers} that a similar phenomenon occurs for the
Potts model on {\em non-planar} graphs when $Q$ is a non-negative 
integer. There are striking parallels between the detailed inclusion-exclusion
scenarios in the two cases, once again provided that one ``translates''
between the two cases by replacing the Beraha numbers by non-negative
integers.

\appendix
%
%
\section{Potts--model partition function for the simple-cubic graphs} 
\label{sec.sc} 

A natural question is that our conclusions are based on a rather particular
family of non-planar graphs. In this appendix we want to examine a more 
``physical'' family of graphs: the graphs $\text{Sc}(L,n)$ formed by a 
simple cubic graph of section of size $L\times L\times n$ with cyclic 
boundary conditions (i.e., free boundary conditions in the transverse 
``space-like'' two--dimensional layers, and periodic boundary conditions
in the longitudinal ``time-like'' direction). The chromatic polynomial
for this family was previously considered in Ref.~\cite{Shrock_01}.

The graph $\text{Sc}(L,n)$ contains $L^2 n$ vertices,
$2(L-1)Ln$ horizontal edges, and $L^2 n$ vertical edges. If we denote
the vertices in $\text{Sc}(L,n)$ as points in $\Z_+^3$, the vertex set
is
\be
V(\text{Sc}(L,n)) \;=\; \{ (x,y,z) \mid  0\le x,y\le L-1\,, \;
                                         0\le z  \le n-1 \} \,,
\ee
and the edge set is the union of the sets
$E(\text{Sc}(L,n))=E_x \cup E_y \cup E_z$, where
\begin{subeqnarray}
E_x &=& \{ ((x,y,z), (x+1,y,z)) \mid 0\le x\le L-2 \,, \, 0\le y\le L-1\,,
                                       \; 0\le z \le n-1 \}\quad\\
E_y &=& \{ ((x,y,z), (x,y+1,z)) \mid 0\le x\le L-1 \,, \, 0\le y\le L-2\,,
                                       \; 0\le z \le n-1 \}\quad\\
E_z &=& \{ ((x,y,z), (x,y,z+1)) \mid 0\le x,y\le L-1 \,, \;
                                       0\le z \le n-1 \}
\end{subeqnarray}
where we have identified $z=n$ and $z=0$.

We can apply the transfer-matrix formalism of Section~\ref{sec.tm} to the 
family $\text{Sc}(L,n)$. If we label the vertices on a horizontal layer as 
$(x,y)$ with $0\le x,y\le L-1$, then the transfer matrix takes the form
${\sf T}_L = {\sf V} \cdot {\sf H}$ with  
\begin{subeqnarray}
 {\sf H}   &=& \prod\limits_{y=0}^{L-1} \prod_{x=0}^{L-2}
               {\sf H}_{(x,y),(x+1,y)} \cdot
               \prod\limits_{x=0}^{L-1} \prod_{y=0}^{L-2}
               {\sf H}_{(x,y),(x,y+1)} \\
 {\sf V}   &=& \prod\limits_{y=0}^{L-1} \prod_{x=0}^{L-1} {\sf V}_{(x,y)}
 \label{TM_decomp_sc}
\end{subeqnarray}
The order of the operators ${\sf H}_{(x,y),(x+1,y)}$ and
${\sf H}_{(x,y),(x,y+1)}$ in (\ref{TM_decomp_sc}a) is of no importance,
as all the horizontal operators commute. The same also holds for the
vertical operators ${\sf V}_{(x,y)}$ in
(\ref{TM_decomp_sc}b). However, the operators ${\sf H}$ and ${\sf V}$ do
not commute.

The computation of the partition function for the graph $\text{Sc}(L,n)$
is quite demanding: the number of partitions we have to deal with is given
by the Bell number $B_{2L^2}$ 
\cite[and references therein]{JS_flow}.\footnote{
  Please, do not confuse this Bell number with a Beraha number \reff{def_Bn}.
} 
This number grows very fast as a function of $L$.
The formula for the partition function for $G=\text{Sc}(L,n)$ is
given in terms of traces of the relevant diagonal blocks of the  
full transfer matrix ${\sf T}_{L}$: 
\be
Z_{\text{Sc}(L,n)}(Q,v) \;=\; \sum\limits_{\ell=0}^{L^2}
   \sum\limits_{\lambda\in S_\ell} \alpha_{\ell,\lambda} \, 
   \tr \left( {\sf T}_{L,\ell,\lambda} \right)^n \,.   
\label{def_Z_Sc}
\ee
Some of the eigenvalues of the transfer matrices 
${\sf T}_{L,\ell,\lambda}$ may coincide.

In practice, we have been able to compute only the exact partition
function for $L=2$. The structural properties of the transfer matrix
${\sf T}_2$ can be summarized as follows:
\begin{itemize}
 \item The trivial eigenvalue $\mu_{2,4}=v^4$ appears for all values of
       the number of links $0\le \ell \le L^2=4$ with multiplicities
       $2,6,12,24$. This implies that the corresponding amplitude is
       $\gamma_4 = Q^2-8Q^3+20Q^2-15Q+1$.
 \item For $\ell=0$ there are two eigenvalues that appear twice; therefore
       in the ``complete'' representation of $Z_{\text{Sc}(2,n)}$, they
       should have the amplitude $2\alpha_0 = 2$.
 \item For $\ell=1$, there are eight eigenvalues that appear twice;
       therefore their amplitude is $2\alpha_1 = 2(Q-1)$.
 \item For $\ell=2$, there are four eigenvalues that appear twice in both
       representations $(2)$, and $(1,1)$. Its amplitude is
       $\alpha_{2,(2)}+\alpha_{2,(1,1)}=2(Q^2-3Q+1)$. In addition, there
       are six eigenvalues in the representation $(2)$ that appear twice;
       and another six eigenvalues in $(1,1)$ that appear twice, too. Their
       corresponding amplitude should be twice the value $\alpha_{1,\lambda}$.
 \item For $\ell=3$, the representations $(3)$ and $(1,1,1)$ have exactly
       the same eigenvalues. Their amplitude is $(Q-1)(Q^2-5Q+3)/3$. Two
       of these eigenvalues appear twice in each representation, therefore
       their amplitude is twice the latter value.
       The representation $(2,1)$ has all its eigenvalues doubled; hence
       their amplitude is $2Q(Q-2)(Q-4)/3$.
\end{itemize}
The number of non-trivial distinct eigenvalues in each
$\ell$ sector are $12, 27, 39, 14$ for $\ell=0,1,2,3$, respectively.
We then obtain a ``complete''  decomposition of $Z_{\text{Sc}(2,n)}$
in terms of distinct eigenvalues $\mu_{L,\ell,\lambda,s}$. The full
expression is rather cumbersome, so we refrain from writing it
explicitly here.

%
%
\begin{figure}
  \centering
  \includegraphics[width=200pt]{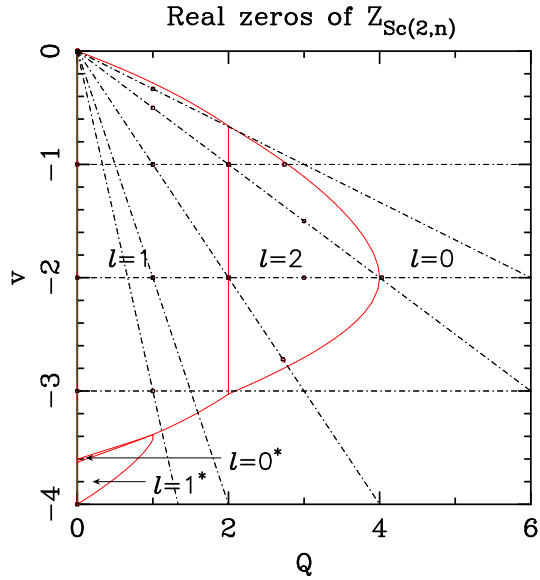}
  \caption{
  Real zeros of the Potts-model partition function
  and limiting curves $\mathcal{B}_2$ in the plane $(Q,v)$  
  for the simple-cubic graphs $\text{Sc}(2,n)$ for 
  $n=9$ (black $\square$), and $n=10$ (red $\circ$). 
  The rest of the notation is as in Figure~\ref{Figures_Z1}.
  }
\label{Figure_sc_L=2}
\end{figure}


We have computed the zeros of the partition function $Z_{\text{Sc}(2,n)}(Q,v)$
along the same lines in the real $(Q,v)$ plane as for the generalized Petersen
graphs. Figure~\ref{Figure_sc_L=2} shows these zeros for
the simple--cubic graphs $\text{Sc}(2,9)$ and $\text{Sc}(2,10)$.

We have also computed the limiting curve $\mathcal{B}_2$ in the real
$(Q,v)$ plane; this is depicted in Figure~\ref{Figure_sc_L=2}.
The phase structure for this family of graphs is similar to that described
for the generalized Petersen graphs. In particular, we can observe two
phases enclosed by the limiting curve $\mathcal{B}_2$ and characterized
by $\ell=1$ and $\ell=2$ respectively. The former contains the interval
$Q\in(0,2)$, and the latter, the interval $Q\in(2,4)$. In each phase, the
dominant eigenvalue comes from the fully symmetric representation of the
corresponding group $S_\ell$. Therefore, as the expressions for these
eigenvalues are the same as for the generalized Petersen graphs, the
values $Q=1$ and $Q=3$ are isolated limiting curves. Notice that there is no
isolated limiting point at $Q=B_5$. 

Finally, we observe close to the lower boundary of the BK phase, two regions
with dominant pairs of complex-conjugate eigenvalues. This feature was 
absent in the Petersen graphs. 

In conclusion, we find a phase diagram for this family of non-planar graphs
which is qualitatively similar to that of the Petersen graphs. This supports
our belief that the results we have found for the Petersen graphs are general
for at least all families of recursive non-planar graphs.

\section*{Acknowledgments}

We thank Gordon Royle for suggesting us to study the generalized Petersen
graphs, and Alan Sokal for many illuminating discussions.  

Both authors also thank the Isaac Newton Institute for Mathematical
Sciences, University of Cambridge, for hospitality during the
programme on Combinatorics and Statistical Mechanics (January--June
2008), where this project started.  J.S. also warmly thanks the \'Ecole 
Normale Sup\'erieure for hospitality in June 2009, June 2011, and December 
2012.

The research of J.L.J. was supported in part by the Agence Nationale
de la Recherche (grant ANR-10-BLAN-0414: DIME) and the Institut
Universitaire de France.
The research of J.S. was supported in part by Spanish MINECO grants
FIS2012-34379 and MTM2011-24097, and by US National 
Science Foundation grant PHY--0424082.

\end{document}